\pdfoutput=1 
\documentclass{jfm}

\usepackage{graphicx}
\usepackage{newtxtext}
\usepackage{newtxmath}
\usepackage{natbib}
\usepackage{hyperref}
\usepackage{xcolor}
\usepackage[acronym]{glossaries}
\usepackage{subfig}
\hypersetup{
    colorlinks = true,
    urlcolor   = blue,
    citecolor  = black,
    linkcolor  = black
}

\newcommand{\RomanNumeralCaps}[1]
\linenumbers

\newacronym{sbli}{SBLI}{Shock Wave/Boundary Layer Interaction}
\newacronym{rans}{RANS}{Reynolds-Averaged Navier--Stokes Simulation}
\newacronym{piv}{PIV}{Particle Image Velocimetry}
\newacronym{les}{LES}{Large Eddy Simulation}
\newacronym{kh}{K--H}{Kelvin--Helmholtz}
\newacronym{iles}{ILES}{Implicit Large Eddy Simulation}
\newacronym{lda}{LDA}{Laser Doppler Anemometry}
\newacronym{dns}{DNS}{Direct Numerical Simulation}
\newacronym{hpc}{HPC}{High Performance Computing}
\newacronym{vg}{VG}{Vortex Generator}
\newacronym{mvg}{MVG}{Microvortex Generator}
\newacronym{ibm}{IBM}{Immersed Boundary Method}
\newacronym{streams}{STREAmS 2.0}{Supersonic TuRbulEnt Accelerated navier stokes Solver 2.0}
\newacronym{rms}{rms}{root mean square}
\newacronym{psd}{PSD}{Power Spectral Density}
\newacronym{weno}{WENO}{Weighted Essentially Non-Oscillatory}
\newacronym{pdf}{pdf}{Probability Density Function}
\newacronym{gpu}{GPU}{Graphics Processing Unit}
\newacronym{gfm}{GFM}{Gallery of Fluid Motion}
\newacronym{dfd}{DFD}{Division of Fluid Dynamics}

\definecolor{lightgr}{rgb}{0.85,0.85,0.85}
\definecolor{darkgr}{rgb}{0.7,0.7,0.7}
\definecolor{grey1}{rgb}{0.2 0.2 0.2}
\definecolor{grey2}{rgb}{0.45 0.45 0.45}
\definecolor{grey3}{rgb}{0.6 0.6 0.6}

\title{Direct numerical simulation of supersonic boundary layers over a microramp: \\ effect of the Reynolds number}

\author{G. Della Posta\aff{1}
  \corresp{\email{giacomo.dellaposta@uniroma1.it}},
  M. Blandino\aff{1}, 
  D. Modesti\aff{2},
  F. Salvadore\aff{3}
 \and M. Bernardini\aff{1}}

\affiliation{\aff{1}Department of Mechanical and Aerospace Engineering, Sapienza University of Rome, via Eudossiana 18, 00184, Rome, Italy
\aff{2} Aerodynamics group, Faculty of Aerospace Engineering, Delft University of Technology, Kluyverweg 2, 2629 HS Delft, The Netherlands
\aff{3} HPC Department, CINECA, via dei Tizii 6/B, 00185, Rome, Italy}

\begin{document}
\maketitle

\begin{abstract}
Microvortex generators are passive control devices smaller than the boundary layer thickness that energise the boundary layer to prevent flow separation with limited induced drag. In this work, we use direct numerical simulations (DNSs) to investigate the effect of the Reynolds number in a supersonic turbulent boundary layer over a microramp vortex generator. Three friction Reynolds numbers are considered, up to $Re_\tau = 2000$, for fixed free-stream Mach number $M_\infty=2$ and fixed relative height of the ramp with respect to the boundary layer thickness. The high-fidelity data set sheds light on the instantaneous and highly three-dimensional organisation of both the wake and the shock waves induced by the microramp. The full access to the flow field provided by DNS allows us to develop a qualitative model of the near wake, explaining the internal convolution of the Kelvin--Helmoltz vortices around the low-momentum region behind the ramp. 
The overall analysis shows that numerical results agree excellently with recent experimental measurements in similar operating conditions
and confirms that microramps effectively induce a significantly fuller boundary layer even far downstream of the ramp.
Moreover, results highlight significant Reynolds number effects, which in general do not scale with the ramp height. 
Increasing Reynolds number leads to enhanced coherence of the typical vortical structures in the field, faster and stronger development of the momentum deficit region, increased upwash between the primary vortices from the sides of the ramp --- and thus increased lift-up of the wake --- and faster transfer of momentum towards the wall. 
\end{abstract}


%
%
\section{Introduction}
\label{sec:introduction}
%
The interaction between shock waves and boundary layers has 
been widely studied during the last 70 years and is still the subject of extensive research, 
because of its frequent occurrence in many aerospace systems \citep{dolling2001fifty, gaitonde2015progress}. 
\glspl{sbli} can be found in a wide range of flows including transonic airfoils, high-speed inlets, control surfaces of high-speed aircraft, space launchers base flows, and overexpanded nozzles. 
In the interaction region, extreme, unsteady, mechanical and thermal loads are typically observed \citep{clemens20145low}, 
and the adverse pressure gradient imposed by the shock often causes the boundary layer to separate intermittently \citep{bernardini2023unsteadiness}. 
This problem represents, for example, a critical issue for high-speed intakes, since the low-momentum separated flow affects the 
mass flow rate entering the engine, producing flow distortions, total pressure losses, and potential inlet unstarts \citep{herrmann2002experimental}. 

Given its pivotal importance, it is evident that alleviating the detrimental effects of \gls{sbli} 
by means of control devices would have a significant impact on the performance and the reliability of many aerospace engineering systems \citep{babinsky2008sbli}. 
The most widespread solution to delay shock-induced separation in high-speed intakes, and to control \gls{sbli} in turn, 
is the so called boundary layer bleed, which consists in diverting the separated flow near to the wall out of the engine. 
Removing low-momentum fluid makes the boundary layer more resistant to shock-induced separation and reduces the distortion of the 
flow ahead of the engine too. However, since the bled air is not re-injected elsewhere, the flow rate is reduced up to the 15-20\%
\citep{loth2000smart}, requiring an increase in the intake frontal area and, in turn, an increase in the aerodynamic drag of the vehicle. 
Other solutions have thus been explored, like classical \glspl{vg} \citep{davis1968performance}, especially with vane-type form.
Classical \glspl{vg}, whose heights are larger than the boundary layer thickness, generate passively trailing vortices that bring high-momentum flow from outside the boundary layer to the near-wall region. However, despite their benefits, \glspl{vg} generate considerable additional drag.

To leverage the same mechanism of \glspl{vg} and energise the near-wall region of the boundary layer upstream of the interaction, 
\glspl{mvg} have been proposed \citep{wheeler1984means, mccormick1993shock}, with heights smaller than the boundary layer thickness. 
Compared to their traditional counterpart, \glspl{mvg} offer the same control efficiency but with a decreased wave drag, given their reduced
height \citep{lin2002review}. Among the various configurations proposed, although microvanes seem to be slightly more effective, microramps 
are typically preferred because of their superior mechanical robustness. In fact, \glspl{mvg} are usually mounted upstream of the \gls{sbli}, 
and thus, in case of failure, their ingestion would bring catastrophic consequences for the inlet. 

\begin{figure*}
     \centering
     \includegraphics[width=\textwidth]{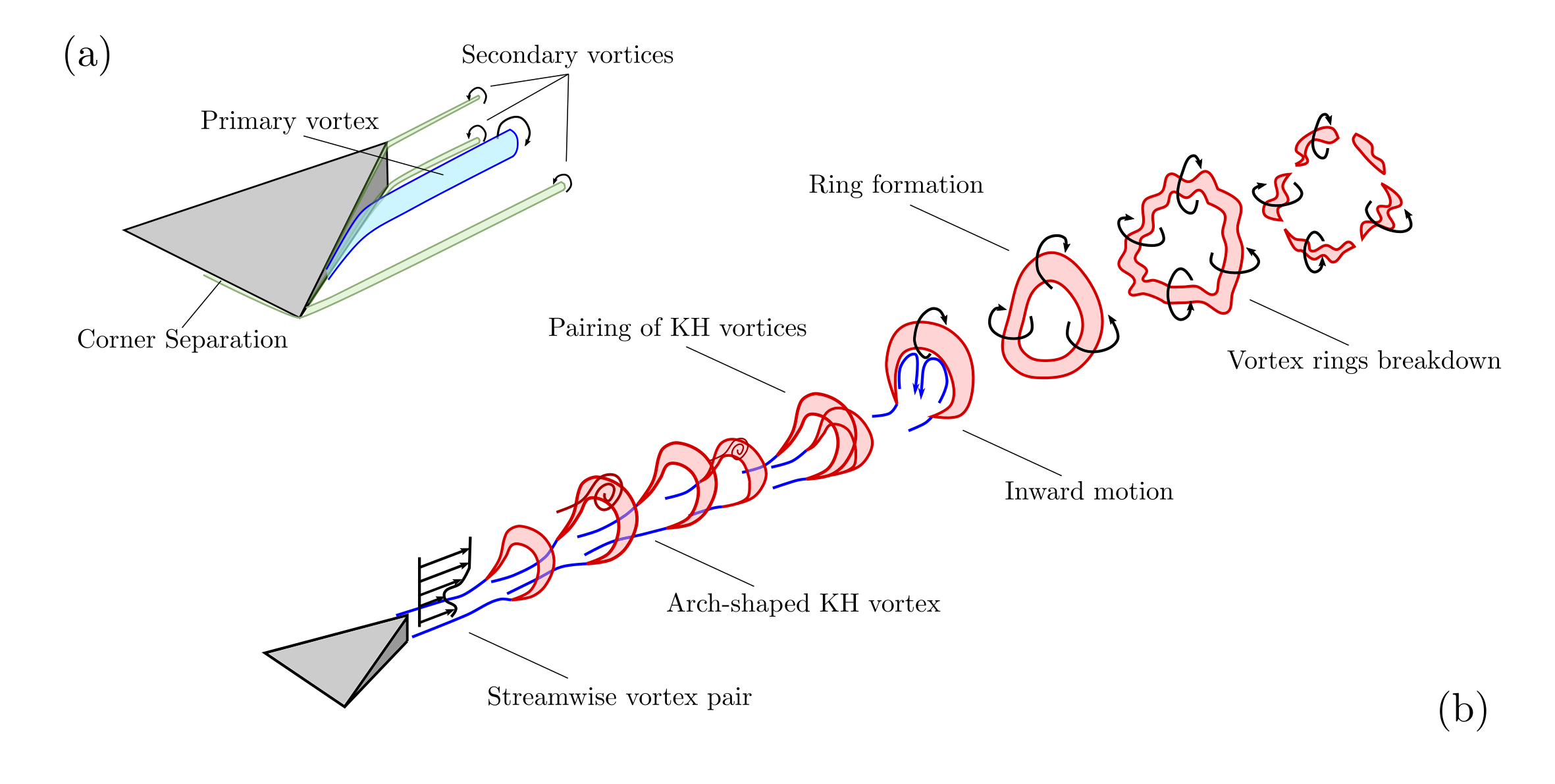}
     \caption{Mean vortical structures in the model of \citet{babinsky2009microramp} (a) and 
     instantaneous flow organisation according to the model of \citet{sun2014decay} (b).}
     \label{fig:intro_fig}
\end{figure*}

During the last twenty years, research focused particularly on the characterisation of the wake downstream of 
microramps and on its influence on \gls{sbli} separation \citep{lu2012microvortex, panaras2015micro}. 
As a matter of fact, the control mechanism of microramps relies on how these change the flow topology upstream of the interaction. 
Therefore, a deep understanding of the complex structure of the wake produced by a microramp in a high-speed boundary layer 
is precious 
and preliminary to the assessment of any effect \glspl{mvg} may have on \gls{sbli}. 
The first extensive description of the mean flow behind an \gls{mvg} is due to \citet{pitt2007micro} and \citet{babinsky2009microramp}. 
The authors analysed through surface oil visualisations the different large scale structures emerging from the 
interaction between a microramp and a supersonic boundary layer. 
Firstly, the incoming flow is subjected to compression on the microramp leading edge, 
which leads to a small pocket of separated flow and to the development of a horseshoe vortex. 
Then, the fluid flows on top of the ramp, turning laterally toward the ramp edges, where it separates from the device 
and generates a pair of counter-rotating vortices on the two ramp sides, the so-called primary vortex pair. 
The vortex pair brings high-momentum flow from the outer region of the boundary layer down towards the wall, inducing 
a fuller boundary layer which is less prone to separation. 
In addition to these primary structures, a series of secondary vortices also originates 
from the bottom and top edges of the device \citep{lu2010experimental}.
Downstream of the microramp trailing edge, the two primary vortices become parallel and auto-induce a vertical 
force in the plane of symmetry
that causes the wake and the vortex pair to lift up. 
Therefore, moving downstream, the wake lift-up, together
with viscous and turbulent dissipation, progressively
weaken the energising mechanism of the wake, which
eventually returns to the undisturbed conditions.
A qualitative representation of the mean vortex structure 
behind the microramp is reported in figure \ref{fig:intro_fig}(a).

Other studies have instead investigated the instantaneous organisation of the flow field behind the microramp.
\citet{blinde2009effects} analysed the effect of an array of microramps on a boundary layer at $M_\infty=1.84$ 
through stereoscopic \gls{piv}. From the distribution of the velocity on two planes parallel to the wall, 
the authors inferred the presence of a train of hairpin vortices surrounding the primary streamwise vortices. 
\glspl{iles} of a microramp in a turbulent boundary layer at $M_\infty=2.5$ by \citet{li2011implicit} revealed instead 
the existence of a train of ring-shaped vortices surrounding the low-momentum region
that induces significant wake meandering. 
According to the authors, these large scale structures are generated by a \gls{kh} instability
originating from the shear layer that envelops the momentum deficit in the microramp wake. 
In their experimental analysis of a boundary layer at $M_\infty=2.7$, \citet{bo2012experimental} observed that these 
\gls{kh} vortices are characterised by a frequency between $50$ KHz and $60$ KHz, which is transferred to the reflected shock donwstream, 
as the intermittent structures pass through the interaction zone and deform the shock system. 
Vortex rings from microramps pass across the \gls{sbli} without losing their coherence and
disrupt the separation bubble according to their shedding frequency, which thus makes them 
a key component in the control mechanism of \gls{sbli} \citep{grebert2018simulations}. 
To better understand the evolution of the wake behind microramps and to solve the controversy in the different vortex models proposed by many authors,
an extensive investigation on the flow generated by a microramp in a turbulent boundary layer at $M_\infty = 2.0$
was carried out through tomographic \gls{piv} and \gls{iles} in \citet{sun2012three, sun2013numerical, sun2014decay}. 
Based on the previous findings and on their results, the authors proposed a conceptual model (see figure \ref{fig:intro_fig}(b)) 
in which arc-shaped \gls{kh} vortices surrounding the primary vortex pair develop at first, right behind the ramp. 
As the lift-up induced by the primary vortex pair acts on the wake, the arc legs are subjected to an inward movement 
that leads to the formation of closed vortex rings. 
These periodic entities modulate then the primary vortex pair and finally undergo turbulent distortion that eventually breaks them down.

Even if comprehensive experimental studies like the ones presented above are paramount, their inherent restrictions in terms of spatial accuracy and data accessibility limit their ability to fully understand how \glspl{mvg} may control \gls{sbli}. For instance, in many studies, flow features are observed only on a limited number of planes, like the vertical symmetry plane. However, the field is highly three-dimensional, and the interpretation of the measurements on these surfaces may be misleading with respect to the global flow structure. High-fidelity numerical simulations represent thus a valid complementary tool that has proved effective to overcome such limitations. As a result, detailed numerical data may help address the physical questions remained in some way open, like the effect of the Mach and Reynolds numbers on the strength and decay of the time-averaged and instantaneous wake features or the precise role that \gls{kh} instability has in the control mechanism of \gls{sbli}. 

To the authors’ knowledge, a limited number of numerical studies has been carried out to fully comprehend 
the flow alteration induced by a microramp, and mainly using \glspl{rans} \citep{ghosh2010numerical}
or \glspl{iles} 
\citep{lee2009supersonic, li2010declining, li2011implicit, lee2011effect, sun2014numerical}, with momentum thickness-based Reynolds numbers $Re_\theta \in [1400, 5800]$. 
Both numerical techniques rely on closure assumptions,
which can lead to model-dependent results, especially
in the case of \gls{rans} of separated flows.
Despite the higher uncertainty of \gls{rans}, however, 
this tool remains the only possibility to explore a large parameter space.
For example, \citet{anderson2006optimal} carried out a multi-objective optimisation of the microramp 
geometry using \gls{rans}, with the aim of minimising the total pressure loss across the 
\gls{sbli} and the boundary layer shape factor $H$. 
The result has been deliberately taken as the baseline, optimal case by the majority of the cited studies.

Although similar methodologies allowed researchers to better describe the evolution of the main vortical structures and to refine models based on experiments, they are not fully able to capture the unsteady and multiscale nature of the flow under consideration.  However, thanks to the technological developments in \gls{hpc} and to the advent of the \gls{gpu} technology for scientific simulations, we are today able to perform \glspl{dns} with parameters that are comparable with those of real experiments. 

For this reason, we present here the results of a set of \glspl{dns} 
of a supersonic turbulent boundary layer over a microramp.
%
%
%
The geometry considered is the one of the optimal microramp defined by \citet{anderson2006optimal}, 
and free-stream Mach number $M_\infty=U_\infty/c_\infty$ is equal to 2, 
where $U_\infty$ and $c_\infty$ are the free-stream 
velocity and speed of sound, respectively.

We especially focus our attention on the effects of the Reynolds number.
Three flow cases are examined at friction Reynolds numbers 
$Re_\tau =\delta_{99}/\delta_v = 500$, $1000$, and $2000$, 
based on the properties of the undisturbed boundary layers at the ramp location,
where $\delta_{99}$ is the boundary layer thickness, $\delta_v=\nu_w/u_\tau$ is the viscous length scale, 
$u_\tau=\sqrt{\tau_w/\rho_w}$ is the friction velocity, $\rho_w$ and $\nu_w$ are the density 
and kinematic viscosity at the wall, respectively. 
The topic has been scarcely considered in the literature. 
In particular, only \citet{giepman2016mach} described the streamwise evolution of the main features 
of the mean wake velocity for Reynolds number based on the boundary layer thickness in the range 
$4.9 \cdot 10^5 \div 10.5 \cdot 10^5$ 
but the results did not show any relevant difference in the interval considered. 
However, despite the potential importance, no study is available at present concerning the effects of 
microramps at small or moderate Reynolds numbers. 

The paper is organised as follows: Section \ref{sec:method} presents the 
methodology and numerical setup of the simulations; Section \ref{sec:dataset} describes the 
database generated and the validation carried out; Section \ref{sec:results} 
presents the results of the analysis; finally, Section \ref{sec:conclusions} 
reports some final comments.
%
%
\section{Methodology and numerical setup}
\label{sec:method}
The \gls{dns} database considered in this work has been generated using
STREAmS 2.0\footnote{\url{https://github.com/STREAmS-CFD/STREAmS-2}}  \citep{bernardini2021streams,bernardini2023streams, sathyanarayana2023highspeed}.
STREAmS is an open-source finite-difference compressible flow solver developed by our group, designed to solve 
the compressible Navier--Stokes equations for a perfect, 
heat conducting gas. 
The solver targets canonical wall-bounded turbulent high-speed flows, 
and it is oriented to modern \gls{hpc} platforms with
the capability to run on both NVIDIA~\citep{bernardini2021streams, bernardini2023streams}
and AMD~\citep{sathyanarayana2023highspeed} GPUs. 

The discretisation of the convective terms adopts a hybrid energy-conservative/shock-capturing scheme 
in locally conservative form~\citep{pirozzoli2010generalized}.
In smooth regions, a sixth-order, central, energy-preserving flux formulation ensures stability 
without the addition of numerical diffusivity. 
Close to the shock waves, a fifth-order, \gls{weno} reconstruction~\citep{jiang1996efficient}
is used to compute numerical fluxes at the cell faces, using 
a Lax--Friedrichs flux vector splitting.
A shock sensor evaluates the local smoothness of the solution and identifies the regions where 
discontinuities occur, and where the switch between the central and the \gls{weno} scheme takes place. 
Viscous terms are approximated by means of sixth-order, central finite-difference approximations. 
Time advancement is carried out by means of a third-order, low-storage, Runge-Kutta scheme~\citep{spalart1991spectral}.   
\begin{figure*}
     \centering
     \includegraphics[width=\textwidth]{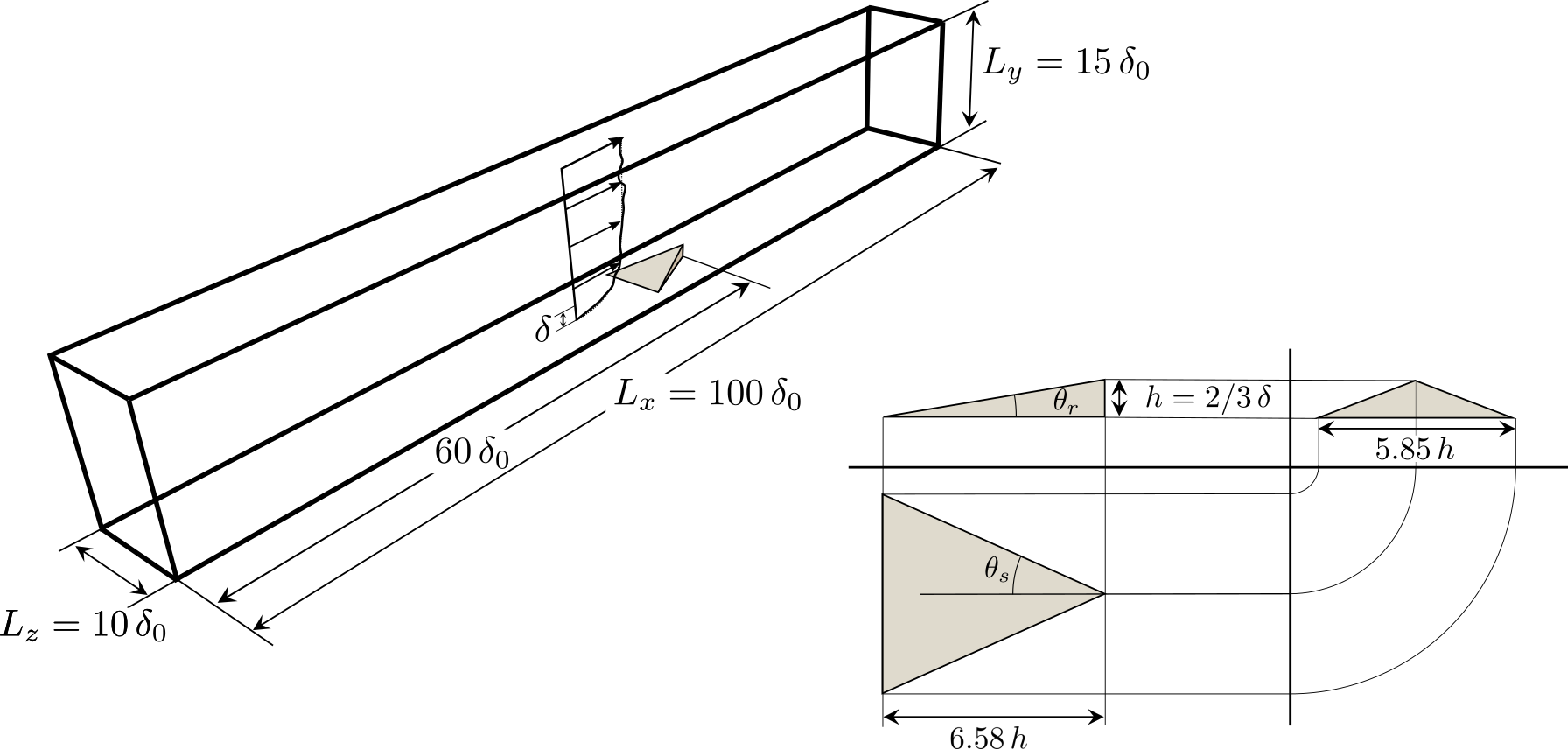} 
     \caption{Domain with sizes and orthogonal projections of the microramp.}
     \label{fig:domain_ramp}
\end{figure*}
The size of the computational domain adopted for the simulations is 
$L_x/\delta_0 \times L_y/\delta_0 \times L_z/\delta_0 = 100 \times 15 \times 10$, 
the reference length $\delta_0$ being the thickness of the boundary layer at the inflow. 
The microramp geometry for the three flow conditions considered is based on the optimal shape defined by \citet{anderson2006optimal} 
(see figure \ref{fig:domain_ramp}), with a constant ratio between the ramp height and the boundary layer thickness of the uncontrolled case, $h/\delta_{99}=2/3$. This configuration has been studied experimentally by \citet{tambe2021relation}, 
at the same Mach number but at a higher Reynolds number ($\Rey_\theta = 2.4 \times 10^4$, $\Rey_\tau \approx 5000$).
The microramp is centred in the cross-stream plane,
and the ramp trailing edge is $60\,\delta_0$ from the inlet section.
The trailing edge of the ramp is also the origin of our reference system.
The mesh is uniform in the wall-parallel directions, 
corresponding to a viscous-scaled spacing of $\Delta x^+ \approx 6.9$ and $\Delta z^+ \approx 5.5$, in the streamwise and spanwise directions, respectively. The mesh is stretched in the wall-normal direction, corresponding to a wall-spacing in the range $\Delta y^+\approx0.75$--$0.95$.
The microramp is simulated by means of a ghost-point-forcing \gls{ibm} \citep{piquet2016comparative} 
already validated in previous works \citep{modesti2022direct}.

The boundary conditions are specified as follows. At the outflow and the top boundaries, non-reflecting
conditions are imposed by performing a characteristic decomposition in the direction normal to the boundaries~\citep{poinsot1992boundary}.
A characteristic wave decomposition is also employed at the bottom no-slip wall, where the wall temperature is set to
the recovery value of the incoming boundary layer.
In order to prescribe suitable turbulent fluctuations at the inflow, 
a recycling-rescaling procedure~\citep{lund1998generation} is used, 
with a recycling station placed at $50\, \delta_0$ from the inflow to guarantee 
a sufficient decorrelation between the inlet and the recycling station.
Finally, the flow is assumed to be statistically homogeneous in the spanwise direction, thus periodic
boundary conditions are applied on the sides. 
Hence, simulations
are representative of an array of microramps  
with lateral spacing $L_z$.
%
%
\section{Numerical data set and validation}
\label{sec:dataset}

The main parameters of the numerical database analysed in this work are reported 
in table~\ref{tab:bl}. 
The data set includes baseline simulations with undisturbed boundary layer, which are 
used as a reference to assess the effects of the microramps. 
We consider three values of friction Reynolds number 
$\Rey_\tau = 500$, $1000$, $2000$,
based on the properties of the undisturbed boundary layers.
The corresponding Reynolds number based on the momentum thickness is reported 
in table ~\ref{tab:bl} together with the number of mesh points. 
The free-stream Mach number $M_\infty$ is equal to 2 for all the flow cases.  
In the following, unless otherwise specified, 
we will indicate with the colours red, blue, and black the cases at 
low, intermediate and high Reynolds number respectively.

%
\begin{table}
 \begin{center}
\def~{\hphantom{0}}
  \begin{tabular}{lccccc}
    \hline\noalign{\smallskip}
 	Case  &  $M_\infty$ & $Re_{\tau}$ & $Re_\theta$ & $N_x \times N_y \times N_z$ & $h/\delta_0$\\
    \noalign{\smallskip}\hline\noalign{\smallskip}
 	{\color{ red}UBL - L} & 2.00 &  500 & 2330 &  {\color{white}a}4096 $\times$ 288 $\times$ {\color{white}a}512 & - \\
 	{\color{blue}UBL - M} & 2.00 & 1000 & 4870 &  {\color{white}a}8192 $\times$ 512 $\times$ 1024 & - \\
 	         UBL - H  & 2.00 & 2000 & 9880 & 16384 $\times$ 896 $\times$ 2048 & - \\
 	{\color{ red}MBL - L} & 2.00 &  500 & 2330 &  {\color{white}a}4096 $\times$ 288 $\times$ {\color{white}a}512 & 1.26 \\
 	{\color{blue}MBL - M} & 2.00 & 1000 & 4870 &  {\color{white}a}8192 $\times$ 512 $\times$ 1024 & 1.16 \\
 	              MBL - H & 2.00 & 2000 & 9880 & 16384 $\times$ 896 $\times$ 2048 & 1.12 \\
    \noalign{\smallskip}\hline
  \end{tabular}
  \caption{Main parameters of the numerical database. Reynolds numbers are evaluated at the trailing edge of the microramp.\\
  UBL: uncontrolled boundary layer, MBL: microramp-controlled boundary layer, L: low Reynolds number, M: intermediate Reynolds number, H: high Reynolds number. \label{tab:bl}}
 \end{center}
\end{table}

\begin{figure*}
     \centering
     \subfloat[]{
     \includegraphics[width=0.45\textwidth]{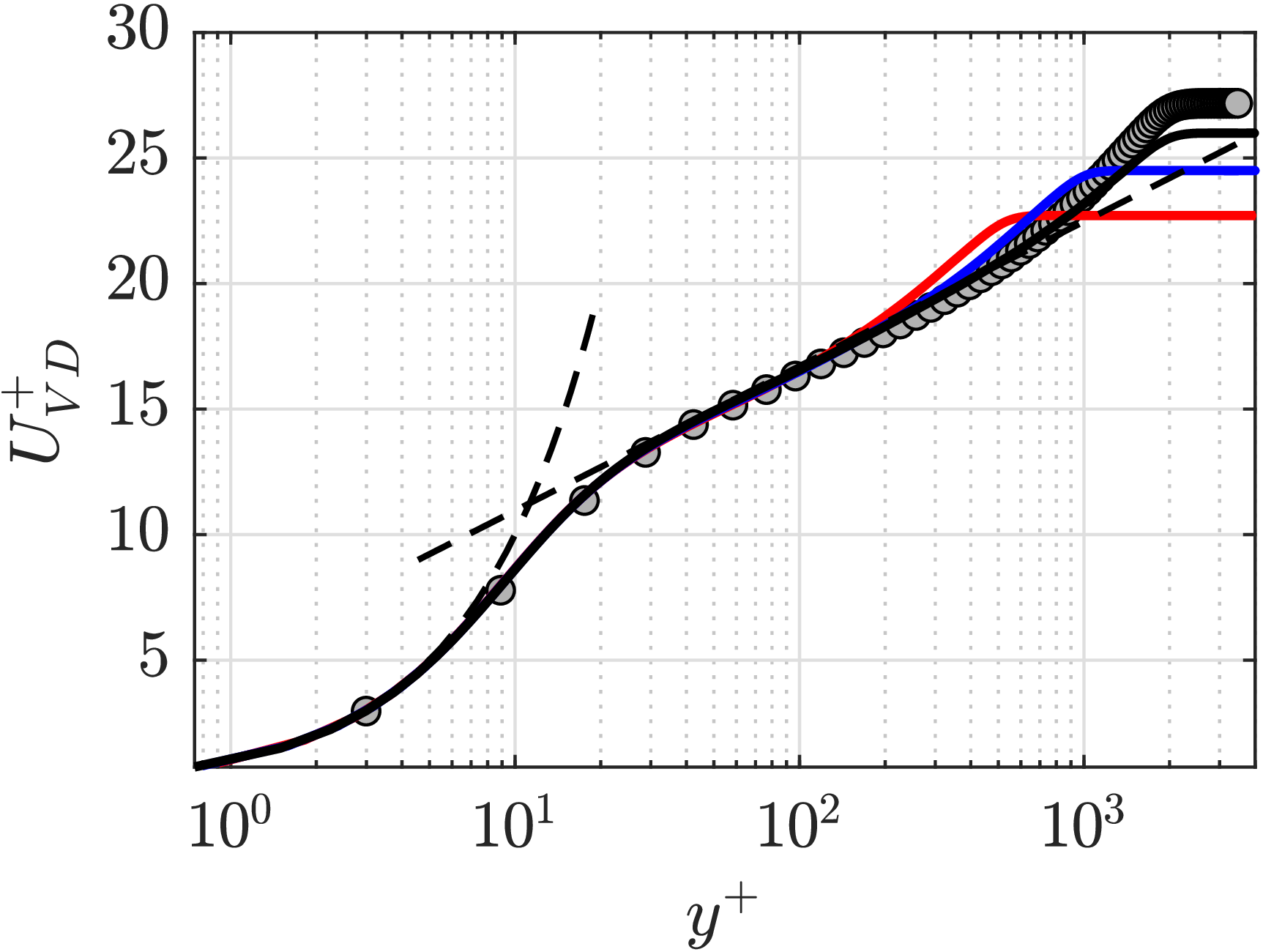}} 
     \qquad
     \subfloat[]{
     \includegraphics[width=0.45\textwidth]{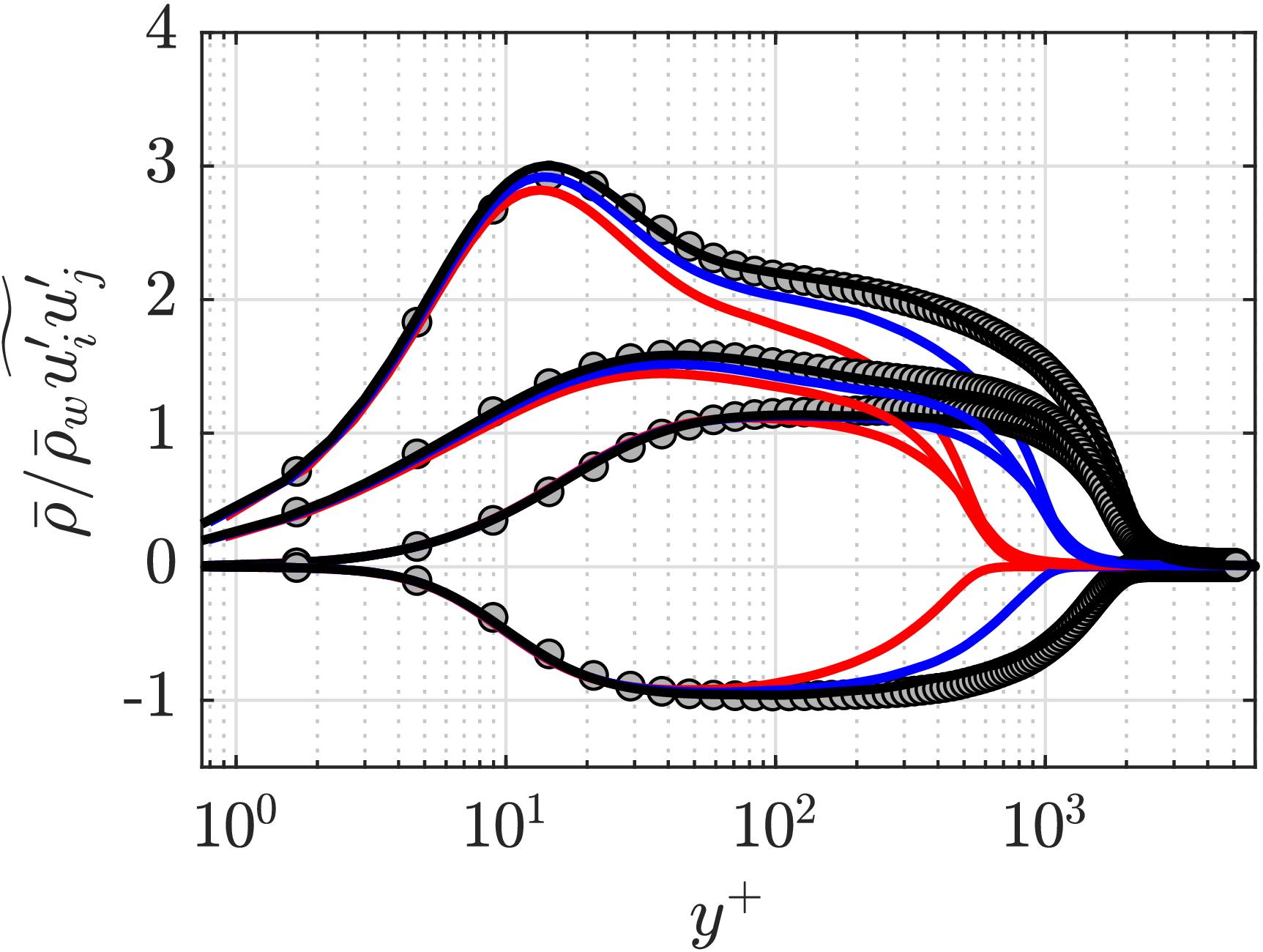}} 
     \caption{Comparison of (a) van Driest–transformed mean velocity profile and (b) density-scaled Reynolds
        stress components for the incoming boundary layer with reference experimental data by \cite{sillero2013one} (circles, $Re_\tau = 1989$).}
     \label{fig:blay_check}
\end{figure*}

In order to characterise the incoming turbulent boundary layer, 
figure \ref{fig:blay_check} shows a comparison of the van Driest transformed mean velocity profiles
and of the density-scaled Reynolds stress components with the incompressible data
from the \gls{dns} database of \cite{sillero2013one}. For the case with the highest Reynolds number, an excellent agreement can be observed for the mean velocity from the viscous sublayer up to the outer region, where the wake component departs from the transformed profile slightly. The Reynolds stress components, represented in wall units, are reported instead in figure \ref{fig:blay_check}(b). Increasing the friction Reynolds number, all the components converge towards the incompressible results.

%
%
\section{Results}
\label{sec:results}

\subsection{Qualitative flow organisation} \label{subsec:qualitative_organisation}
First, we describe the qualitative behaviour of the instantaneous flow in terms of turbulent structures, 
and we outline the organisation of the shock system taking place in the field. 

\subsubsection{Turbulent structures}
\begin{figure*}
     \centering
     \includegraphics[width=0.8\textwidth]{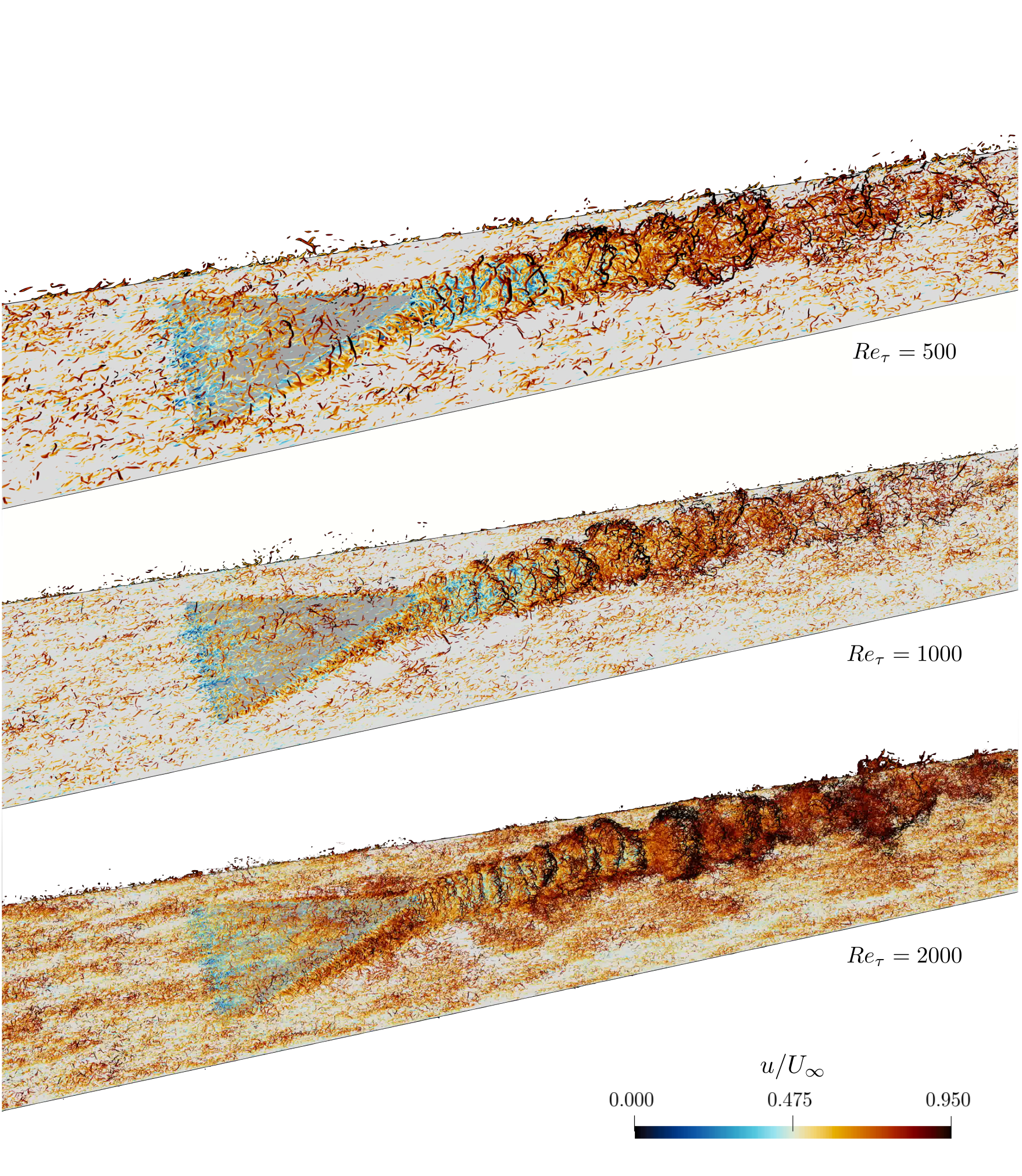} 
     \caption{Isosurfaces of the swirling strength 
     coloured by the streamwise velocity component. 
     From top to bottom: low, intermediate, and high Reynolds number cases.
     }
     \label{fig:3d_qcrit_inst}
\end{figure*}
Figure \ref{fig:3d_qcrit_inst} depicts, for the three 
controlled cases, the instantaneous turbulent structures identified by means of 
isosurfaces of the imaginary part of the velocity gradient
tensor's complex eigenvalues (swirling strength criterion, \citet{zhou1999mechanisms}), 
coloured by the streamwise velocity component\footnote{
A supporting video about the fluid field generated by a supersonic turbulent boundary layer impinging on 
a microramp is available in high resolution at \url{https://youtu.be/o8olmjiWSl8} \citep{bernardini2022video}. 
}.
First of all, it is possible to observe the fine turbulent structures 
upstream of the microramp, which become smaller by increasing the Reynolds number.
At high Reynolds number, the large-scale streamwise streaks 
are evident, corresponding to regions of alternating 
high- and low-speed flow.
When the flow encounters the ramp, the slope of the top surface induces a first shock
that generates a small region with reduced velocity at the foot of the ramp. 
On the top ramp surface, the flow accelerates as streamlines are forced
to converge, then the flow deflects laterally and diverges towards the sides of the ramp, up to its edges. 
From here, the accelerated flow leaves the top surface and naturally generates the primary vortices on the sides of the ramp, 
together with smaller secondary vortices close to the top and bottom edges. 
The primary vortices increase in strength and radius proceeding downstream
until they converge on the trailing edge of the ramp
where they deflect, becoming nearly parallel and aligned 
with the streamwise direction. 
The convergence of the primary vortex pair and the accelerated flow from the 
top edge of the ramp initiate vortex roll-up 
in the arc region above the primary vortices. Soon after the edge of the ramp, 
arc-shaped vortical structures are distinguishable in all the three cases, 
especially for higher Reynolds numbers. 
In accordance with the model proposed by \cite{sun2014decay}, 
the lift-up of the wake and the entrainment of lateral flow in the near-wall region 
tend to close the arc-shaped structures into almost-toroidal vortex rings that enclose
the decaying primary vortex pair. Vortex rings are then convected 
downstream and, 
together with the primary vortex pair inside, induce velocity fluctuations 
that are responsible for the meandering and, eventually, the breakdown of the wake. 

\subsubsection{Shock system}
\begin{figure*}
     \centering
     \includegraphics[width=0.65\textwidth]{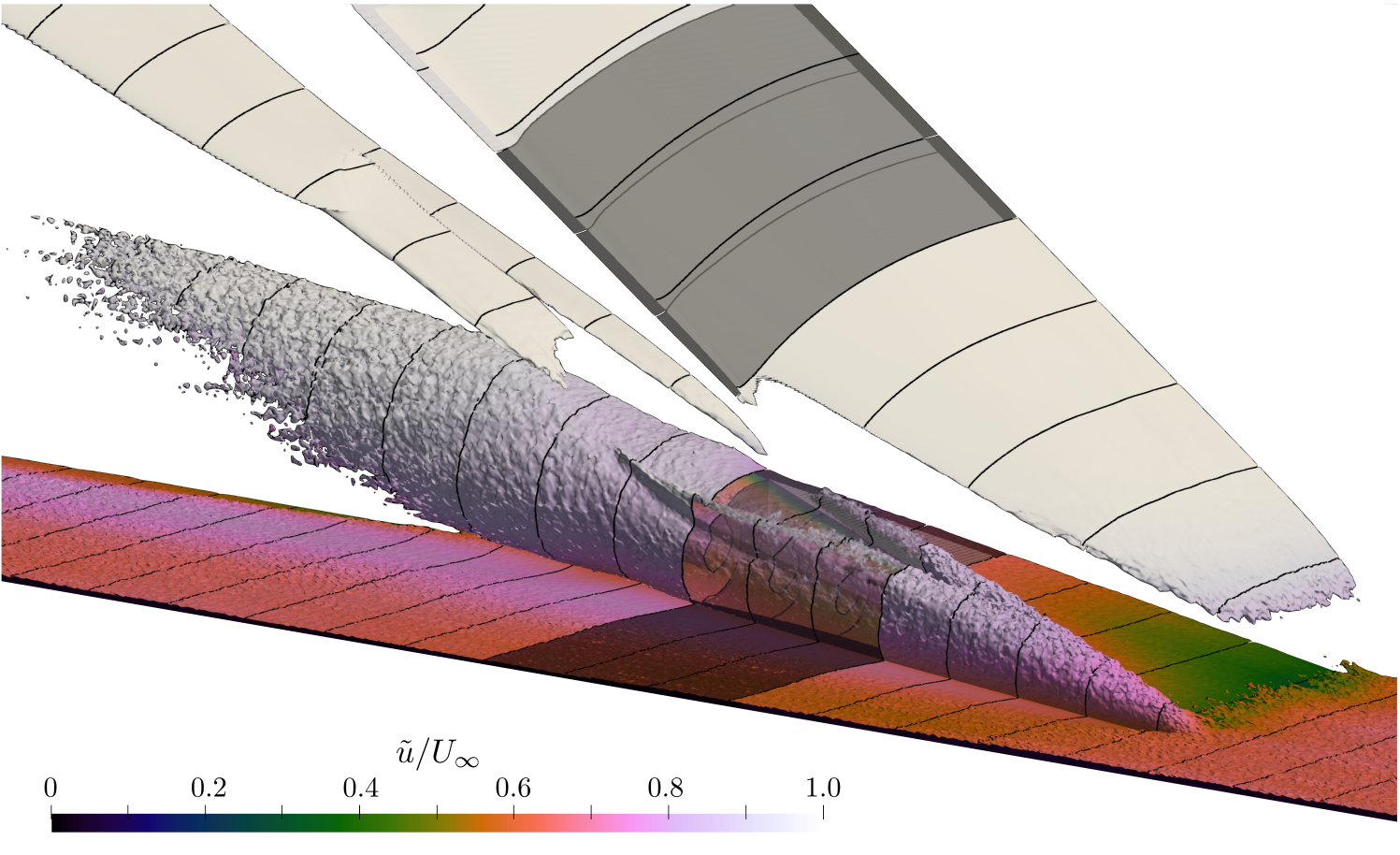}
     \caption{Isosurface of the magnitude of the mean density gradient, 
     coloured by the streamwise velocity component. Only half of the domain is shown for symmetry.}
     \label{fig:3dshock}
\end{figure*}
\begin{figure*}
     \centering
     \includegraphics[width=\textwidth]{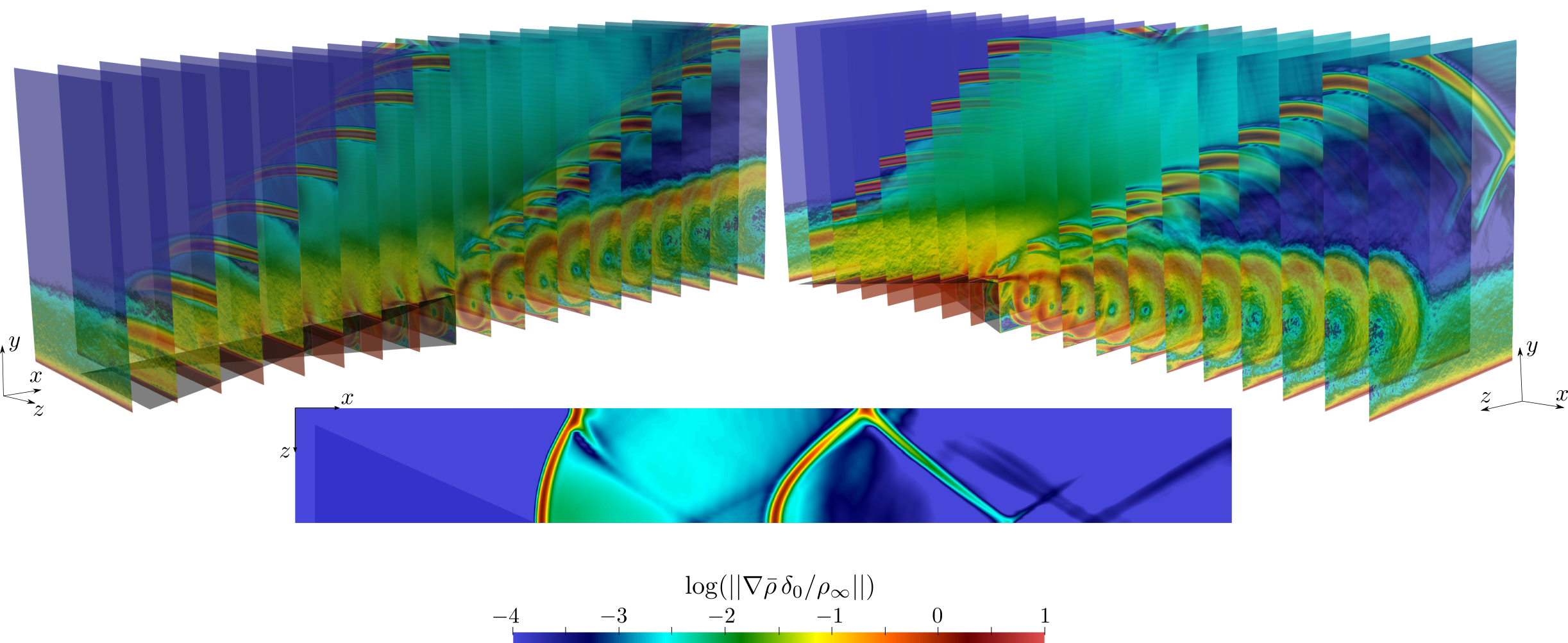} 
     \caption{Three-dimensional views of the contours of the non-dimensional mean density gradient 
     in logarithmic scale on several yz planes at different streamwise locations. 
     Intermediate Reynolds number case. Only half of the domain is shown for symmetry.}
     \label{fig:slicesxy_shocks}
\end{figure*}
Besides the highlighted vortical structures, 
a series of complex, 3D shocks is also generated in the field
as a consequence of the numerous changes in direction of the supersonic flow around the ramp. 
To better understand the complex structure of the shocks,  
we report in figure~\ref{fig:3dshock} 
an isosurface of the magnitude of the mean density gradient, coloured by the streamwise 
velocity for the intermediate Reynolds number. 
Slices at constant $x$ and transparency help understand the 3D curvature 
and interior structure of the field. 
Figure \ref{fig:slicesxy_shocks}, instead, shows the gradient of the mean density in logarithmic scale 
on several planes at constant $x$ from two different 3D views and 
on a wall-parallel plane at the top of the computational domain. 
For symmetry reasons, only half of the domain is shown.

We note that the density gradient is able to highlight better the shocks 
and the dominant vortical structures delimiting the wake.
As mentioned above, the slope of the \gls{mvg} induces a first shock from the
foot of the ramp, which we will refer to as foot shock. 
Since the ramp has a finite extension, the originally planar foot shock
must curve in the spanwise direction while evolving in the longitudinal direction, 
as can be seen from its curvature on the top of the domain. 

After the foot shock, in the final region of the developing primary vortices 
at the sides of the ramp, we observe the formation of an 
oblique shock wave inclined of approximately 45 degrees right on top of the vortices.
These shocks, which we will refer to as vortex shocks, are stronger for flow cases at higher Reynolds numbers. 
Although in part related to the 
reduced viscous dissipation, we will discuss in section \ref{sec:wall_normal_sym}
that if the ramp geometry is fixed with respect to the height of the boundary layer, 
increasing Reynolds number leads to an increase 
in the circulation of the vortex pair and in the velocity magnitude of the flow captured in the 
side vortices in general, which increases the strength of the compression in the end.

At the trailing edge of the ramp, the flow on the top 
surface undergoes an initial vertical deflection in the central plane 
induced by the converging primary vortices, giving rise to a shock from the ramp main edge. 
This flow deflection 
is significant only in the trailing edge region close to the symmetry plane, whereas it decreases laterally,
thus resulting in a conical shock. 
Slightly downstream the ramp trailing edge,
the increased lift-up generated by the mutual interaction of the primary vortex pair develops fully
and pushes the wake further up, inducing a second shock.
We will refer to these shocks as the trailing edge shocks. 
Given the topology of the foot shock and the spanwise variation of the 
flow deviation, also the trailing edge shocks have a conical shape, 
as can be seen from the slice in the xz plane at the top of the computational domain.
After less than one ramp length, the two trailing edge shocks coalesce.  
Below the conical compression, it is possible to appreciate the development 
of the primary vortex pair, 
with two distinct convoluted vortices just after the ramp trailing edge. 
Proceeding downstream, the diffusion and the entrainment of flow from outside of the 
wake enlarges the cores of the vortices and ``unroll'' the inner convolution:
after slightly more than one ramp length from the main edge, 
the low-momentum region has an almost circular shape.
Finally, we also observe the shocks coming from the 
side ramps as a result of the periodic boundary conditions. 
\begin{figure*}
     \centering
     \subfloat{
     \includegraphics[width = 0.33\textwidth]{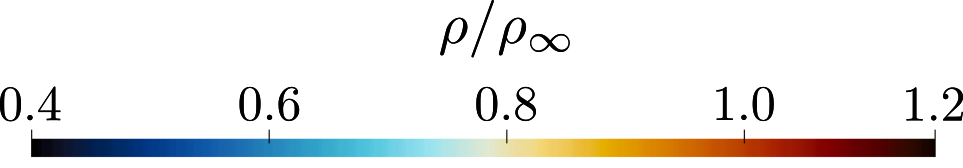} 
     }
     \\
     \setcounter{subfigure}{0}
     \subfloat[ \label{fig:scene02401}]{
     \includegraphics[width=0.49\textwidth]{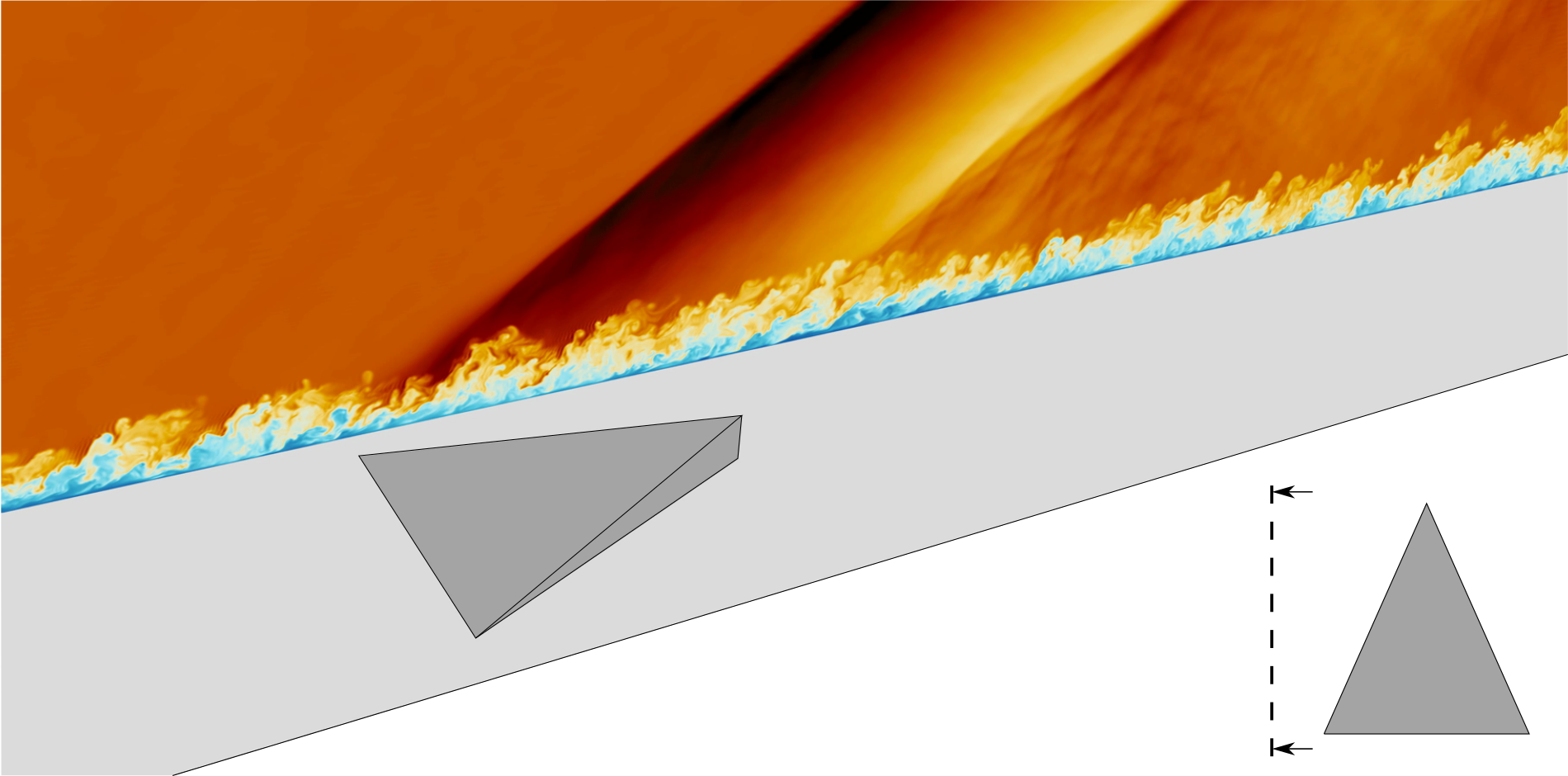} 
     }
     \subfloat[ \label{fig:scene02251}]{
     \includegraphics[width=0.49\textwidth]{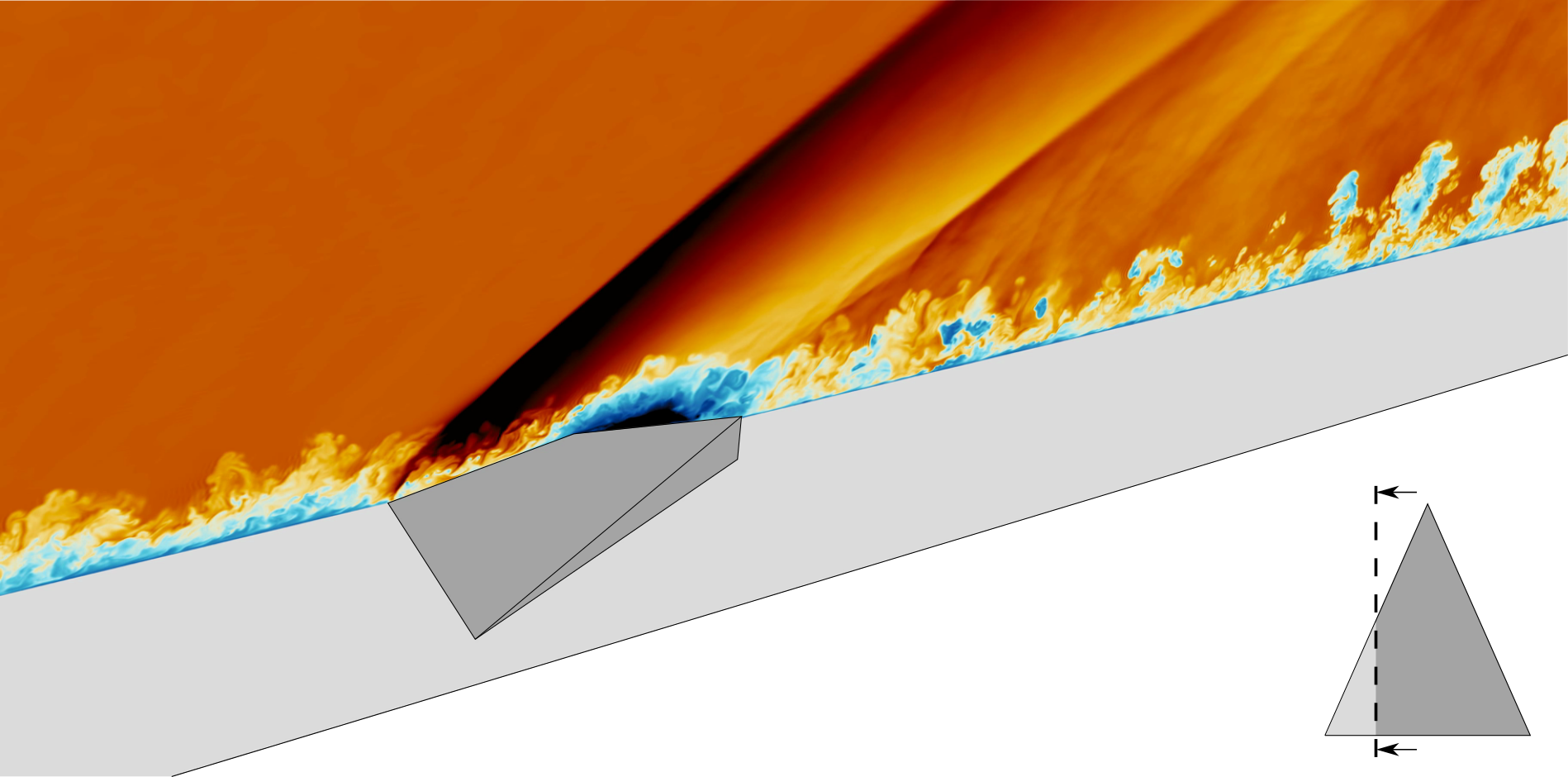} 
     }
     \\
     \subfloat[ \label{fig:scene02101}]{
     \includegraphics[width=0.49\textwidth]{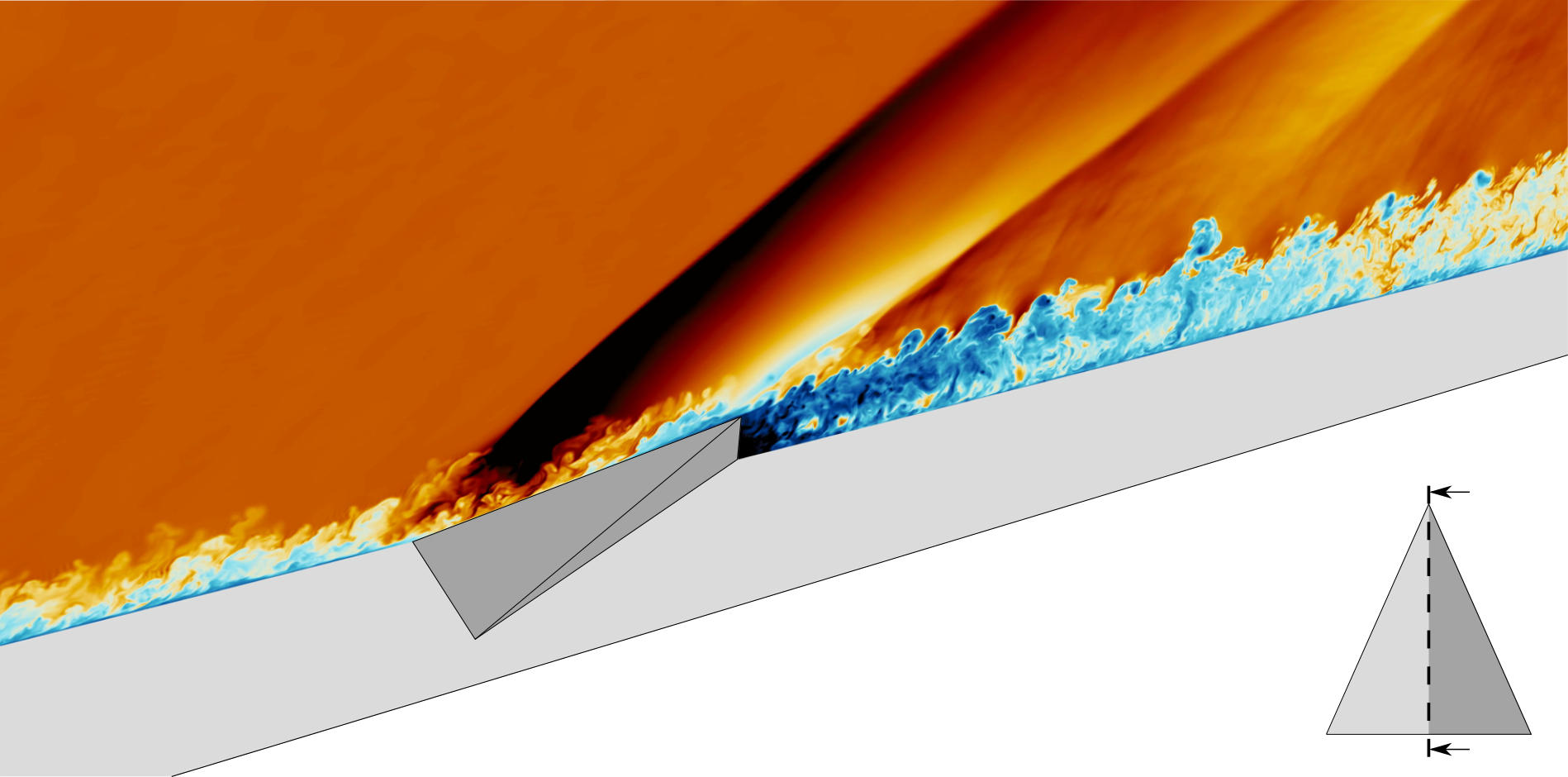} 
     }
     \caption{Contours of the instantaneous density on vertical slices at different spanwise sections: (a) lateral wall ($z/h = - 0.5 L_z/h$), 
     (b) intermediate spanwise location ($z/h = -0.5 b/h$), 
     (c) centre of the ramp ($z/h = 0$). Low Reynolds number case.}
     \label{fig:slicesxy_span}
\end{figure*}

In order to obtain a clearer picture of the shock system, we also report three contours 
of the instantaneous density in the xy planes at $z/h = -0.5 L_z/h$, $-0.5 b/h$, and $0$ 
for the case at high Reynolds number, where $b$ is the spanwise half-width of the ramp.
At the side surfaces of the domain, it is possible to 
see only the foot shock and the trailing edge shock. 
From the intermediate spanwise position, we can observe the significant compression 
induced by the foot shock. The core of the primary vortex is then highlighted 
by the region with low density close to the edge of the ramp. Slightly after, on top 
of the trace of the primary vortex pair, the trailing edge shock occurs, followed 
 by a milder compression due to the shock coming from the side ramps. 
 Further downstream, low-density regions suggest the presence of the vortex rings 
observed in figure \ref{fig:3d_qcrit_inst}. 
Finally, the slice at the symmetry plane of the ramp shows all the main shocks of the flow 
field clearly. The foot shock and the trailing edge shocks have here their 
maximum intensities, and behind the trailing edge shocks, 
the shock stemming from the side ramp is also visible.
The instantaneous density on this plane also makes 
evident how, after less than two ramp heights, \gls{kh} instabilities 
generate vortical structures over the primary vortices, with cores highlighted by
spots of reduced density.

Concerning the effect of the Reynolds number on the shock structure, a comparison among the cases
did not reveal relevant qualitative differences with respect 
to the described picture.

\subsection{Streamwise evolution of the wake}

In the following, we discuss various aspects concerning the streamwise development 
of the wake, focusing in particular on the 
Favre-averaged streamwise and vertical velocity components, 
for which also experimental data are available. Favre-averaging of a generic
variable $f$ is indicated with the tilde and is defined as 
$\tilde{f} = \overline{\rho f}/\bar{\rho}$, where the bar indicates 
Reynolds-averaging instead.

\subsubsection{Wall-normal profiles of velocity on the symmetry plane} \label{sec:wall_normal_sym}

\begin{figure*}
     \centering
     \subfloat[]{
     \includegraphics[width=0.43\textwidth]{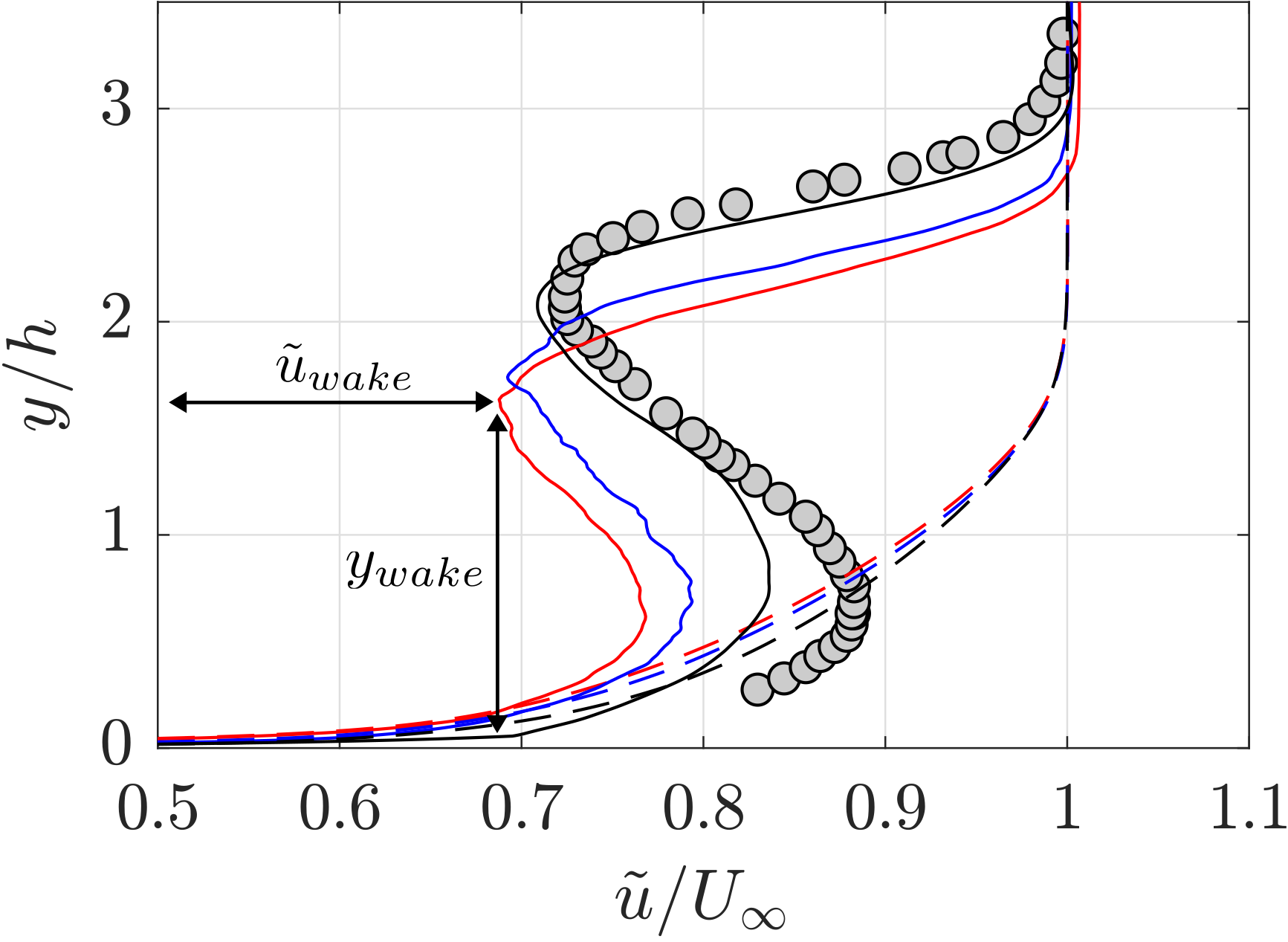}} 
     \qquad
     \subfloat[]{
     \includegraphics[width=0.43\textwidth]{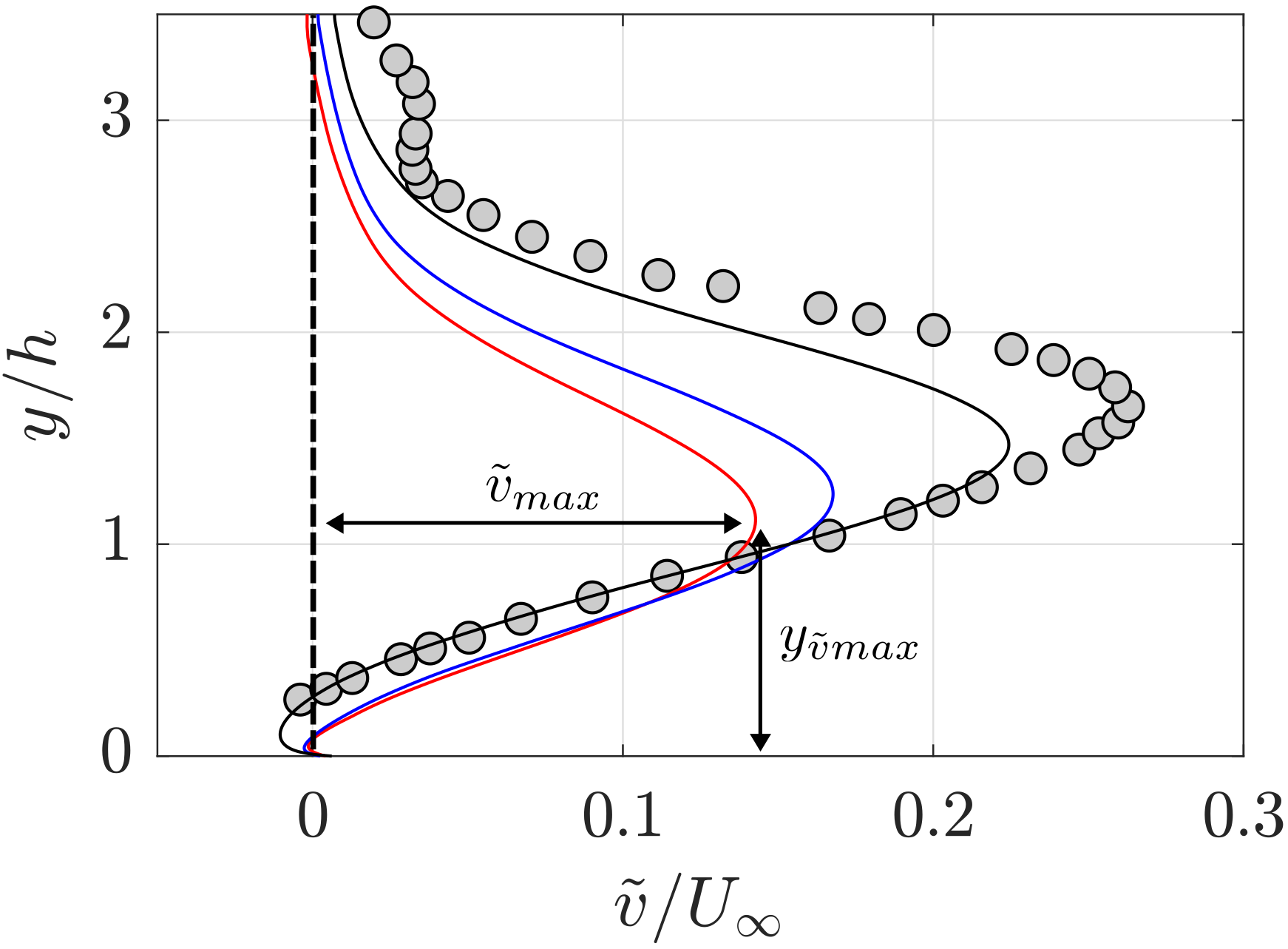}}
     \caption{Velocity profiles at $x/h=10.7$: (a) streamwise and (b) wall-normal components. Dashed lines denote the velocity profiles in undisturbed conditions. Symbols denote
        experiments by \cite{tambe2021relation} (circles, $M_\infty = 2.0$, $Re_\theta = 2.4 \times 10^4$). }
     \label{fig:profiles_xh10.7}
\end{figure*}

Figure \ref{fig:profiles_xh10.7} shows the velocity profiles on the symmetry plane at $x/h = 10.7$ for the three different Reynolds numbers, together with the corresponding uncontrolled cases and the experimental results of \cite{tambe2021relation}. 
From the streamwise velocity component $\tilde{u}/U_\infty$ 
in figure \ref{fig:profiles_xh10.7}a, it is evident that for increasing $\Rey_\tau$, 
the microramp is more efficient, and is thus able to produce 
fuller streamwise velocity profiles in the near-wall region (see also section \ref{sec:close_wall}).
Moreover, the region in the wake characterised by low momentum with respect to the undisturbed profiles rises to higher relative wall-normal locations. 
An explanation to this phenomenon is given by the profiles of the 
wall-normal velocity reported in figure \ref{fig:profiles_xh10.7}b, 
where peak values of $\tilde{v}/U_\infty$ increase markedly with the Reynolds number.
The numerical results of both the velocity components tend to
the experimental results of \citet{tambe2021relation}, which have been performed at $\Rey_\tau \approx 5000$. 

\begin{figure*}
     \centering
     \subfloat[]{
     \includegraphics[width=0.9\textwidth]{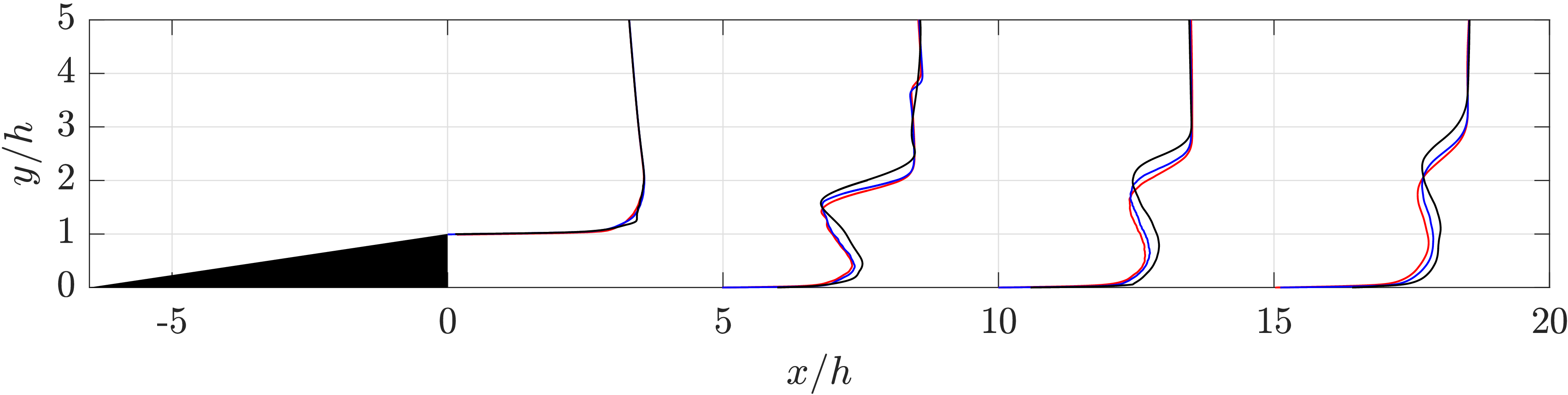}} 
     \\
     \subfloat[]{
     \includegraphics[width=0.9\textwidth]{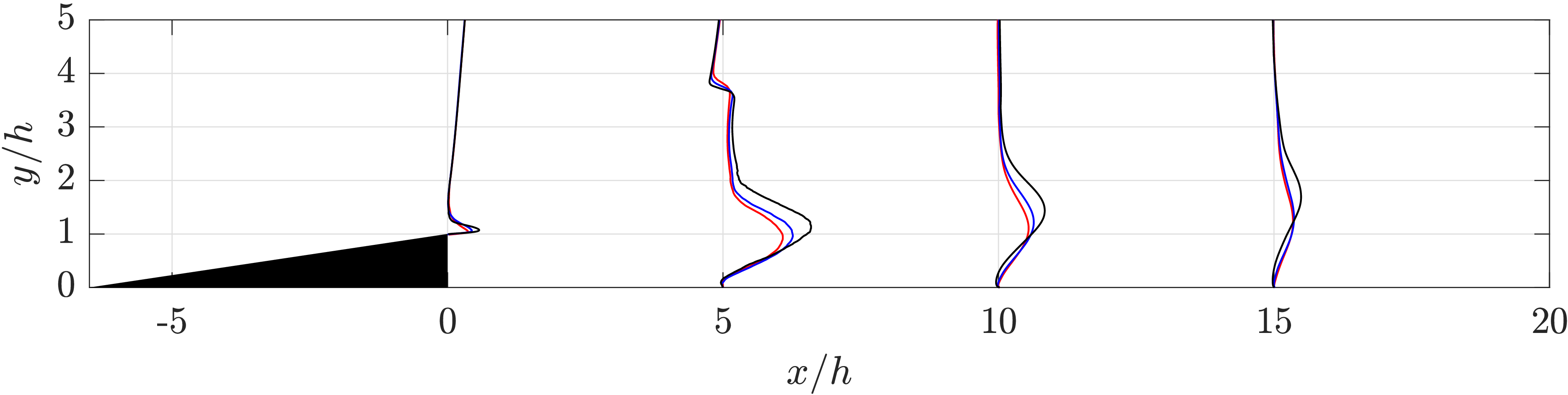}} 
     \caption{Streamwise evolution of the velocity profile at four different stations $x/h=0, \, 5, \, 10, \, 15$. (a) streamwise component and (b) wall-normal component.}
     \label{fig:profiles_streamwise_development}
\end{figure*}

Observing the longitudinal and vertical components of the velocity at four sections 
on the symmetry plane in figure \ref{fig:profiles_streamwise_development}, 
we appreciate the 
lifting up effect on the wake due to the longitudinal primary vortex pair. 
The trace of the shock wave generated by the microramp trailing edge is also 
 visible in the profiles at section $x/h = 5$. 
Farther donwstream from the microramp, the profiles of both velocity components
tend to return to the undisturbed conditions
due to the turbulent mixing and flow entrainment from outside the wake. 

%
\begin{figure*}
     \centering
     \subfloat{
     \includegraphics[width = 0.35\textwidth]{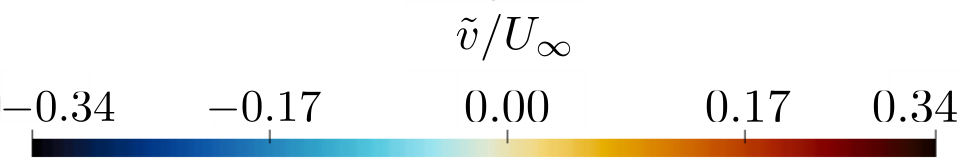} 
     }
     \\
     \setcounter{subfigure}{0}
     \subfloat[\label{fig:toy_model_batchelor}]{
     \includegraphics[height=0.3\textwidth]{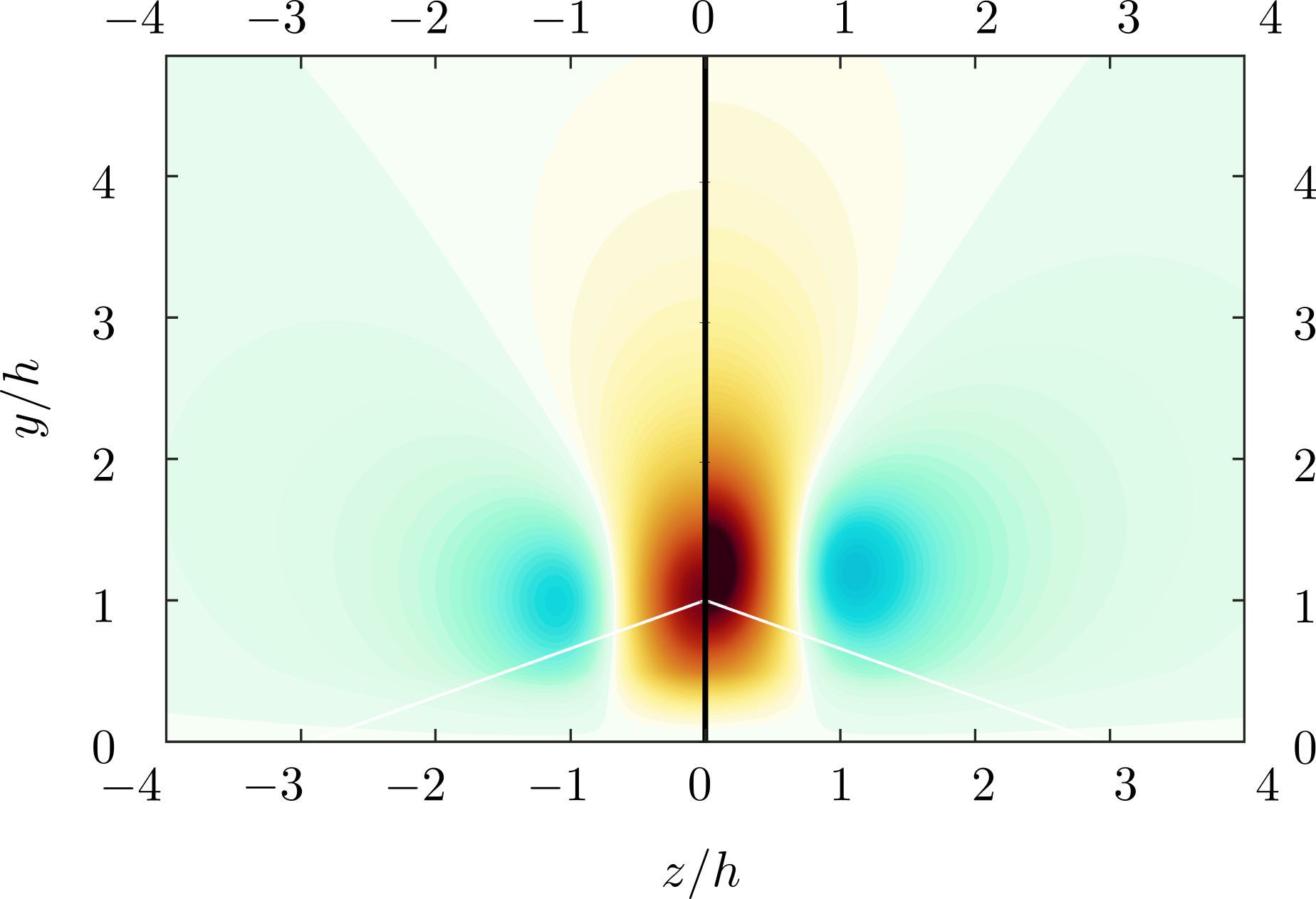}} 
     \,
     \subfloat[\label{fig:vertical_velocity_comparison}]{
     \includegraphics[height=0.3\textwidth]{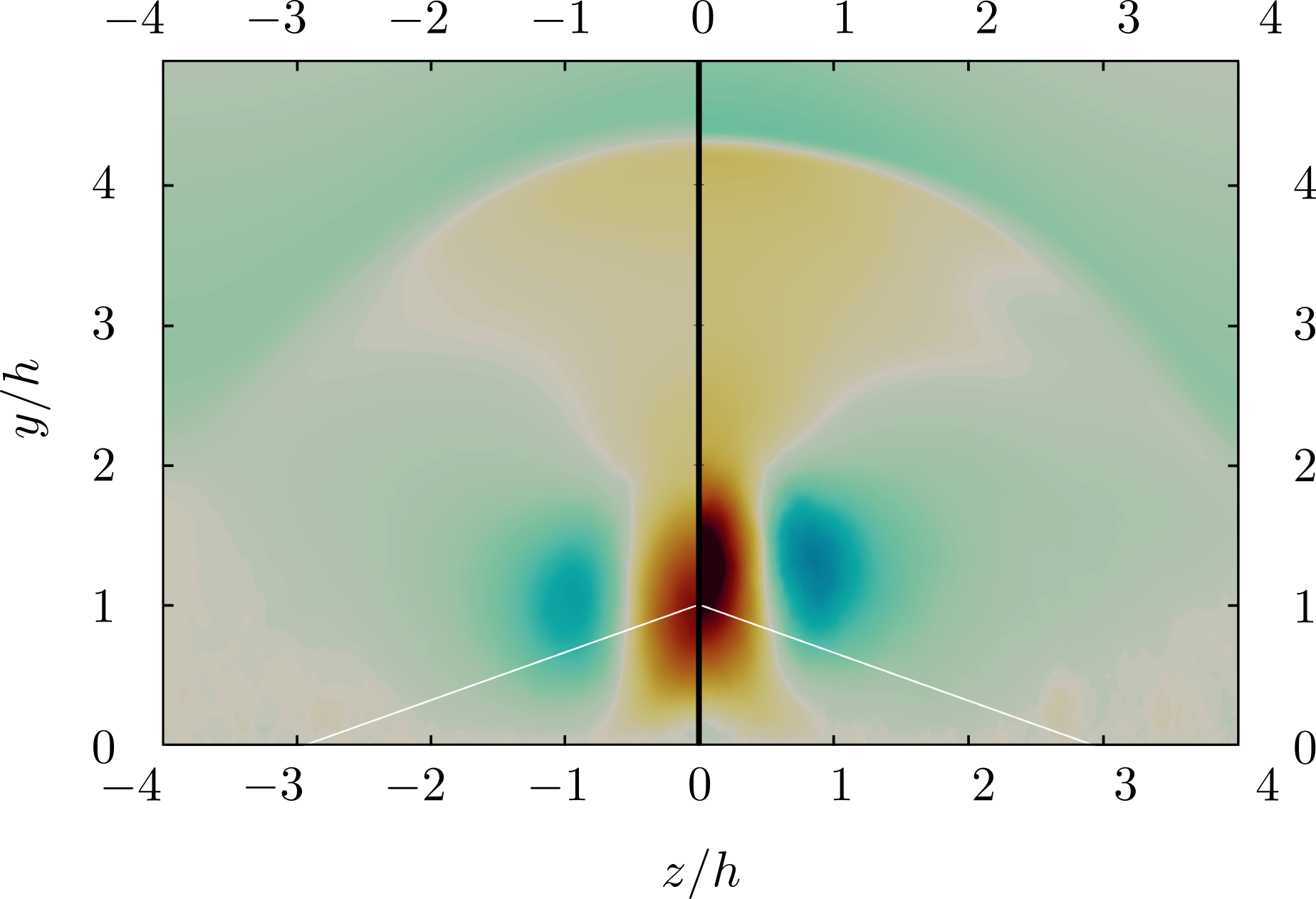}}
     \caption{Analytical and numerical vertical velocity. 
     (a) Field generated by two parallel Batchelor vortices 
     on a yz plane. Vortex strength $q_1$ at negative $z$, vortex strength $q_2 \approx 1.25\,q_1$, 
     at positive $z$. (b) Favre-averaged vertical velocity at $x/h = 6$ for low (negative $z$) and 
     high (positive $z$) Reynolds numbers simulations.  
     }
     \label{fig:toy_dns_model_batchelor}
\end{figure*}


The increased vertical velocity at the symmetry plane suggests that the circulation of the 
two parallel primary vortices increases with the Reynolds number. 
Following for example the approach used in \cite{mole2022interaction}, 
the velocity field in the yz plane produced by the primary vortex pair from a microramp
can be modelled schematically by means of a couple of 
counter-rotating Batchelor vortices \citep{batchelor1964axial}, with two additional mirror
vortices with opposite circulation below the wall, to impose wall impermeability, and 
a linear damping of the azimuthal velocity close to the wall, to impose no-slip condition. 
The azimuthal velocity $V_\theta$ of an isolated Batchelor vortex is defined as
\begin{equation}
    V_\theta(r) = \frac{q\,R}{r} \left[ 1 - \mathrm{exp}\left( - \frac{r^2}{R^2} \right) \right]
\end{equation}
where $q$ is the vortex strength, $r$ is the distance from the centre of the vortex, and
$R$ is the radius of the vortex core. 
Figure \ref{fig:toy_model_batchelor} shows only the vertical motion induced by a 
toy model with two parallel vortices of constant circulation on a slice at constant x 
for two cases where the vortex strength -- and hence the circulation -- is increased ($q_2/q_1 = 1.25$). 
Except for the confinement induced by the trailing edge shocks and for the 
overall downward motion induced by the 3D flow organisation at the observed station (which makes the \gls{dns}
results more ``greenish''), 
the distribution is very similar to 
the one observed in the wake of the simulated cases reported in figure~\ref{fig:vertical_velocity_comparison}, 
confirming that the toy model is able to reasonably represent the real case. 
From the comparison, we note that the effect of increasing the Reynolds 
number in the numerical simulations is equivalent to increasing the 
circulation in the analytical model. 
%


\subsubsection{Boundary layer thickness and shape factor}
\begin{figure*}
     \centering
     \subfloat[]{
     \includegraphics[width=0.43\textwidth]{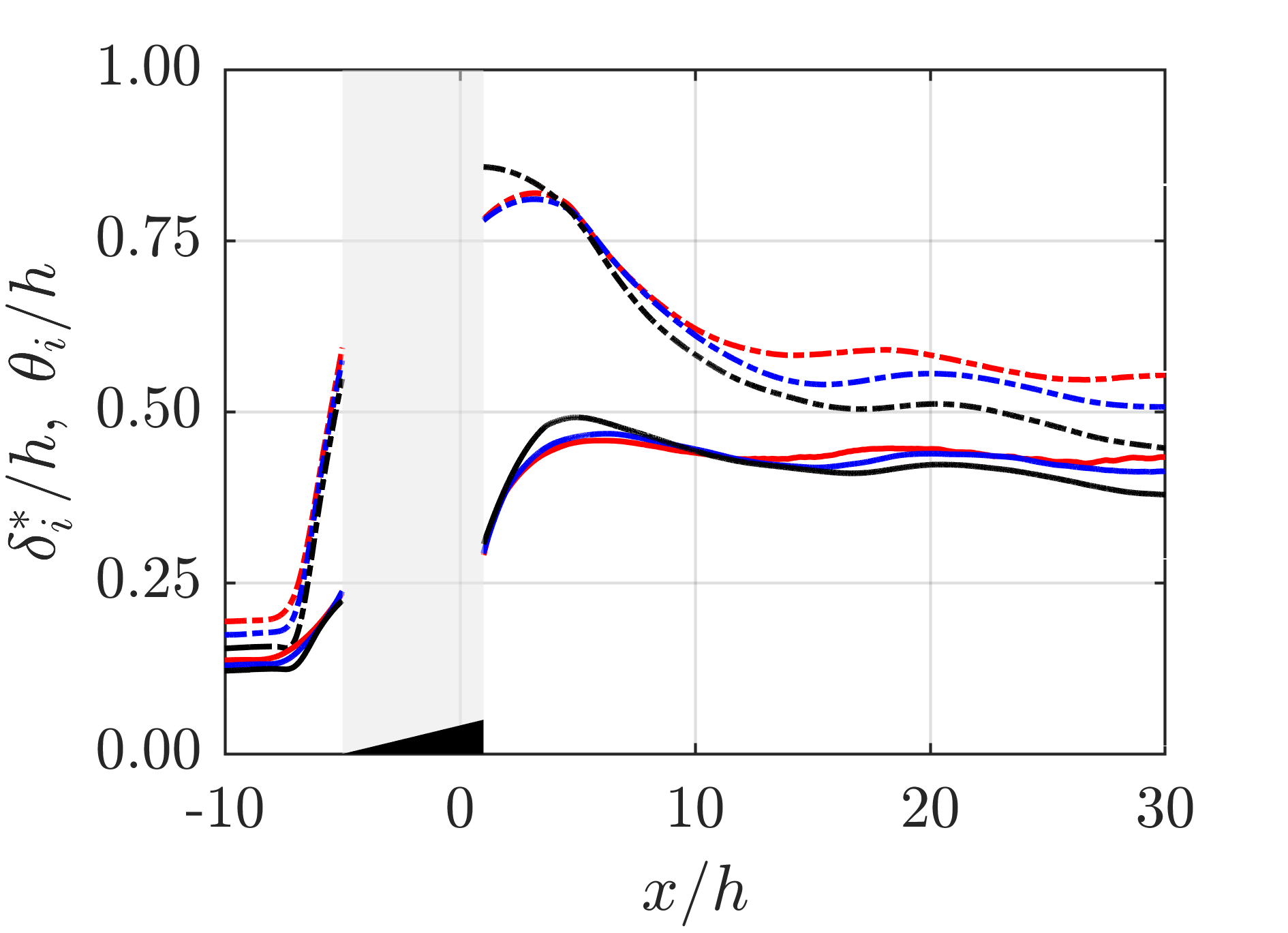}} 
     \qquad
     \subfloat[]{
     \includegraphics[width=0.43\textwidth]{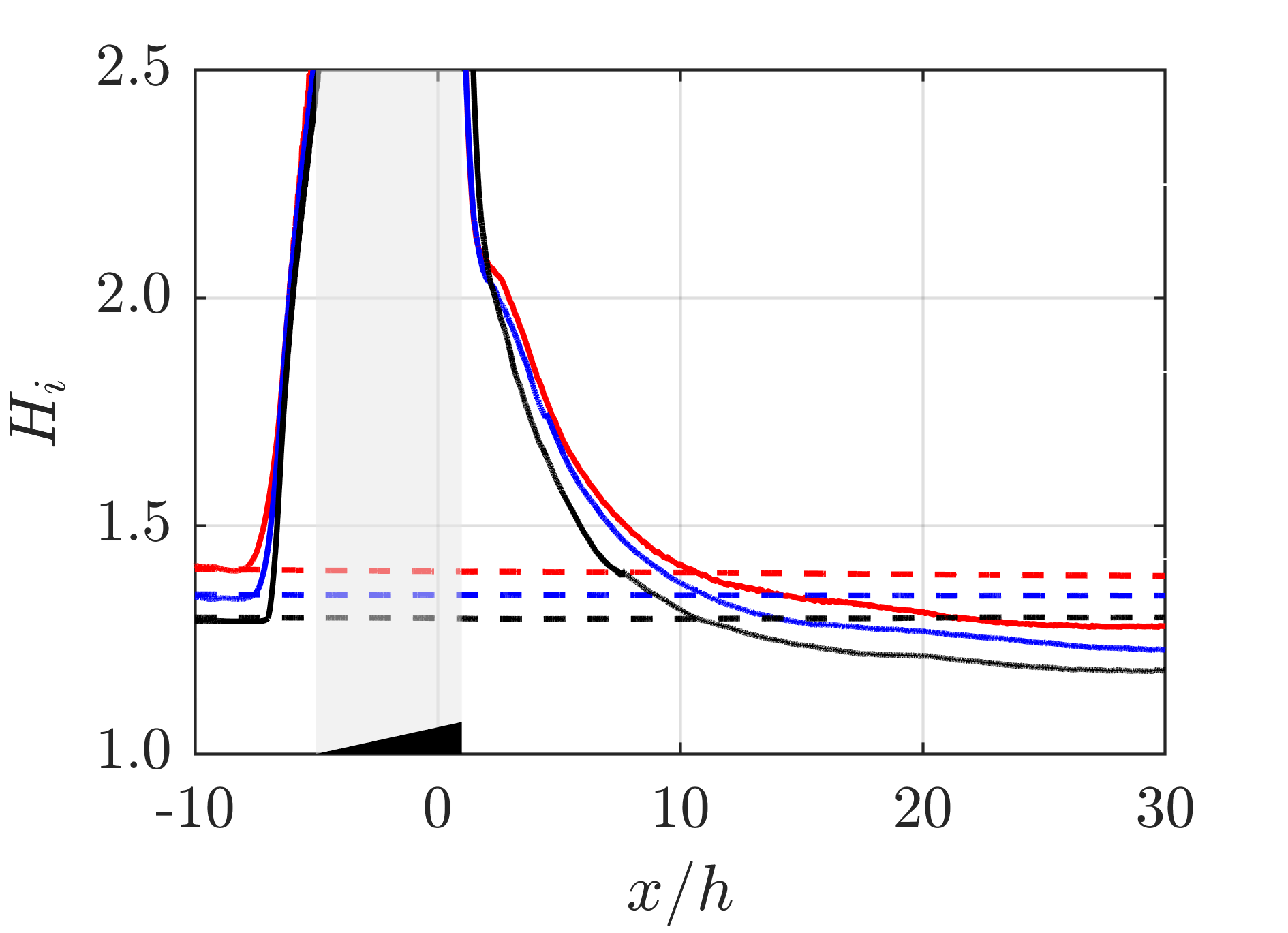}} 
     \caption{Streamwise evolution of (a) boundary layer displacement thickness (-.), momentum thickness  (-) and (b) incompressible shape factor for the\\uncontrolled case (- -) and with the microramp (-).}
     \label{fig:displacement_momentum_shape}
\end{figure*}

In order to evaluate the global properties of the boundary layer before and after 
the microramp, at the symmetry plane, we use the incompressible displacement and momentum thicknesses, 
\begin{equation}
\delta_i^* = \int_0^\infty \left(1-\frac{u}{U_\infty}\right) \mathrm{d}y \quad \mathrm{and}\quad \theta_i = \int_0^\infty \frac{u}{U_\infty} \left(1-\frac{u}{U_\infty}\right) \mathrm{d}y
\end{equation}
which are reported in figure \ref{fig:displacement_momentum_shape}a.
Both $\delta_i^*$ and $\theta_i$ grow remarkably after the microramp, but at different rates, 
and thus the incompressible shape factor $H_i = \delta_i^*/\theta_i$ becomes lower
than the upstream condition and also than the uncontrolled cases (figure \ref{fig:displacement_momentum_shape}b),
indicating fuller boundary layers close the wall, 
which would be less prone to separation in the presence of \glspl{sbli}.
\subsubsection{Wake velocity and maximum upwash}
\begin{figure*}
     \centering
     \subfloat[\label{fig:u_wake}]{
     \includegraphics[width=0.43\textwidth]{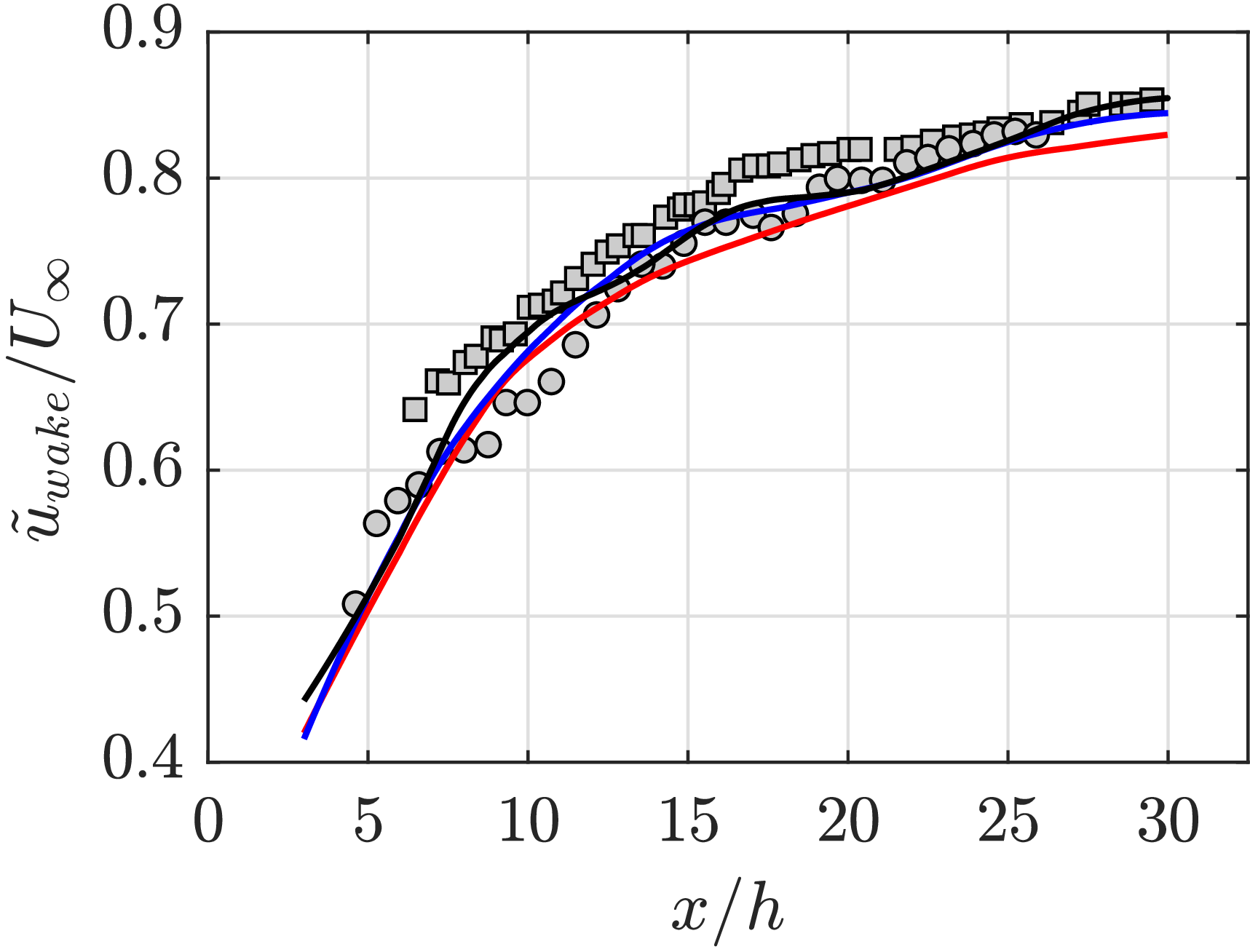}} 
     \qquad
     \subfloat[\label{fig:y_wake}]{
     \includegraphics[width=0.43\textwidth]{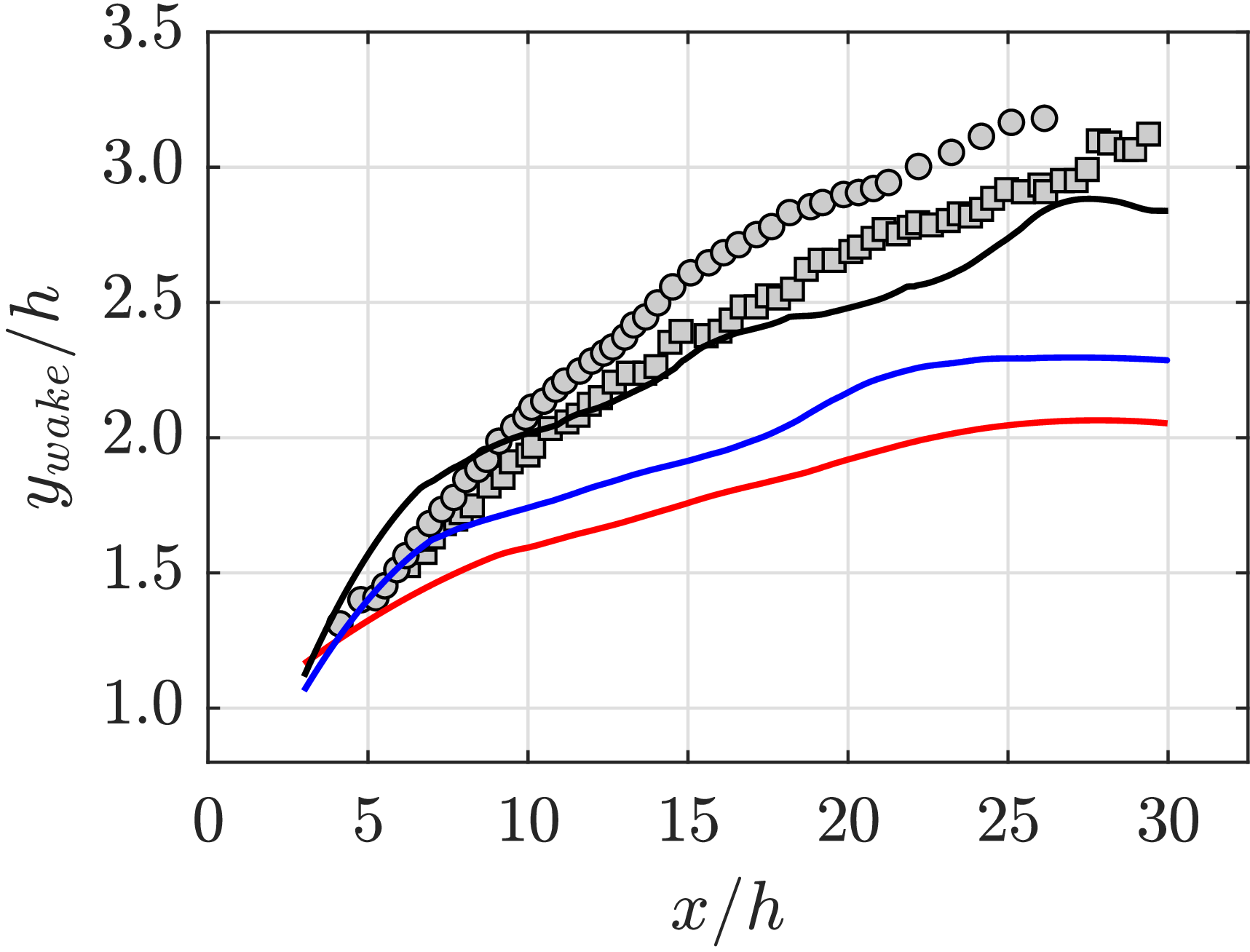}} 
     \\
     \subfloat[\label{fig:v_max}]{
     \includegraphics[width=0.43\textwidth]{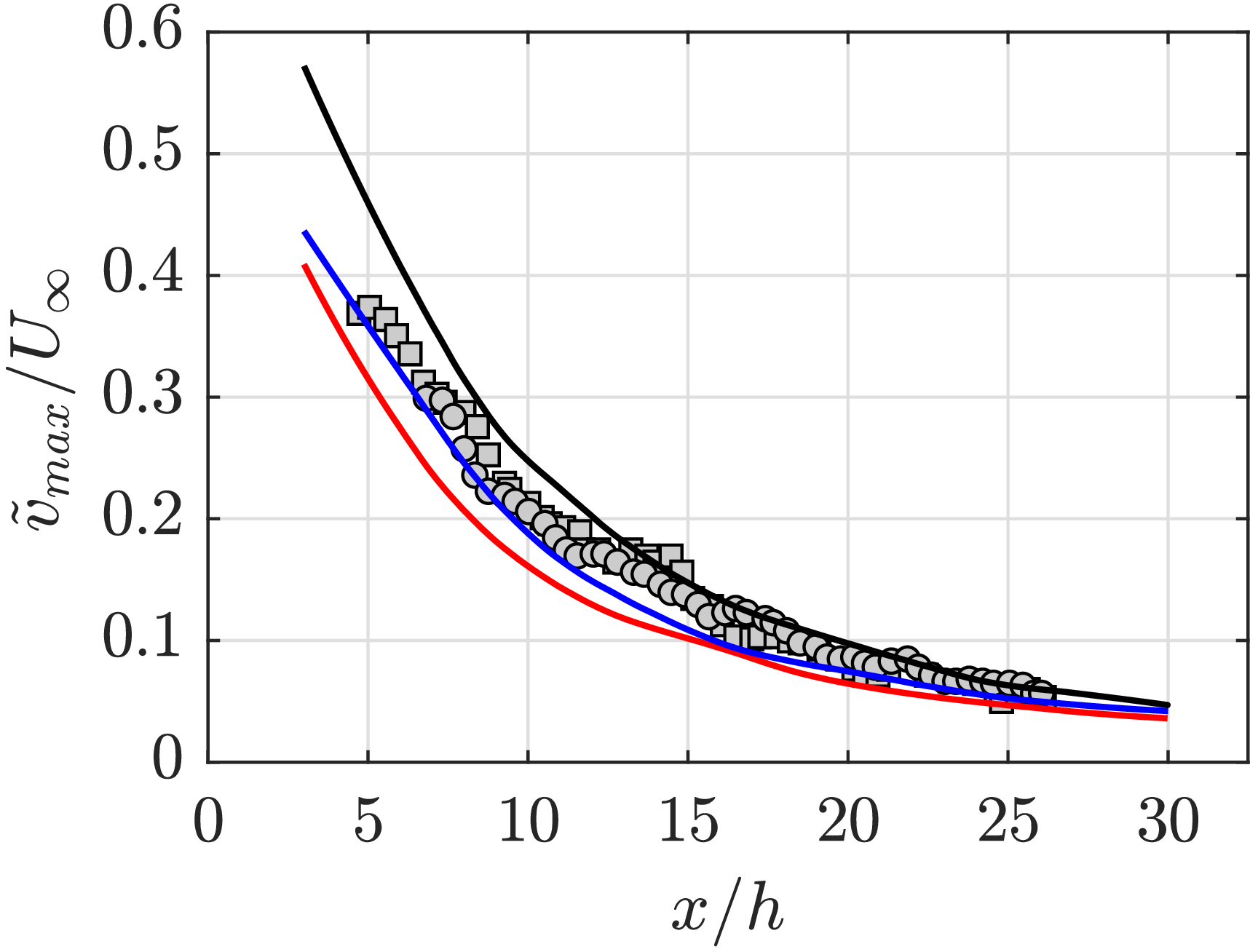}} 
     \qquad
     \subfloat[\label{fig:yv_max}]{
     \includegraphics[width=0.43\textwidth]{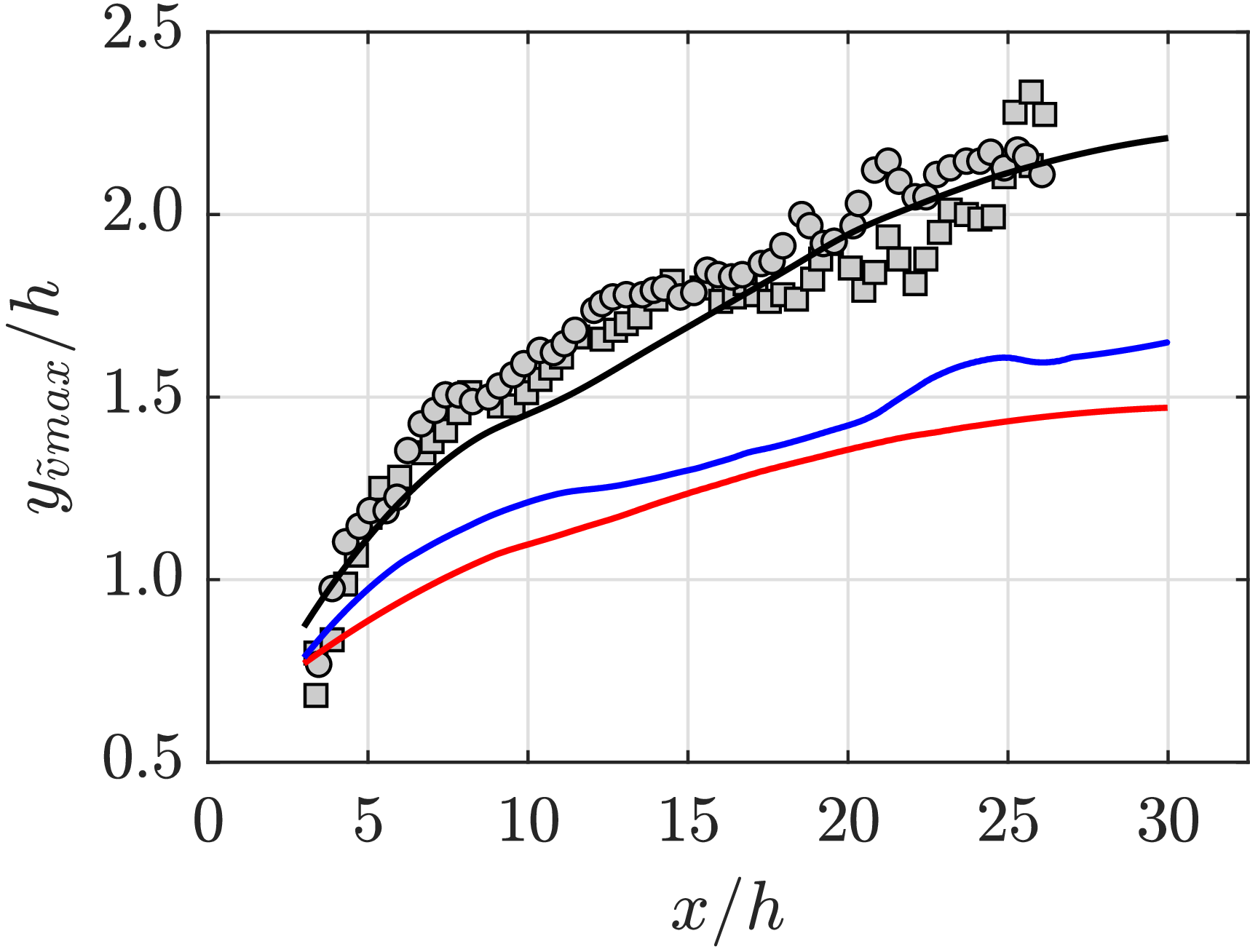}} 
     \caption{
     Streamwise evolution of (a) wake velocity and (b) wake location, (c) maximum upwash velocity, (d) maximum upwash location for different $Re_\tau$, 
     scaled by $h$. \\
     Symbols in (a) and (b) denote experiments 
     by \cite{tambe2021relation} 
     (circles, $M_\infty = 2.0$, $Re_\theta = 2.4 \times 10^4$) 
     and by \cite{giepman2014flow} 
     (squares, $M_\infty = 2.0, Re_\theta = 2.1 \times 10^4$).\\
     Symbols in (c) and (d) denote experiments 
     by \cite{giepman2016mach} 
     (circles, $M_\infty = 2.0$, $Re_\theta = 5.0 \times 10^4$ and squares, $M_\infty = 2.0$, $Re_\theta = 9.9 \times 10^4$).
     \label{fig:wake_develop}}
\end{figure*}

Another key element in the control efficiency of microramps is the strength and decay of the 
low-momentum deficit in the wake. 
In fact, in addition to the vertical motion induced by the primary vortices,
the continuous entrainment of fluid with higher momentum from outside the wake, 
the interaction between the toroidal vortical structures and the primary vortex pair, 
and the effect of viscous and turbulent dissipation, gradually enlarge and restore the region with low momentum, 
leading the wake to disappear after some distance from the ramp.

The wake development along the $x$ direction is typically analysed 
by considering the following quantities on the basis of the 
vertical velocity profiles at the symmetry plane \citep{giepman2016mach}: 
the wake velocity $\tilde{u}_{wake}$, defined as the minimum streamwise velocity in the 
low momentum region, and its corresponding vertical position $y_{wake}$; 
the maximum upwash velocity $\tilde{v}_{max}$ and its vertical position $y_{\tilde{v}max}$ (see figure \ref{fig:profiles_xh10.7}). 

\citet{babinsky2009microramp} reported the streamwise evolution of the 
momentum deficit on several cross-stream planes and, 
inspired by the work of \citet{ashill2005review} in the incompressible regime, 
hypothesised that the flow development
varies with the height of the device (for fixed boundary layer thickness). 
To test this hypothesis for geometrically 
similar ramps with different heights, the authors 
reported the wake location as a function of the streamwise distance from the 
main edge, both normalised with the ramp height. 
The collapse between the curves of the different cases, 
later confirmed also by other studies \cite{giepman2014flow}, 
indicated that, 
at least for similar inflow conditions, the ramp height is able to scale 
the wake features successfully. 
For this reason, in the following, we try also in our case to scale wake lengths by the ramp height $h$. 

Figure \ref{fig:u_wake} shows that the streamwise evolution of $\Tilde{u}_{wake}$, 
normalised by the undisturbed velocity $U_\infty$, collapses reasonably on a single curve for the 
three cases, following the experimental results of \citet{tambe2021relation} and 
\citet{giepman2014flow}. 
The wake position is much more sensitive to the Reynolds number
and numerical results converge towards the experimental ones only for the case 
at high Reynolds number, which suggests that after a certain threshold the flow becomes 
independent from the Reynolds number. 
A similar trend is also observed for the wall-normal upwash velocity $\Tilde{v}_{max}$
and its location in figure~\ref{fig:v_max} and \ref{fig:yv_max}.

\subsubsection{Self-similarity of velocity profiles} \label{sec:self_similarity}
\begin{figure*}
     \centering
     \subfloat{
     \includegraphics[width=0.5\textwidth]{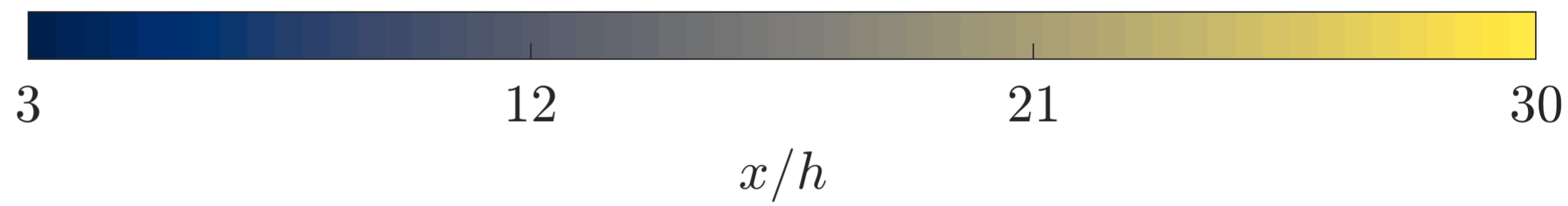}}
     \\
     \setcounter{subfigure}{0}
     \subfloat[]{
     \includegraphics[width=0.495\textwidth]{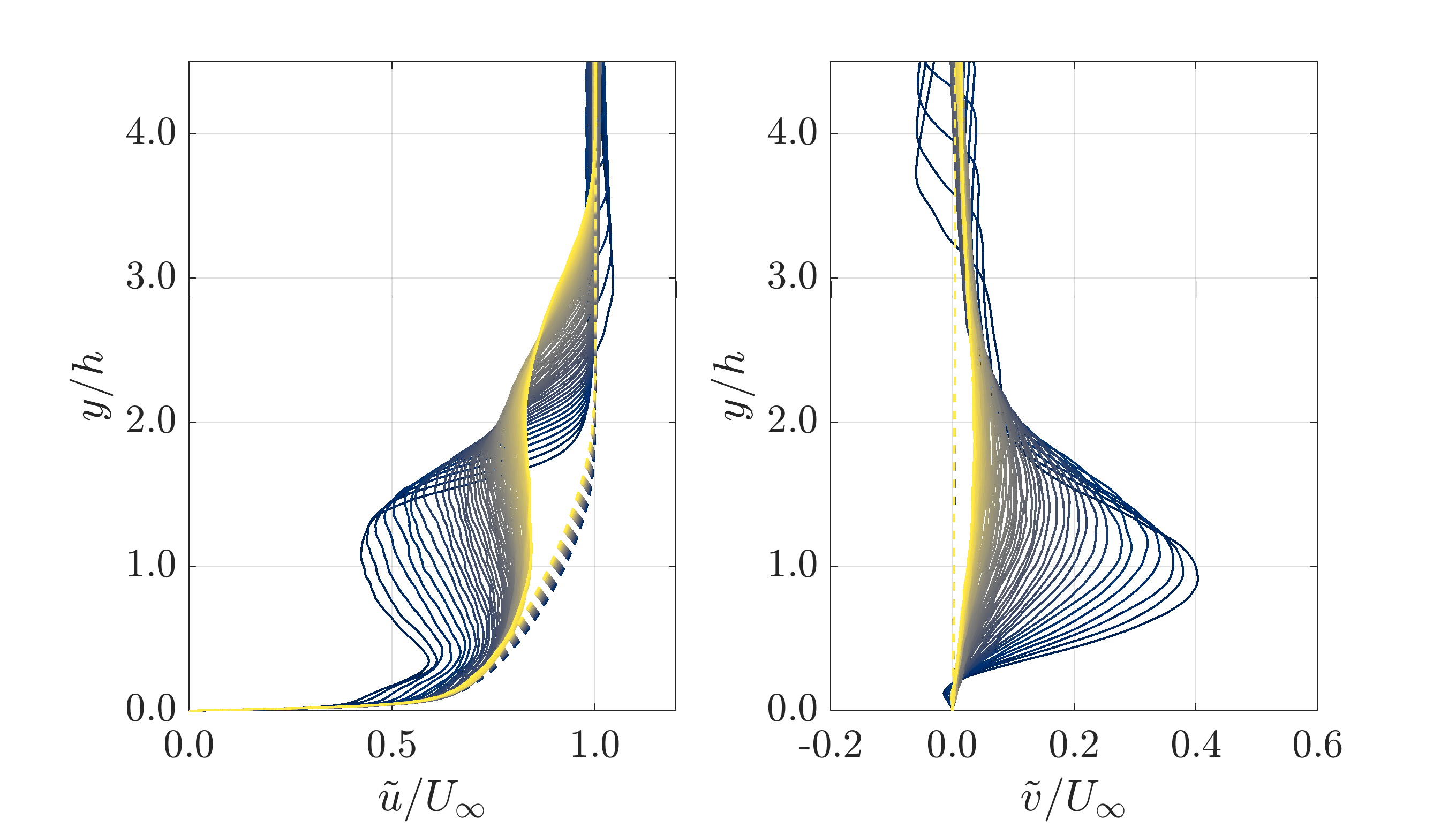}
     \includegraphics[width=0.495\textwidth]{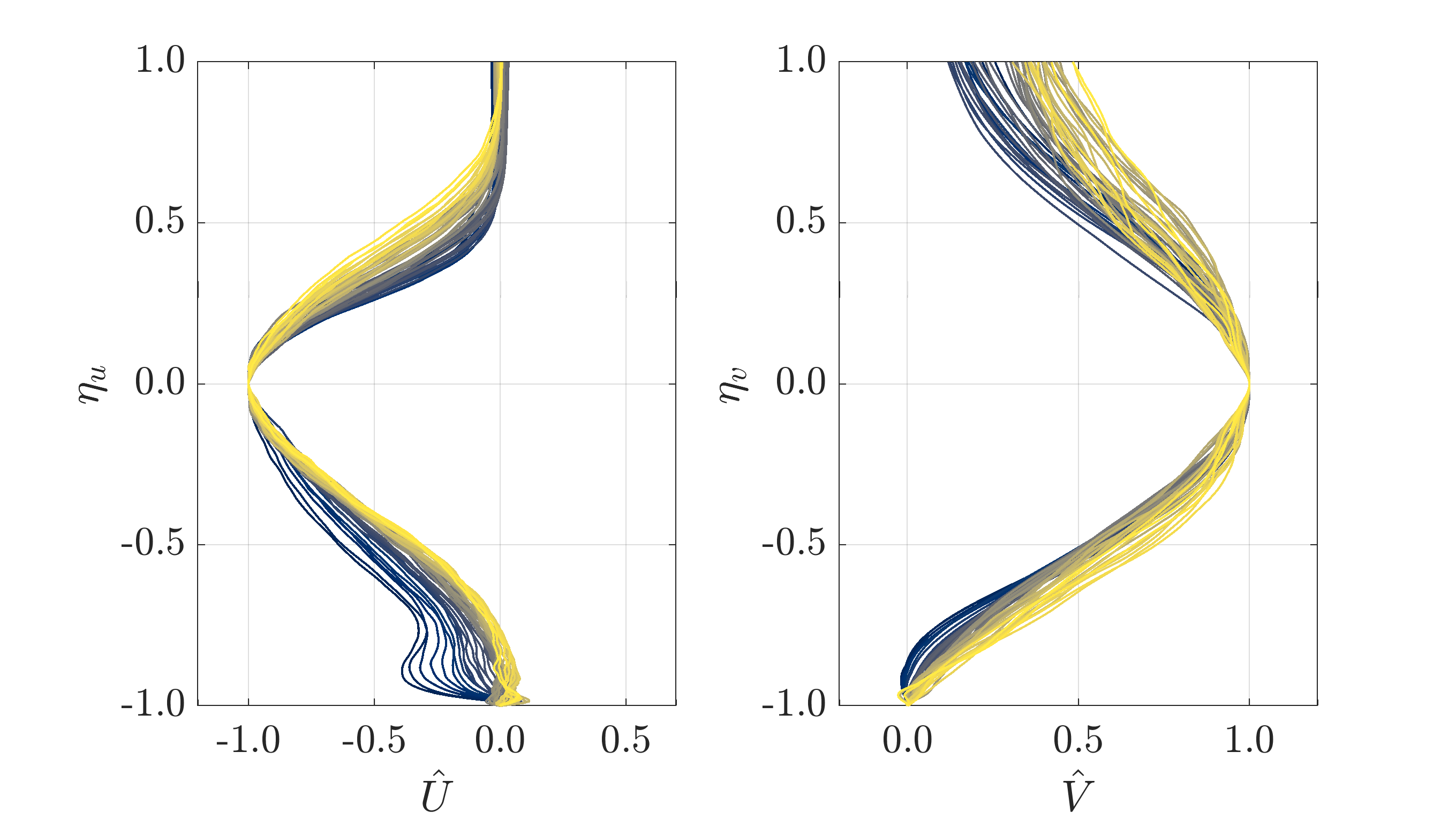}}
     \\
     \subfloat[]{
     \includegraphics[width=0.495\textwidth]{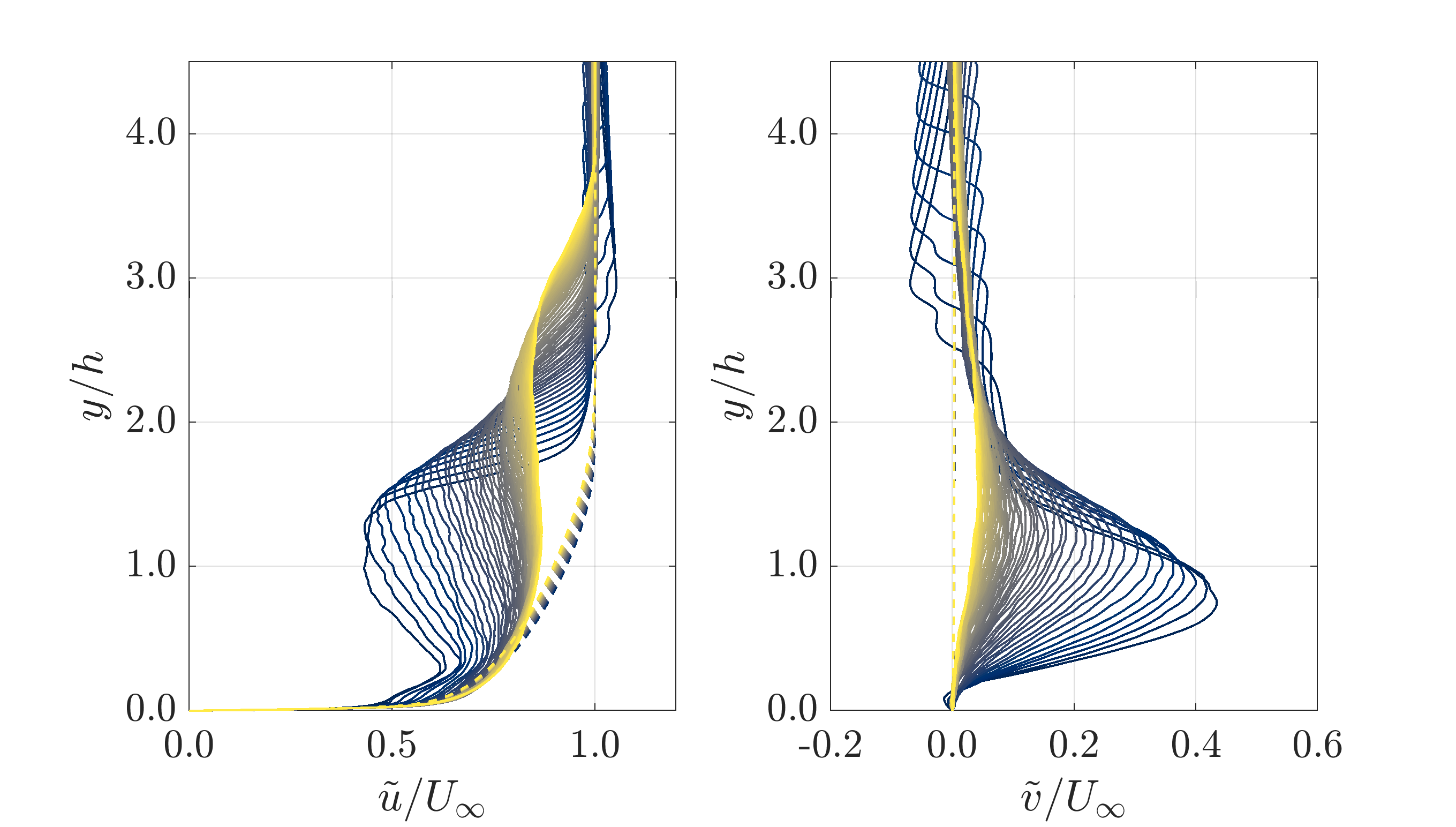}
     \includegraphics[width=0.495\textwidth]{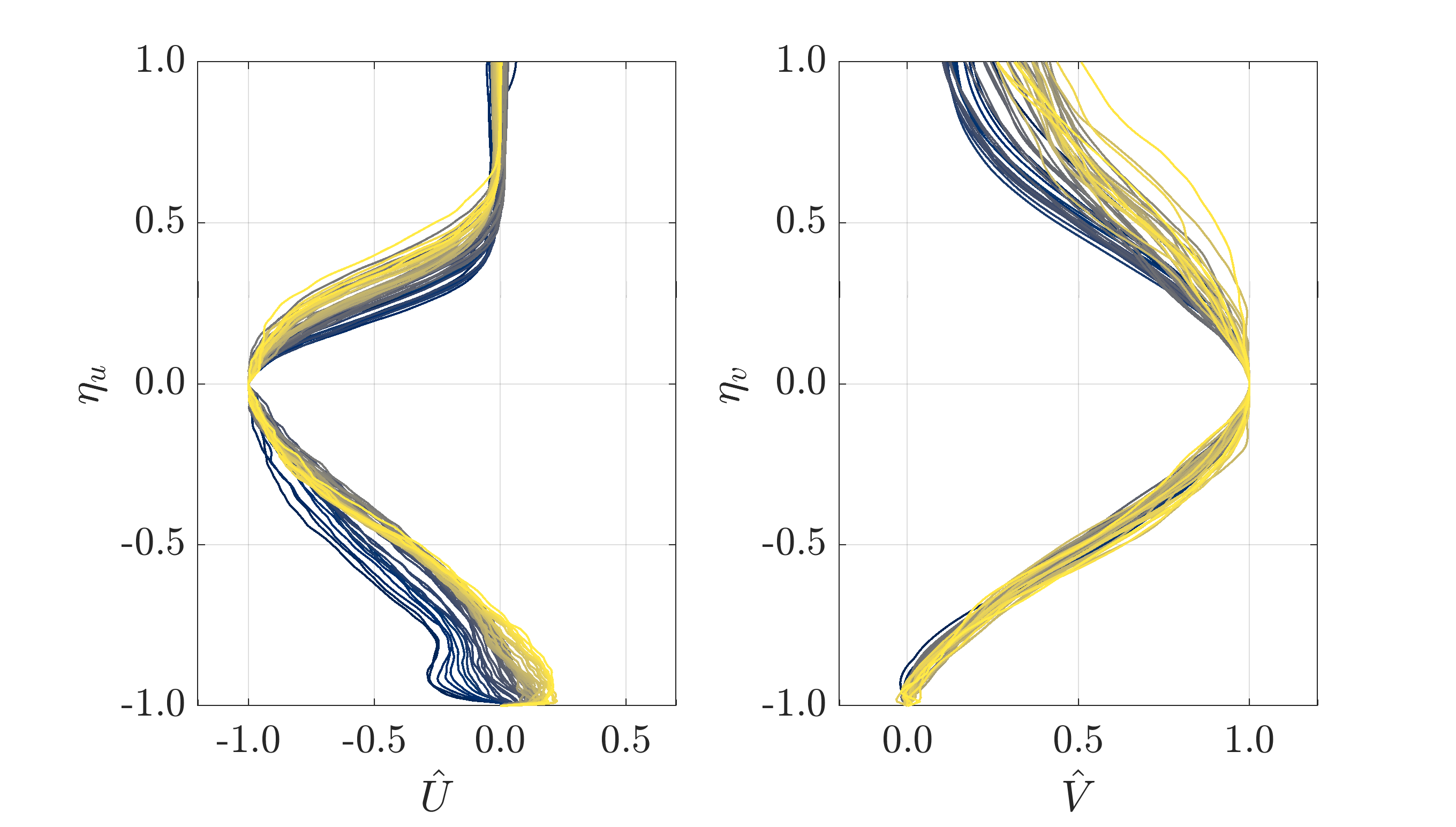}}
     \\
     \subfloat[]{
     \includegraphics[width=0.495\textwidth]{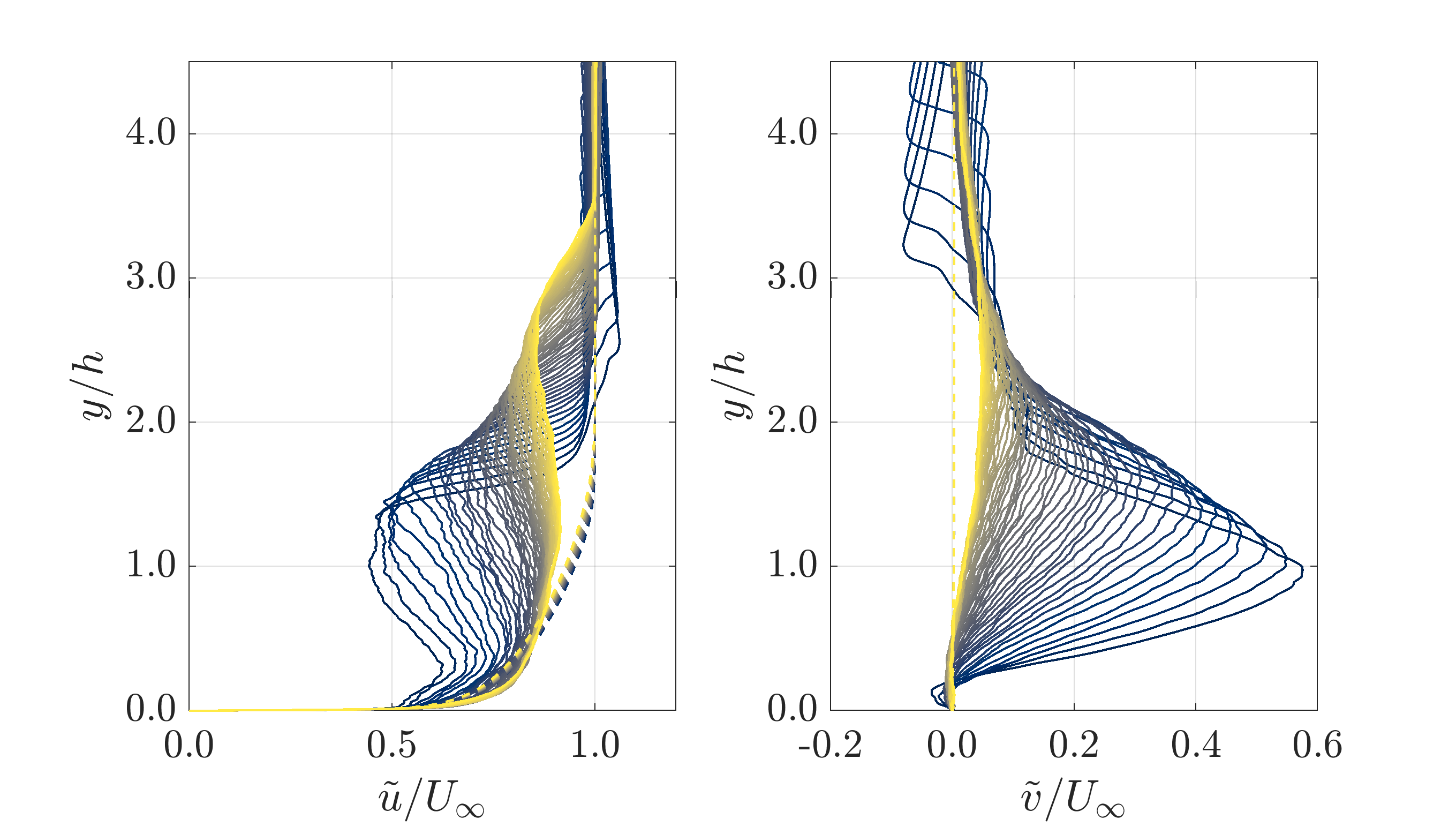}
     \includegraphics[width=0.495\textwidth]{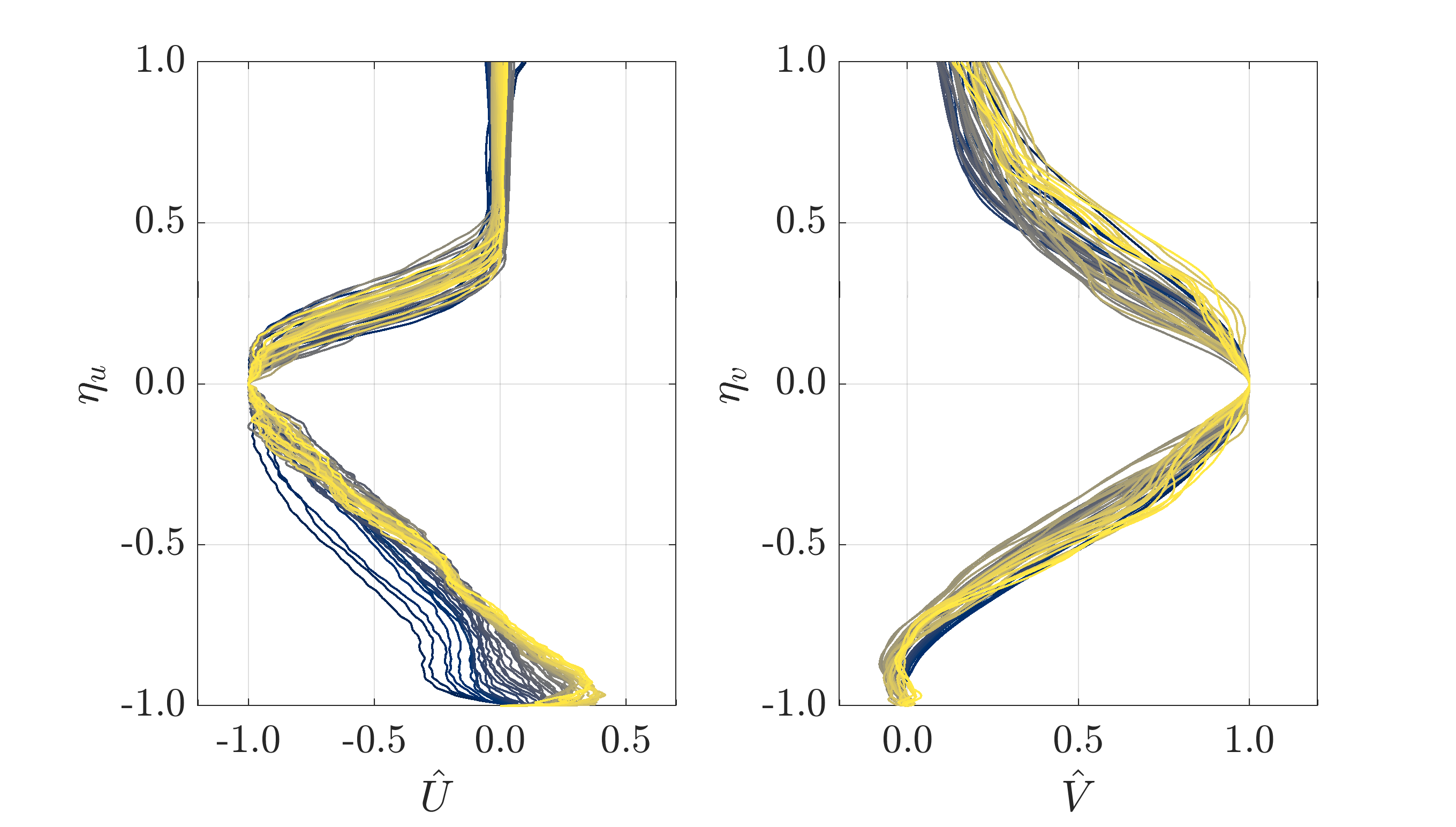}}
     \caption{Velocity profiles at: (a) $Re_\tau=500$, (b) $Re_\tau=1000$, (c) $Re_\tau=2000$. 
     Non-normalised (left) and normalised (right) streamwise and wall-normal velocity.
     Colour indicates the streamwise location according to the colorbar.  Dashed lines indicate
     the uncontrolled boundary layer.}
     \label{fig:selfsimilarity}
\end{figure*}
\begin{figure*}
     \centering
     \subfloat[]{
     \includegraphics[width=0.3\textwidth]{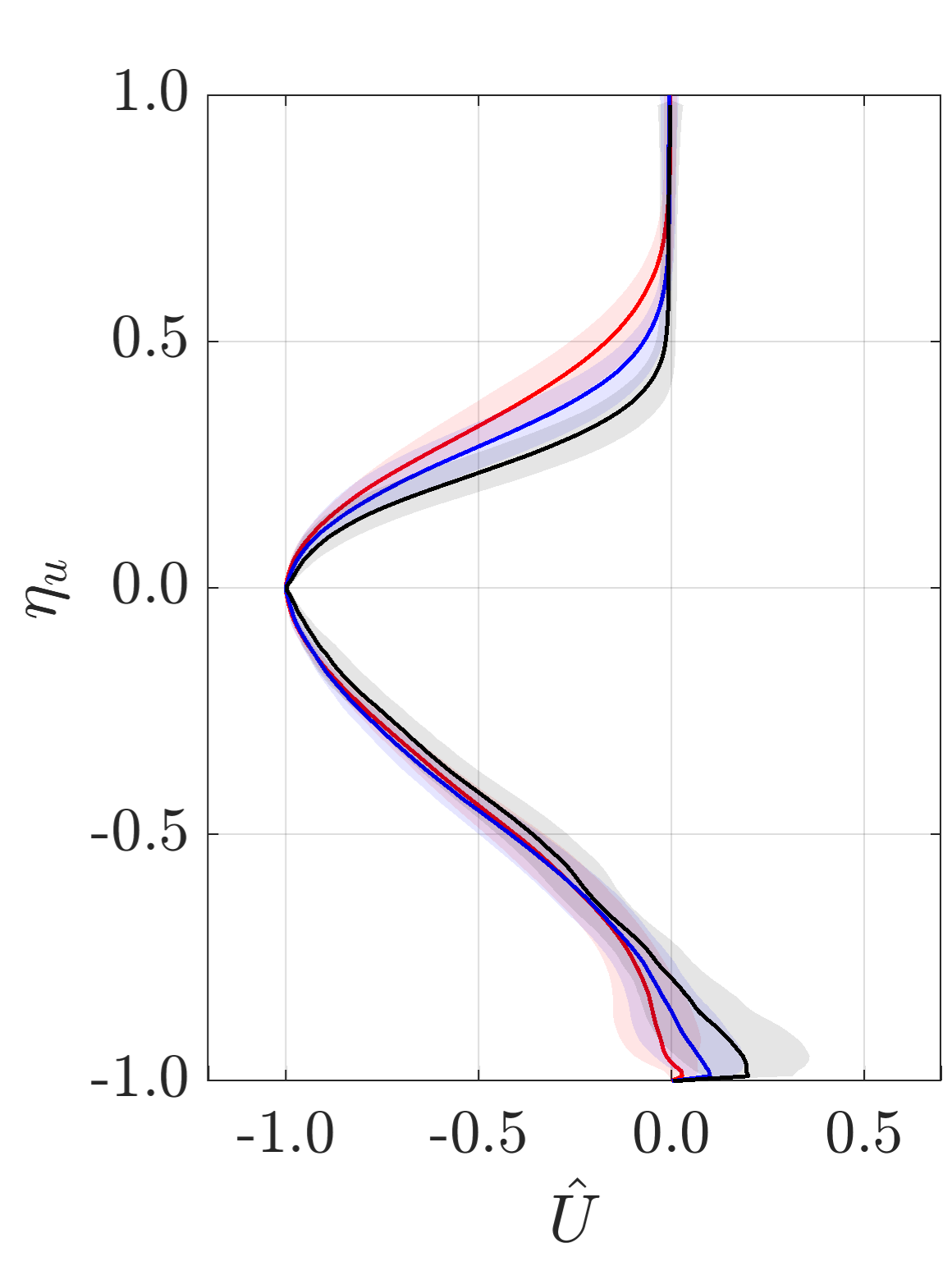}}
     \qquad \qquad
     \subfloat[]{
     \includegraphics[width=0.3\textwidth]{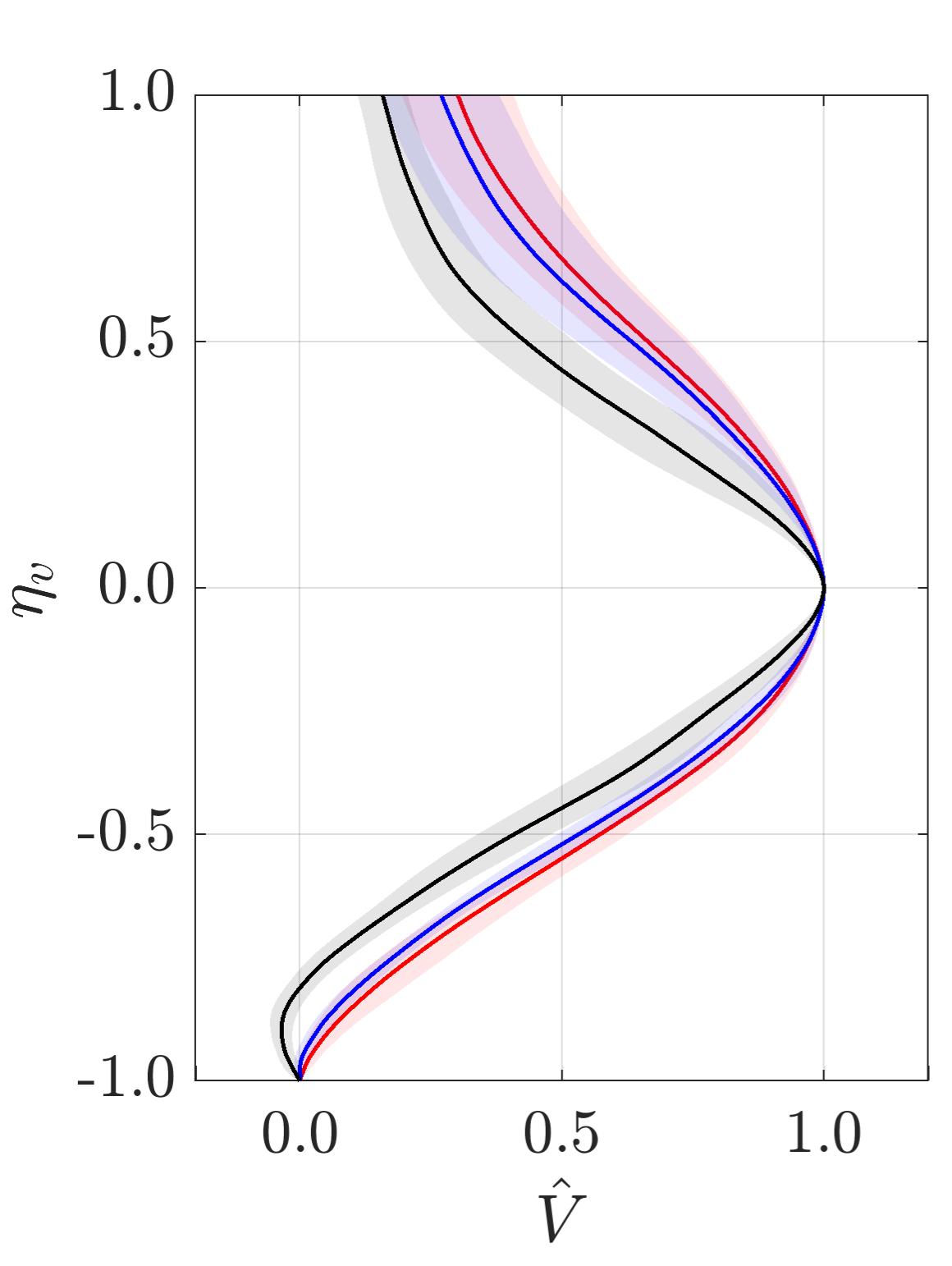}}
     \caption{Normalised velocity profiles for different $Re_\tau$: (a) streamwise component and (b) wall-normal component. The normalised profiles are averaged 
     in the streamwise direction. Banded regions indicate the corresponding
     standard deviation from the streamwise-averaged profile. }
     \label{fig:selfsimilarity_comparison}
\end{figure*}

In this section, we also examine the self-similarity of the velocity profiles following 
the formulation proposed by \citet{sun2014decay}. 
For the profiles of the streamwise component, the velocity is normalised by 
the quantity $\Delta \Tilde{u}_{deficit}$, defined as the maximum difference between the 
velocity profile of the controlled case and that of the undisturbed boundary layer $u_{BL}$, 
while the vertical coordinate is normalised by the wake location $y_{wake}$.
Thus, we have that
\begin{equation}
    \hat{U}(\eta_u) = \frac{\Tilde{u}(\eta_u) - u_{BL}(\eta_u)}{\Delta \Tilde{u}_{deficit}}\,,
    \qquad \eta_u = \frac{y - y_{wake}}{y_{wake}}
\end{equation}
On the other hand, for the wall-normal component, the velocity is normalised by the maximum upwash velocity $\Tilde{v}_{max}$, while the vertical coordinate is normalised by $y_{\tilde{v}max}$, so that
\begin{equation}
    \hat{V}(\eta_v) = \frac{\Tilde{v}(\eta_v)}{\Tilde{v}_{max}} \,, 
    \qquad\eta_v = \frac{y - y_{\tilde{v}max}}{y_{\tilde{v}max}}
\end{equation}
%
%
As also demonstrated by \citet{sun2020wake}, velocity
profiles exhibit self-similarity, especially in the central region of the wake, whereas 
discrepancies are expected in the near wake and in the 
near-wall region, where the bottom secondary vortices
induce a region of low velocity and flow reversal can also occur.

The normalised velocity profiles at several equispaced 
streamwise locations 
in the region $x/h = 3$ to $x/h = 30$ are shown in 
figure \ref{fig:selfsimilarity}. 
A very good agreement is observed for the
scaled profiles, which is quantified in
figure~\ref{fig:selfsimilarity_comparison}
by reporting the streamwise average of
the scaled profiles along with the corresponding standard
deviation, for the three Reynolds numbers.

Excluding the very near-wall region where
viscous effects are dominant,
the maximum standard deviation for the streamwise 
velocity component $\sigma_{\hat{U}}$ is 0.078, 0.082, 0.091 for 
the low, intermediate, and high Reynolds numbers respectively, 
which confirms that $\tilde{u}$ is self-similar.
Differences among the three flow cases are also
of the same order of the standard deviation, and 
therefore are considered satisfactory.

Regarding the normalised vertical velocity profiles, 
minor deviations 
are visible in the lower region of the wake, and the standard
deviations is generally very small.
Vertical self-similarity is more troublesome to infer 
in the upper part of the wake and has a 
different behaviour especially 
for higher Reynolds number. 
In particular, the outer region
is also influenced by the trailing edge shock, and  
thus perfect self-similarity is difficult to achieve.
Additionally, the increased lift-up and the faster rise of the wake for the higher Reynolds number case,
is strong enough to induce a region of negative $\tilde{v}$ at the symmetry plane 
(see figure~\ref{fig:vertical_velocity_comparison}), which further 
aggravates the differences with respect to the other two cases.



\subsection{Mean wake organisation}

In the following, we consider the evolution of the time-averaged flow quantities 
in the very near ($x/h < 2$) and far wake ($x/h > 4$) behind the microramp. 

\subsubsection{The near wake}
\begin{figure*}
     \centering
     \subfloat[ \label{fig:vel_near}]{
     \includegraphics[width=0.83\textwidth]{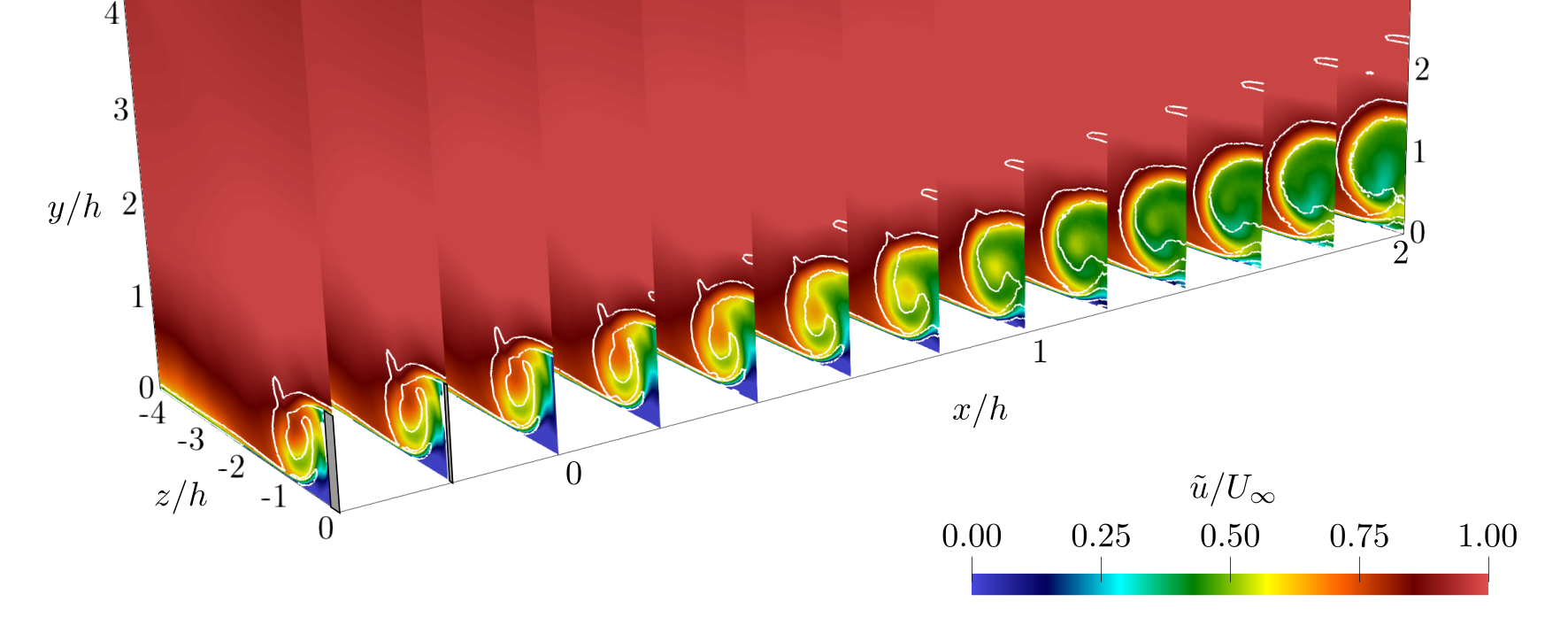} 
     }\\
     \vspace{-10pt}
     \subfloat[ \label{fig:vortx_near}]{
     \includegraphics[width=0.83\textwidth]{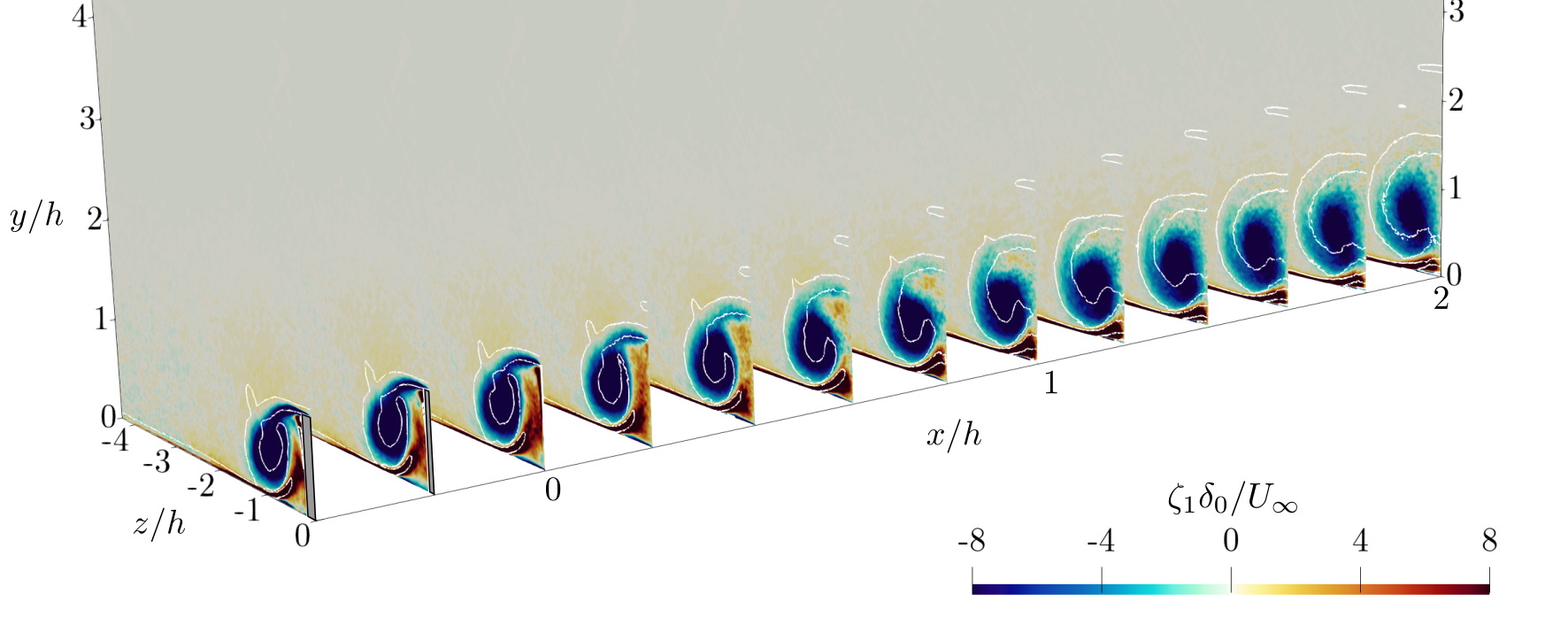}
     }\\
     \vspace{-10pt}
     \subfloat[ \label{fig:vorty_near}]{
     \includegraphics[width=0.83\textwidth]{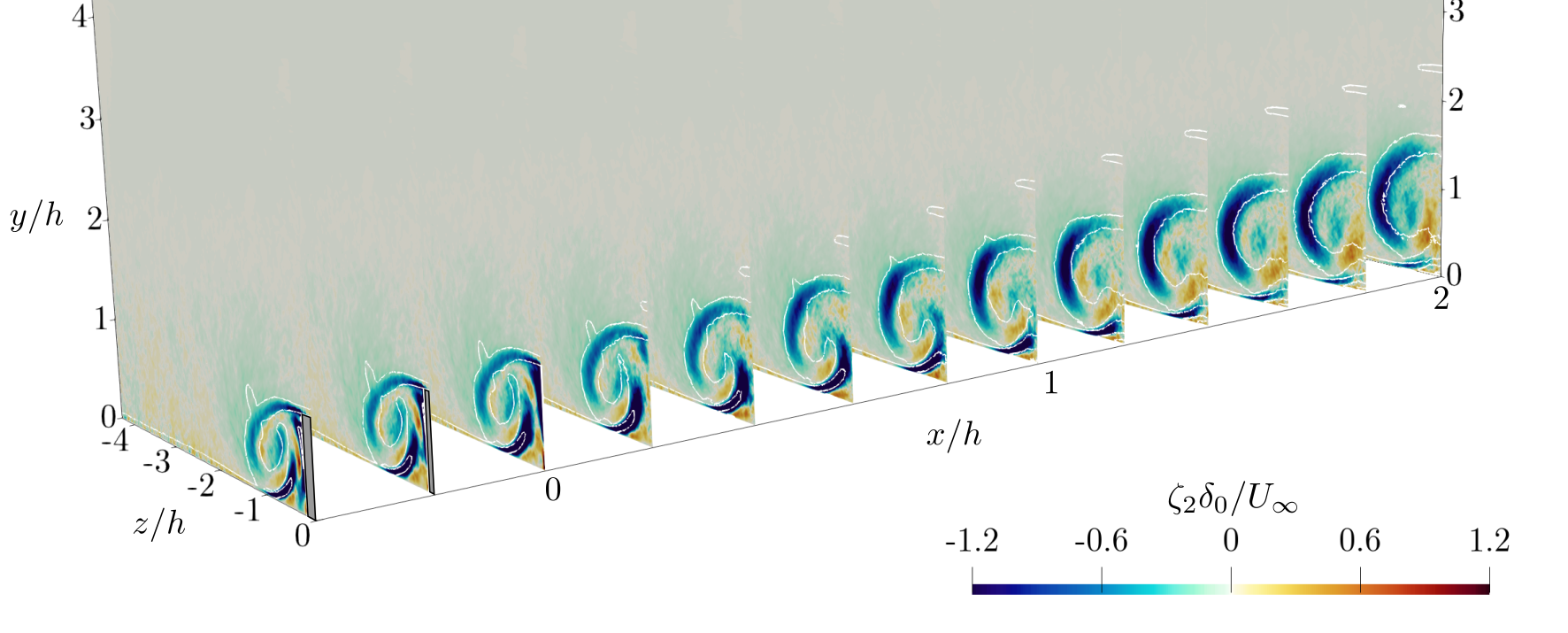}
     }\\
     \vspace{-10pt}
     \subfloat[ \label{fig:vortz_near}]{
     \includegraphics[width=0.83\textwidth]{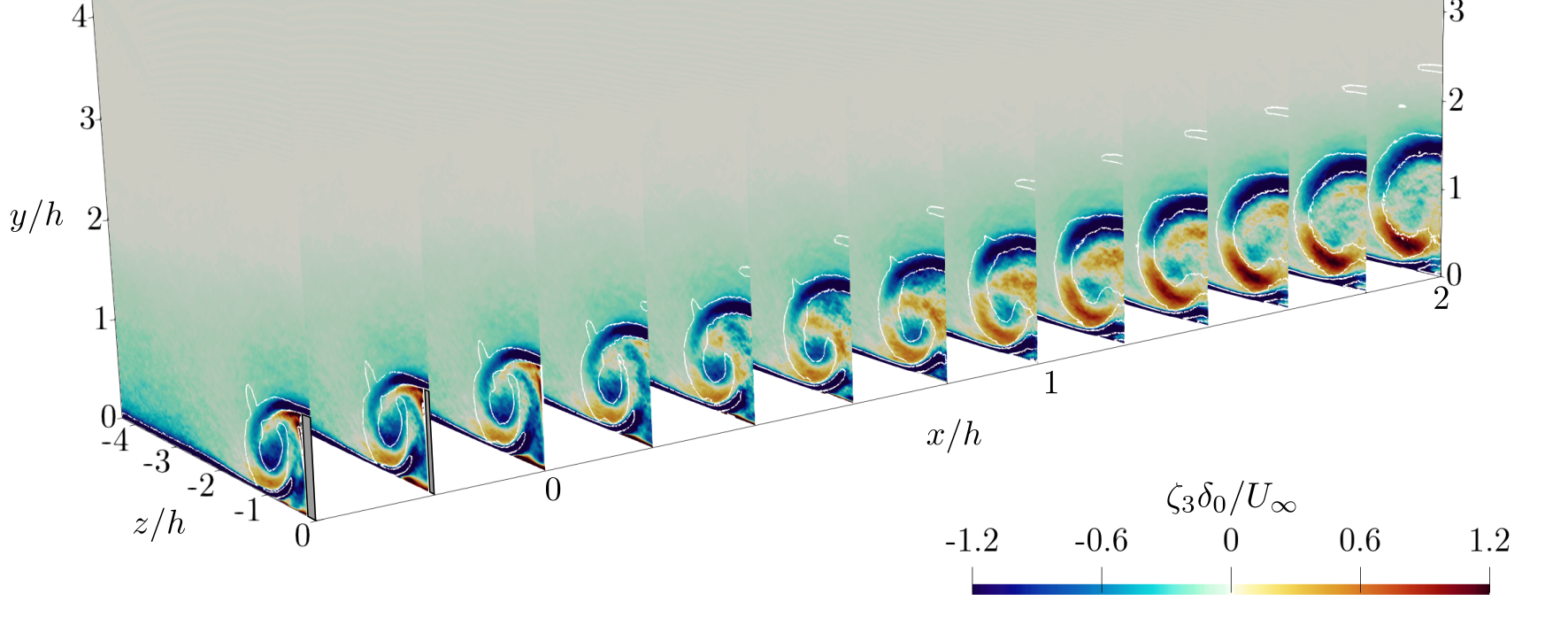} 
     }
     \caption{Multiple vertical slices in the longitudinal range $x/h \in [-0.4, 2]$ for the 
     case at intermediate Reynolds number. Favre-averaged: (a) streamwise velocity, 
     (b) streamwise, (c) wall-normal, (d) spanwise vorticity components.}
     \label{fig:near_field_slices}
\end{figure*}


We have seen in the previous sections that the captured flow from the incoming boundary layer 
generates two symmetric vortical structures at the sides of the ramp, which converge at 
the trailing edge and then proceed approximately in parallel.  
The behaviour and decay of the primary vortex pair far from the edge has been widely considered 
in several experimental studies~\citep{sun2014decay, giepman2014flow, giepman2016mach, tambe2021relation}, and although some aspects still remain unclear, 
several studies provided valuable insights into the development of the far wake. 

On the other hand, 
less is known about the near wake, as this region
is difficult to study experimentally due 
to the presence of the solid walls.
Despite its reduced extent, in this critical region important changes take place
that influence the evolution of the entire wake downstream. 
In order to draw a qualitative picture
of the flow in the near wake, we report in
figure \ref{fig:near_field_slices} the Favre-averaged streamwise velocity component 
and the three orthogonal components of the Favre-averaged vorticity $\boldsymbol{\zeta}$ 
on equispaced cross-stream planes in the region $x/h \in [-0.4, 2]$ for the flow case at intermediate Reynolds number. 
White isolines indicate a constant value of the density gradient, 
highlighting the edges of the main vortices and shocks, as reported in 
figure \ref{fig:slicesxy_shocks}. 
Due to symmetry, only half of the domain is reported and discussed.

The velocity field in figure~\ref{fig:near_field_slices}(a) well highlights the core of the primary vortex,
from which we can infer its internal organization.
Close to the walls, especially right after the ramp, 
we note an extended region of low-speed reversed flow, 
corresponding to a small separation
and a near-wall secondary vortex. 
This region shrinks progressively downstream and flow reversal is 
almost absent at the end of the considered interval. 
The isolines of the density gradient 
on top of the main vortex show 
the oblique shock on the side vortex propagating downstream
and disappearing 
approximately one $h$ from the ramp. 
The isoline of the density gradient
also highlights the spreading and internal convolution of the
main vortex, 
represented by the negative values
of $\zeta_1$ in figure~\ref{fig:near_field_slices}(b).


The distributions of the wall-normal and spanwise vorticity components in 
figure~\ref{fig:near_field_slices}(c)-(d) complements the 
understanding of the above described vortices. 
In particular, before the main edge, the two components help understand 
the orientation of the primary and secondary vortices
and are able to highlight the internal convolution of the side vortices, 
superposed to the main streamwise helical motion. 
After the main edge, where secondary vortices become less important, 
the transversal components $\zeta_2$ and $\zeta_3$ are particularly useful to 
capture the trace of the vortical structures inside the wake. 

\begin{figure*}
     \centering
     \includegraphics[width=0.9\textwidth]{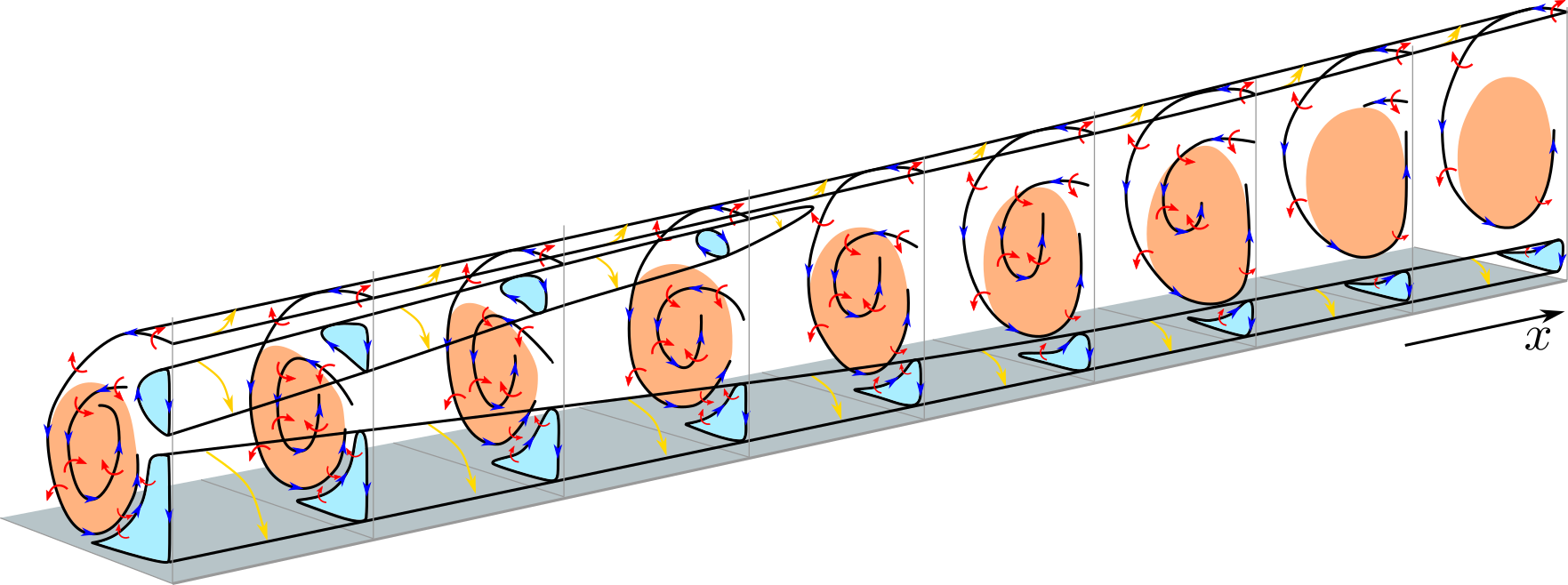} 
     \caption{Model sketch of the vortical structures in the near wake ($x/h \in [0.4, 2]$). \\
     Yellow arrows indicate the orientation of the helical flow in the primary and secondary vortices;
     blue arrows indicate the flow rotation in the yz planes; 
     red arrows indicate the orientation of the transversal rotation of the 
     vortical structures superposed to the main helical motion; 
     black lines indicate the approximate boundaries of the primary, secondary 
     and inner vortices. 
     Finally, the region with negative (positive) streamwise vorticity related to 
     the primary vortex (secondary vortices) is indicated in orange (light blue).}
     \label{fig:model_near_field}
\end{figure*}
On the basis of the behaviour of the observed quantities, in 
figure~\ref{fig:model_near_field} we propose
a qualitative description summarising 
the evolution of the vortical structures in the near wake ($x/h \in [0.4, 2]$). 
Before the trailing edge, the presence of the vertical lateral wall of the ramp 
and the inclination of the primary vortices imposes the formation of two asymmetric 
secondary vortices rotating in the opposite direction of the primary vortices.
As soon as the fluid from the top of the ramp flows 
over the converging vortex pair, the strong shear induces 
the roll-up of \gls{kh} instabilities. The \gls{kh} vortices on top of the wake
thus join laterally the counter-rotating helical motion of the two primary vortices, 
and the typical almost-toroidal vortices are generated as a result.
Given its reduced intensity and the significant activity in the top part of the wake, 
the upper secondary vortex first moves upwards and laterally, 
following the rotation of the primary vortex, and then disappears soon after the ramp. 
The bottom secondary vortex instead persists for a longer distance.
However, once 
the primary vortices align and the mutual interaction between them is fully established, 
the upward vertical motion in between the vortex pair reduces drastically the intensity 
of the bottom vortex.\\
While the wake expands, the lift-up in the central region 
also affects the internal convolution of the primary vortices. 
The combined distributions of $\zeta_2$ and 
$\zeta_3$ 
makes it possible to 
identify an internal ring-like structure with tangential vorticity oriented in the 
opposite direction compared to the external almost-toroidal vortices.
This structure is superposed to the primary vortices and 
is progressively pushed upwards by the increased vertical motion induced by the 
primary vortex pair.
In the space where the wake gradually starts to lift up, the 
intensity of these internal structures decreases and, as a result, 
there is no evident sign of them after less than $2\,h$ from the ramp 
trailing edge anymore.
Close to the symmetry plane, on the other hand, 
convolution is still visible from $\zeta_2$ (around $ z/h \approx -0.5$ in figure~\ref{fig:near_field_slices}(c)), consistently with the 
observations of secondary vortices inside the wake in the \gls{piv} measurements of \cite{sun2012three} and in the numerical results of \cite{sun2014numerical}. 
Contrarily to what proposed by \citet{sun2014numerical}, however, 
our observations suggest that the ``curved legs'' of the \gls{kh} vortices are 
already present from the start and are  
a consequence and evolution of the initial convolution of the primary vortices. 
Only far from the ramp (approximately $x/h > 15$), 
the tangential vorticity of the inner, vertical legs is almost completely 
dissipated and only a mild trace is visible inside the wake from $\zeta_2$ 
close to the symmetry plane. 
This suggests that in the first part of the wake, the \gls{kh} vortices are not closed 
near the wall but are instead strictly tied to the helical motion of the 
primary vortices they surround.
A stricter closure of the tangential vorticity takes place only far downstream, 
where instead we observed a significant decrease in coherence of the 
azimuthal structures (see figure~\ref{fig:3d_qcrit_inst}),
at least at the present Reynolds numbers.

\subsubsection{The far wake}
\begin{figure*}
     \centering
     \subfloat{
     \includegraphics[width=0.19\textwidth]{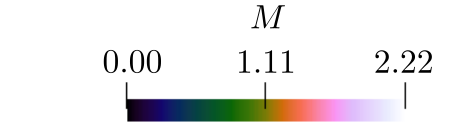} }
     \subfloat{
     \includegraphics[width=0.19\textwidth]{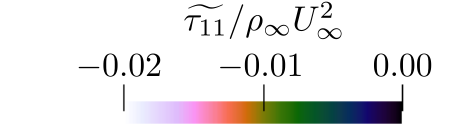} } 
     \subfloat{
     \includegraphics[width=0.19\textwidth]{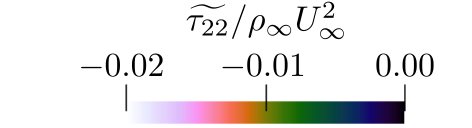} }
     \subfloat{
     \includegraphics[width=0.19\textwidth]{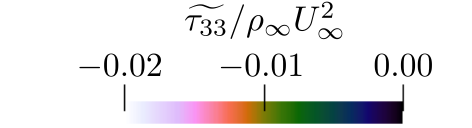} }
     \subfloat{
     \includegraphics[width=0.19\textwidth]{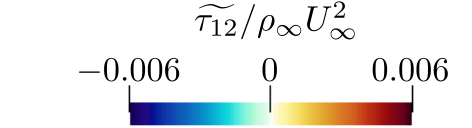} }
     \\
     \setcounter{subfigure}{0}
     \subfloat[]{
     \includegraphics[width=0.19\textwidth]{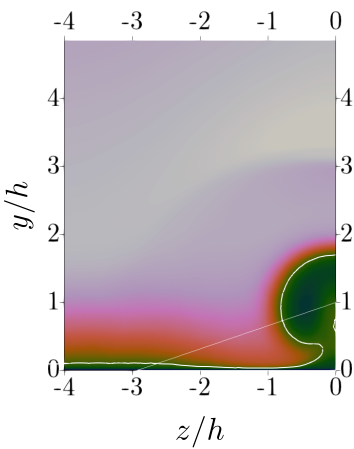}
     }
     \subfloat[]{
     \includegraphics[width=0.19\textwidth]{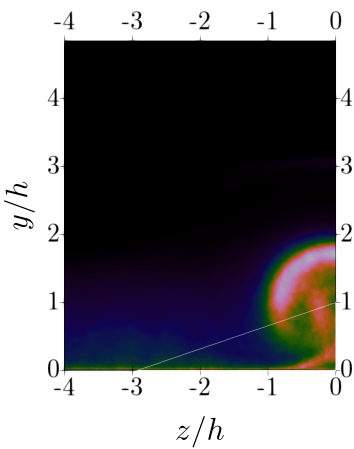}
     }
     \subfloat[]{
     \includegraphics[width=0.19\textwidth]{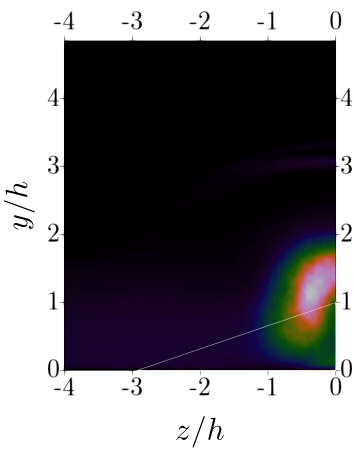}
     }
     \subfloat[]{
     \includegraphics[width=0.19\textwidth]{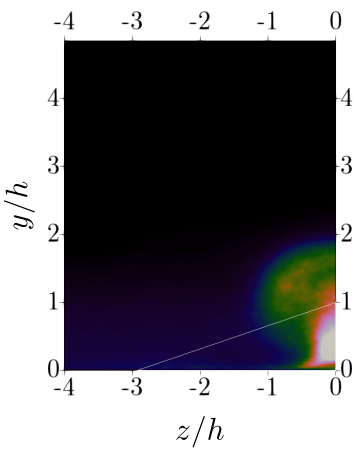}
     }
     \subfloat[]{
     \includegraphics[width=0.19\textwidth]{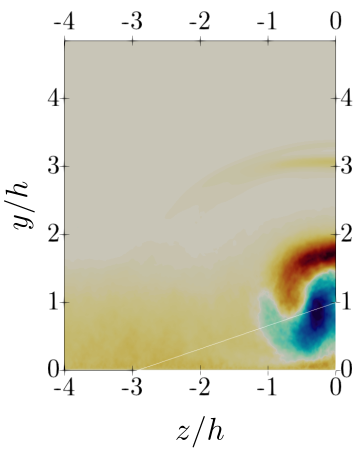}
     }
     \\
     \vspace{-5pt}
     \subfloat[]{
     \includegraphics[width=0.19\textwidth]{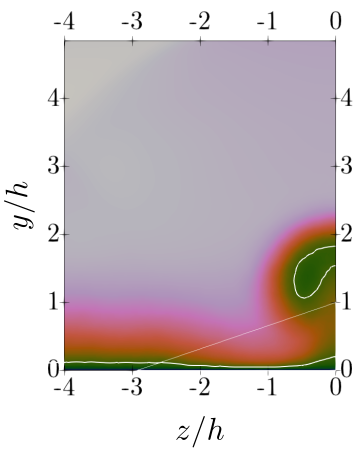}
     }
     \subfloat[]{
     \includegraphics[width=0.19\textwidth]{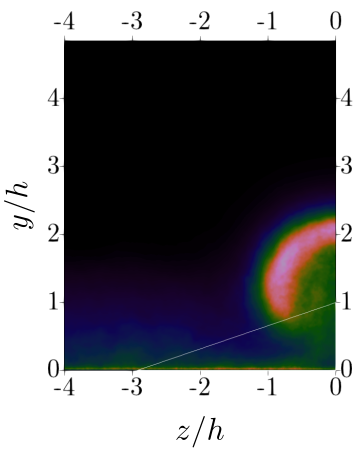}
     }
     \subfloat[]{
     \includegraphics[width=0.19\textwidth]{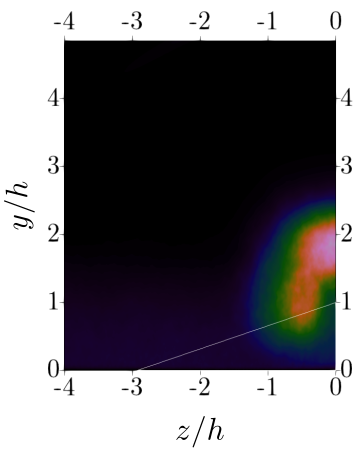}
     }
     \subfloat[]{
     \includegraphics[width=0.19\textwidth]{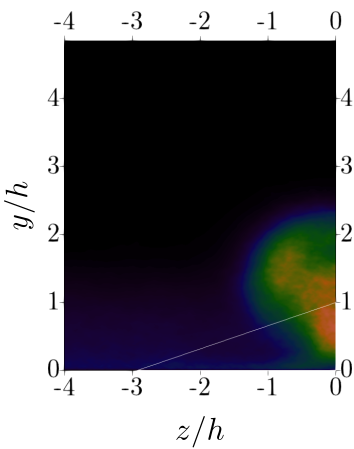}
     }
     \subfloat[]{
     \includegraphics[width=0.19\textwidth]{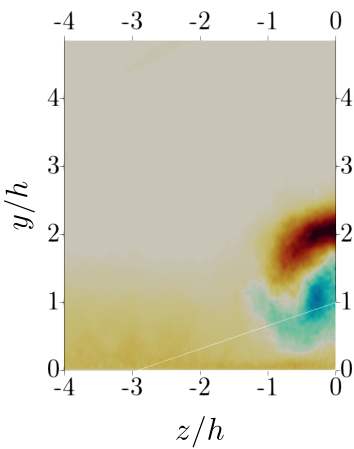}
     }
     \\
     \vspace{-5pt}
     \subfloat[]{
     \includegraphics[width=0.19\textwidth]{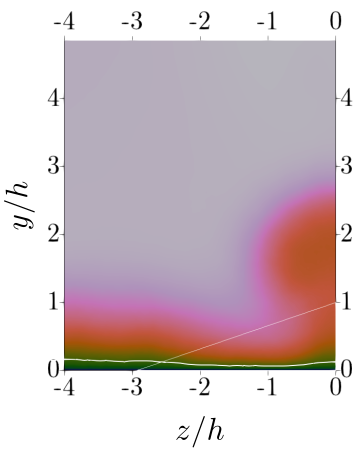}
     }
     \subfloat[]{
     \includegraphics[width=0.19\textwidth]{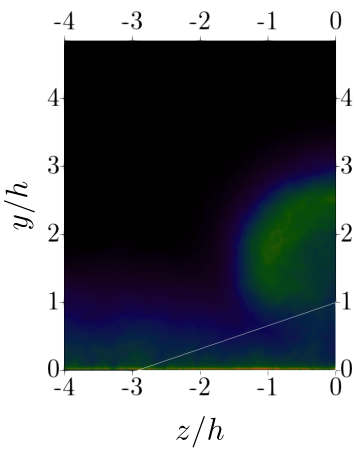}
     }
     \subfloat[]{
     \includegraphics[width=0.19\textwidth]{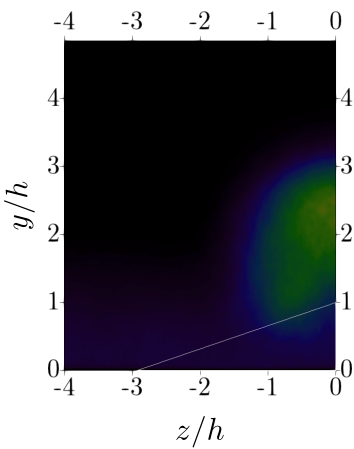}
     }
     \subfloat[]{
     \includegraphics[width=0.19\textwidth]{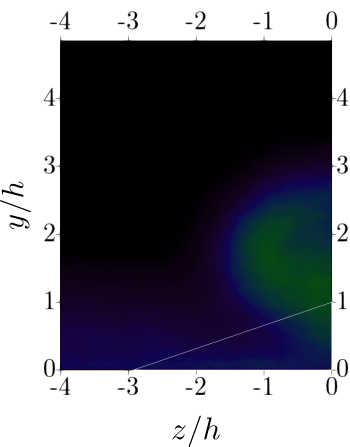}
     }
     \subfloat[]{
     \includegraphics[width=0.19\textwidth]{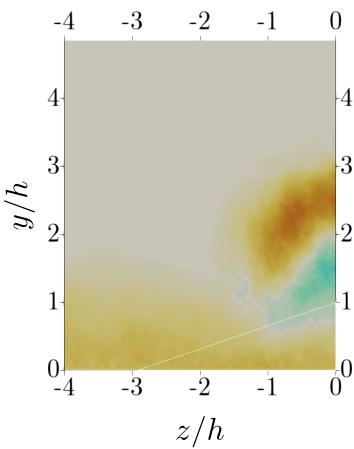}
     }
     \caption{Contours of the Favre-averaged Mach number $\widetilde{M}$ (first column),
      and of the Favre-averaged Reynolds stress components $\widetilde{\tau_{11}}/\rho_\infty U_\infty^2$ (second column), 
      $\widetilde{\tau_{22}}/\rho_\infty U_\infty^2$ (third column), $\widetilde{\tau_{33}}/\rho_\infty U_\infty^2$ (fourth column),
      $\widetilde{\tau_{12}}/\rho_\infty U_\infty^2$ (fifth column) for the intermediate Reynolds number 
      at $x/h = 4$ (first row), $8$ (second row), and $16$ (third row). 
      White isolines in the Mach contours indicate the points with $\widetilde{M} = 1$.
      }
     \label{fig:far_field}
\end{figure*}
\begin{figure*}
     \centering
     \subfloat{
     \includegraphics[width=0.22\textwidth]{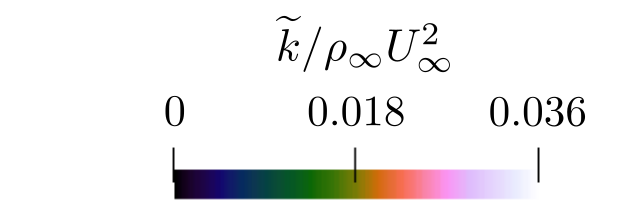}}
     \,
     \subfloat{
     \includegraphics[width=0.22\textwidth]{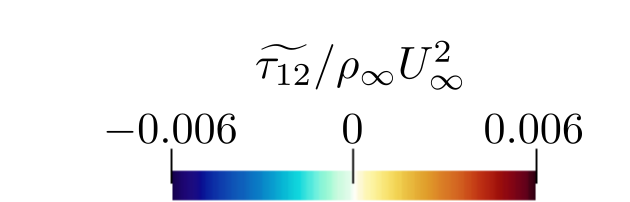}} 
     \,
     \subfloat{
     \includegraphics[width=0.22\textwidth]{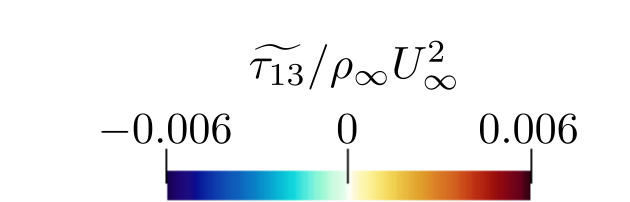}}
     \,
     \subfloat{
     \includegraphics[width=0.22\textwidth]{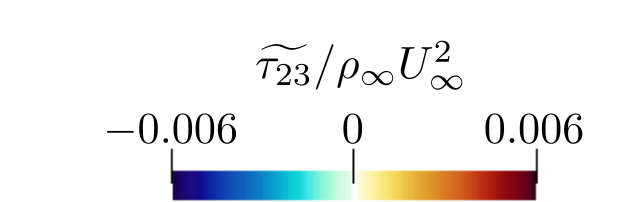}}
     \\
     \setcounter{subfigure}{0}
     \subfloat[]{
     \includegraphics[width=0.22\textwidth]{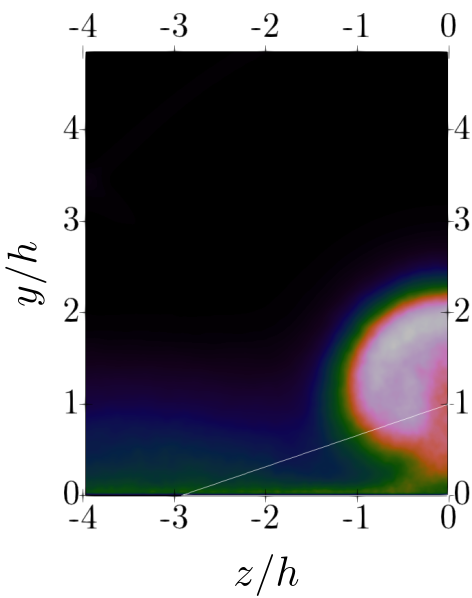}} 
     \,
     \subfloat[]{
     \includegraphics[width=0.22\textwidth]{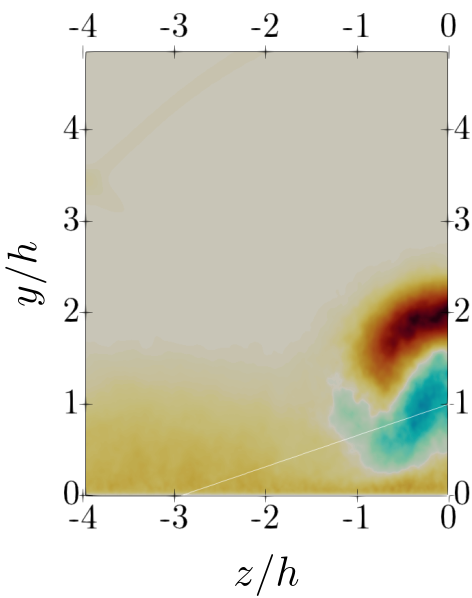}}
     \,
     \subfloat[]{
     \includegraphics[width=0.22\textwidth]{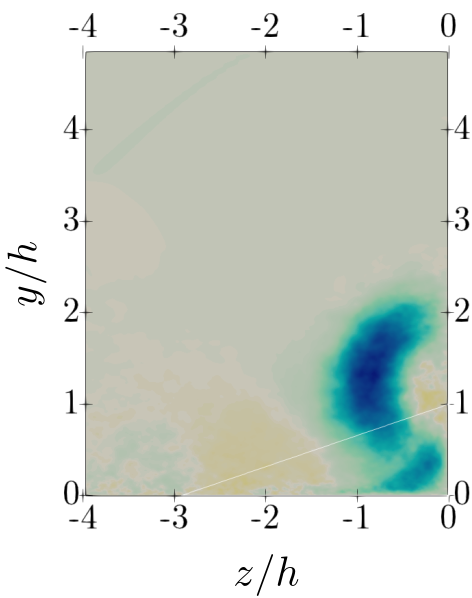}}
     \,
     \subfloat[]{
     \includegraphics[width=0.22\textwidth]{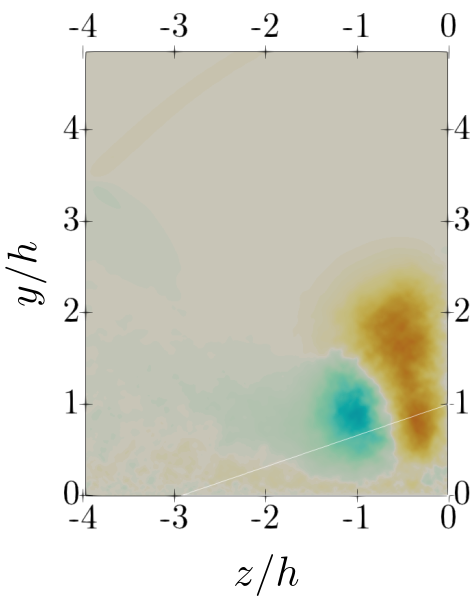}}
     \\
     \setcounter{subfigure}{3}
     \subfloat[]{
     \includegraphics[width=0.22\textwidth]{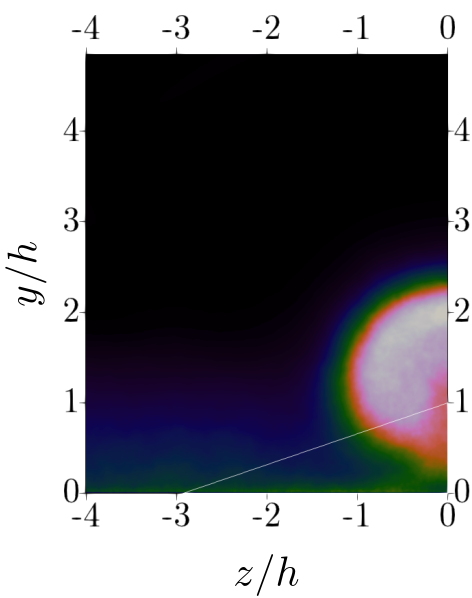}} 
     \,
     \subfloat[]{
     \includegraphics[width=0.22\textwidth]{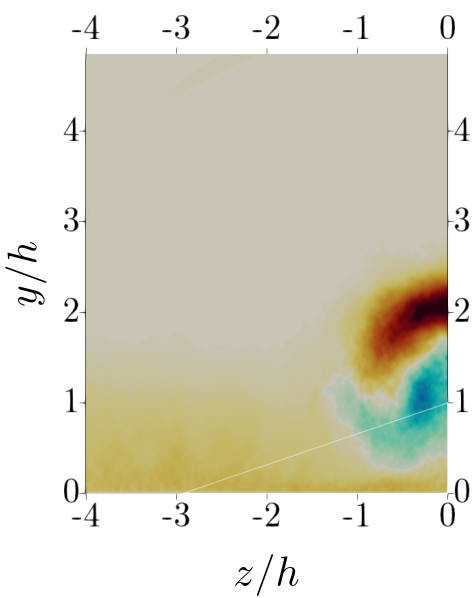}}
     \,
     \subfloat[]{
     \includegraphics[width=0.22\textwidth]{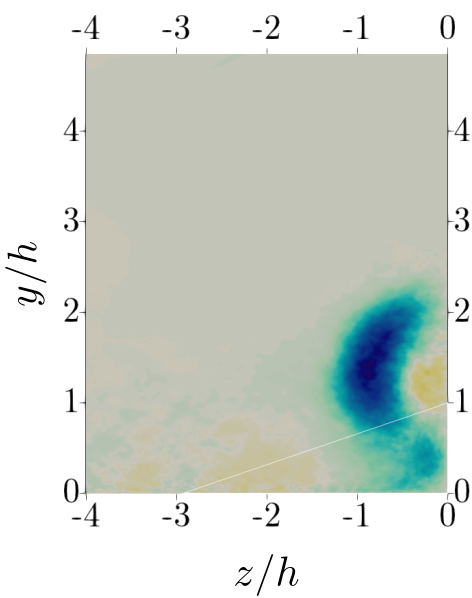}}
     \,
     \subfloat[]{
     \includegraphics[width=0.22\textwidth]{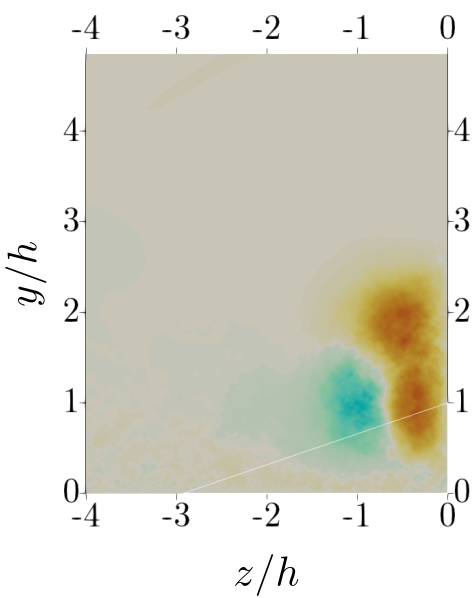}}
     \\
     \setcounter{subfigure}{3}
     \subfloat[]{
     \includegraphics[width=0.22\textwidth]{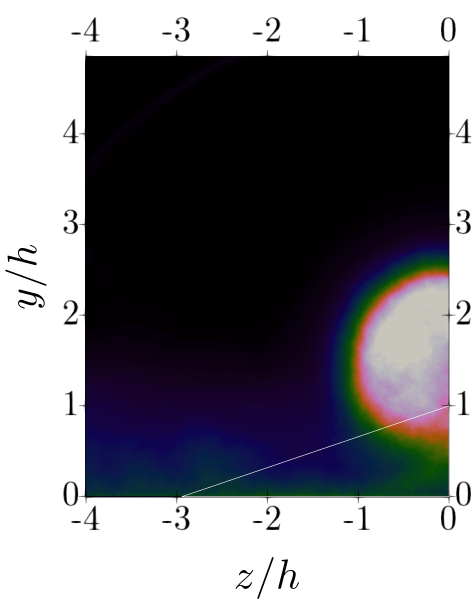}} 
     \,
     \subfloat[]{
     \includegraphics[width=0.22\textwidth]{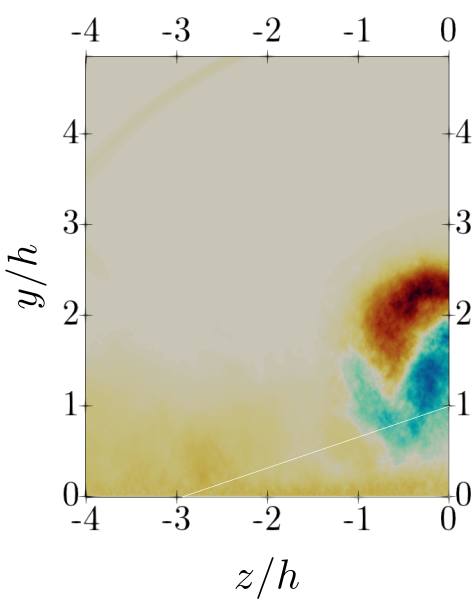}}
     \,
     \subfloat[]{
     \includegraphics[width=0.22\textwidth]{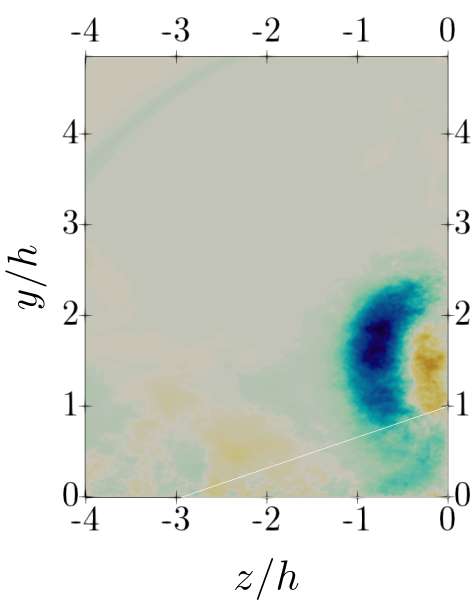}}
     \,
     \subfloat[]{
     \includegraphics[width=0.22\textwidth]{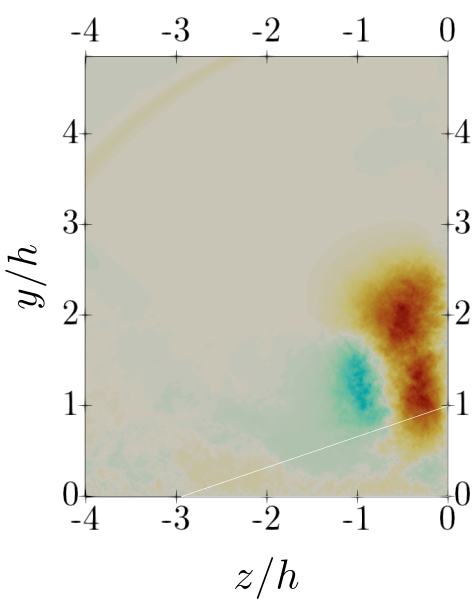}}
     \caption{Turbulent kinetic energy $\widetilde{k}/\rho_\infty U_\infty^2$ (first column) 
     $\widetilde{\tau_{12}}/\rho_\infty U_\infty^2$ (second column), $\widetilde{\tau_{13}}/\rho_\infty U_\infty^2$ (third column), 
     $\widetilde{\tau_{23}}/\rho_\infty U_\infty^2$ (fourth column), 
     at $x/h = 8$ for low (first row), intermediate (second row), 
     high (third row) Reynolds number. }
     \label{fig:tke_tau12_reynolds_comparison}
\end{figure*}

Once the primary vortices become parallel and the wake starts to lift up, 
the low-momentum region spreads and the intensity of the wake deficit 
progressively decays. In order to describe the evolution of the wake far from 
local effects due to the presence of the ramp, we report in figure \ref{fig:far_field}
the behaviour of the Favre-averaged Mach number ($\widetilde{M}=\sqrt{\tilde{u}^2 + \tilde{v}^2+ \tilde{w}^2}/\sqrt{\gamma R \tilde{T}}\,$),
and of the Favre-averaged Reynolds stress components $\widetilde{\tau_{11}} = - \overline{\rho u'' u''}$,
$\widetilde{\tau_{22}} = - \overline{\rho v'' v''}$,
$\widetilde{\tau_{33}} = - \overline{\rho w'' w''}$,
$\widetilde{\tau_{12}} = - \overline{\rho u'' v''}$. 
The figure shows the above-mentioned quantities on three yz slices 
at $x/h = 4$, $8$, and $16$, for the intermediate Reynolds case only. 
Since we observed decay properties similar to the $\widetilde{\tau_{12}}$ component
for the other two transversal components, $\widetilde{\tau_{13}}$ and 
$\widetilde{\tau_{23}}$, we do not report them in this figure
and we report only their distributions at $x/h = 8$ (figure~\ref{fig:tke_tau12_reynolds_comparison}).

The distribution of the Mach number highlights the wake region behind the microramp, 
which is initially characterised by subsonic conditions, given the lower velocity -- and the higher 
temperature -- experienced by the momentum deficit area. 
In fact, the fluid in the wake is the one captured directly 
(fluid from the top of the ramp captured by the side vortices) or indirectly 
(fluid entrained in the near-wall region from outside of the primary vortices) 
by the \gls{mvg} flow structures. 
In both cases, fluid inside the wake was previously close to the wall.
However, as a consequence of the imposed adiabatic wall conditions, 
the aerodynamic heating raises the temperature of this fluid, 
which thus results in a wake hotter than the corresponding undisturbed flow.
The sonic line indicated in white also suggests the thinning of the boundary layer below the  
wake, due to the energising exchange of momentum promoted by the microramp. 
Proceeding downstream, the entrainment of colder and faster flow from outside induces an increase  
in the Mach number, which tends to the undisturbed conditions far from the ramp. 
In the wake region, even after $15\,h$ from the trailing edge, we can see that the subsonic region --
and the wake in general -- is not circular and homogeneous inside. As a matter of fact, 
the lift-up close to the symmetry plane and the external flow summoning due to the 
primary vortices makes the magnitude of the Mach number at the centre higher than 
elsewhere in the wake. 

The principal Reynolds stress components highlight instead that the wake region, 
with its shear layers and vortical structures, is characterised as expected
by an increased turbulent kinetic energy that decays as soon as the wake spreads. 
Given the larger exposure to the incoming flow, the magnitude of the 
streamwise velocity fluctuations is more significant in the top part of the wake. 
At the same time, vertical fluctuations fed by the lift-up of the primary vortex pair
decay slowly and are more important at the symmetry plane, where the upwash is maximum. 
Other regions of intense turbulent activity are noticeable in the field, especially at $x/h = 4$, 
and represent a trace of the near wake features that rapidly disappears. 
In particular, initially $\widetilde{\tau_{11}}$ exhibits
large values in magnitude also close to the symmetry plane, probably due to the 
observed volutes, with their secondary \gls{kh} vortices inside the wake. 
Moreover, in the lower part at the spanwise centre, 
strong spanwise fluctuations are highlighted by 
negative peak values of $\widetilde{\tau_{33}}$, likely caused by the 
convergence of the primary vortex pair and by the 
last effects of the bottom secondary vortices.

The Reynolds shear stress $\widetilde{\tau_{12}}$ confirms that the top part of the annular shear 
layer is related to the most important turbulent mixing, 
where the maximum value is very close to the experimental measurements of \citet{sun2012three} ($-\overline{u'v'}/U_\infty^2 \approx 0.007$). 
Furthermore, the negative regions in the bottom part of the wake suggest that the bottom 
shear layer is weaker than the top one, in accordance with 
the previous observations in section \ref{sec:self_similarity} about the 
difference in the vertical gradients close and far from the wall.

Figure \ref{fig:tke_tau12_reynolds_comparison} shows instead a comparison 
of the Favre-averaged turbulent kinetic energy 
($\widetilde{k} = - ( \widetilde{\tau_{11}} + \widetilde{\tau_{22}} + \widetilde{\tau_{33}})$) 
and of the transversal shear stress components between the 
three cases at different Reynolds numbers on an intermediate yz plane at $x/h = 8$.
Increasing the Reynolds number does not affect qualitatively the distribution and  
has more of a quantitative effect on the evolution of the far wake. 
In particular, the magnitude of the turbulent kinetic energy and of the transversal turbulent 
stress components in the wake increases with the Reynolds number, indicating an 
increased mixing especially at the external and internal edges of the wake, 
where velocity gradients become stronger.
 
\subsection{Near-wall behaviour} \label{sec:close_wall}
As we demonstrated throughout the previous discussion, the wake behind a microramp 
is highly three-dimensional and, as a result, streamwise and spanwise
modulation of the flow behaviour close to the wall is expected. In order 
to discuss these modifications, we consider in the following the 
distribution on the wall-parallel plane of the added momentum, introduced by \citet{giepman2014flow}, and 
of the difference of the friction coefficient between the controlled and uncontrolled cases.

\subsubsection{Added momentum}
\begin{figure*}
     \centering
     \subfloat{
     \includegraphics[width=0.33\textwidth]{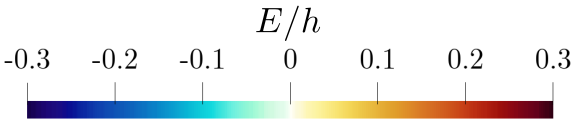} 
     }\\
     \setcounter{subfigure}{0}
     \subfloat[ \label{fig:L_Eadd}]{
     \includegraphics[width=0.9\textwidth]{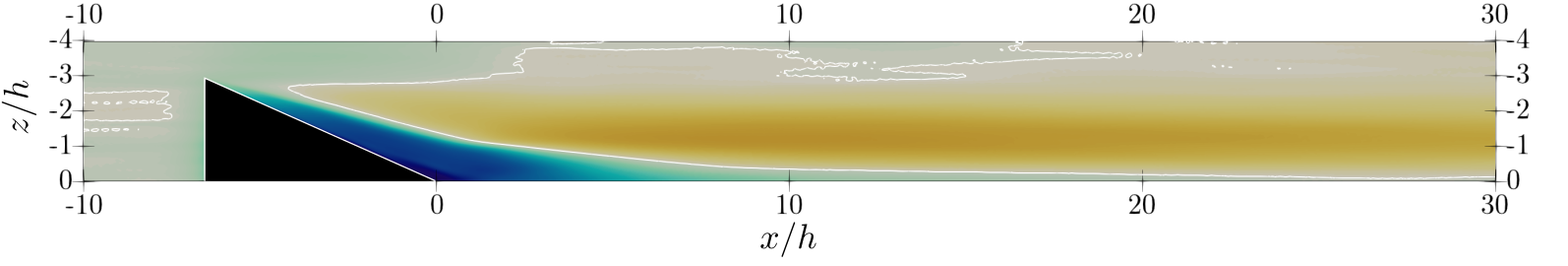} 
     }\\
     \subfloat[ \label{fig:M_Eadd}]{
     \includegraphics[width=0.9\textwidth]{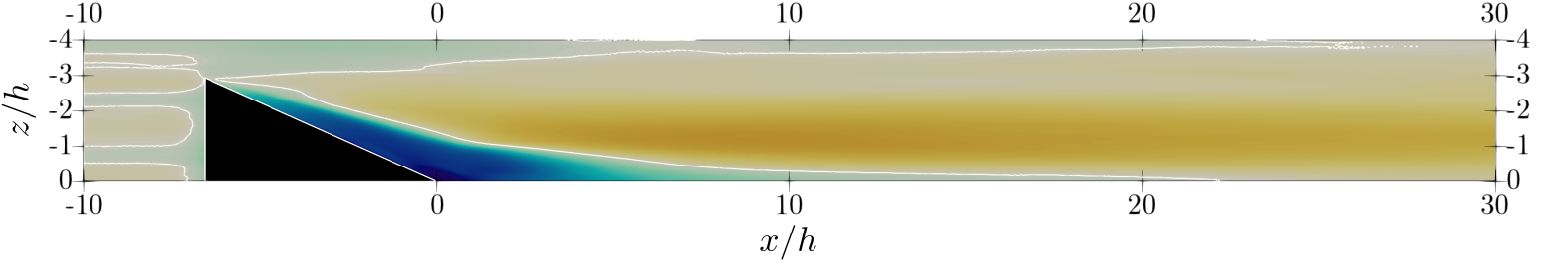} 
     }\\
     \subfloat[\label{fig:H_Eadd}]{
     \includegraphics[width=0.9\textwidth]{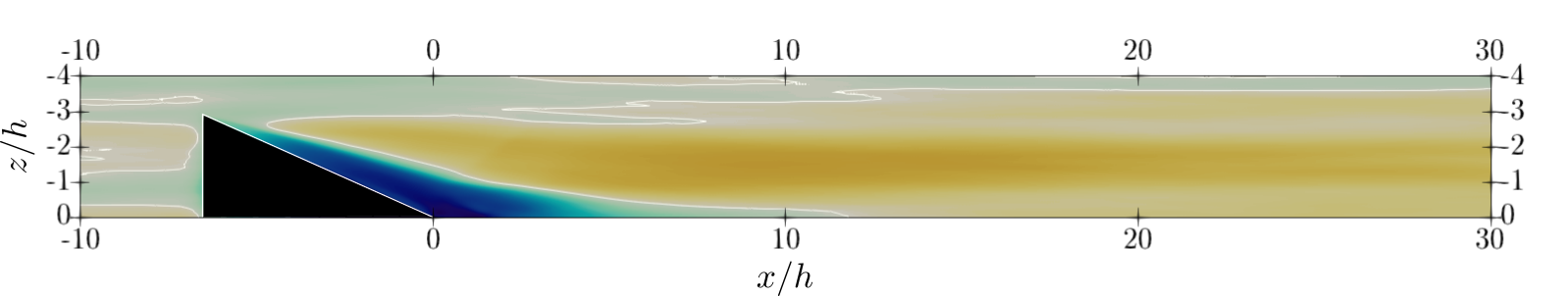} 
     }
     \caption{Normalised added momentum in the wall-parallel plane. 
     White lines indicate the points for which $E/h = 0$.
     }
     \label{fig:E_add_contours}
\end{figure*}
\begin{figure*}
     \centering
     \subfloat[ \label{fig:Eadd_span}]{
     \includegraphics[width=0.47\textwidth]{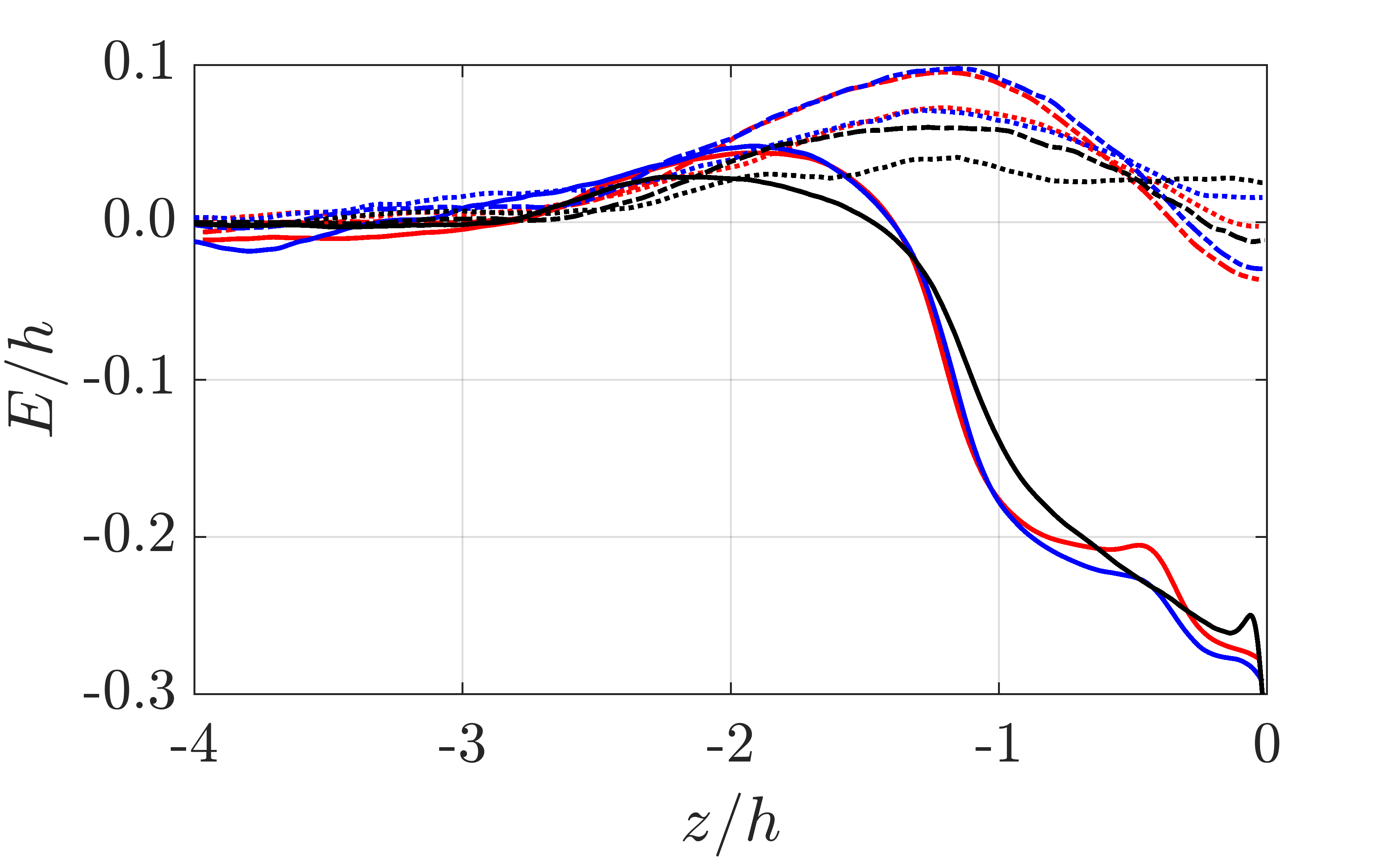} 
     }
     \subfloat[ \label{fig:Eadd_stream}]{
     \includegraphics[width=0.485\textwidth]{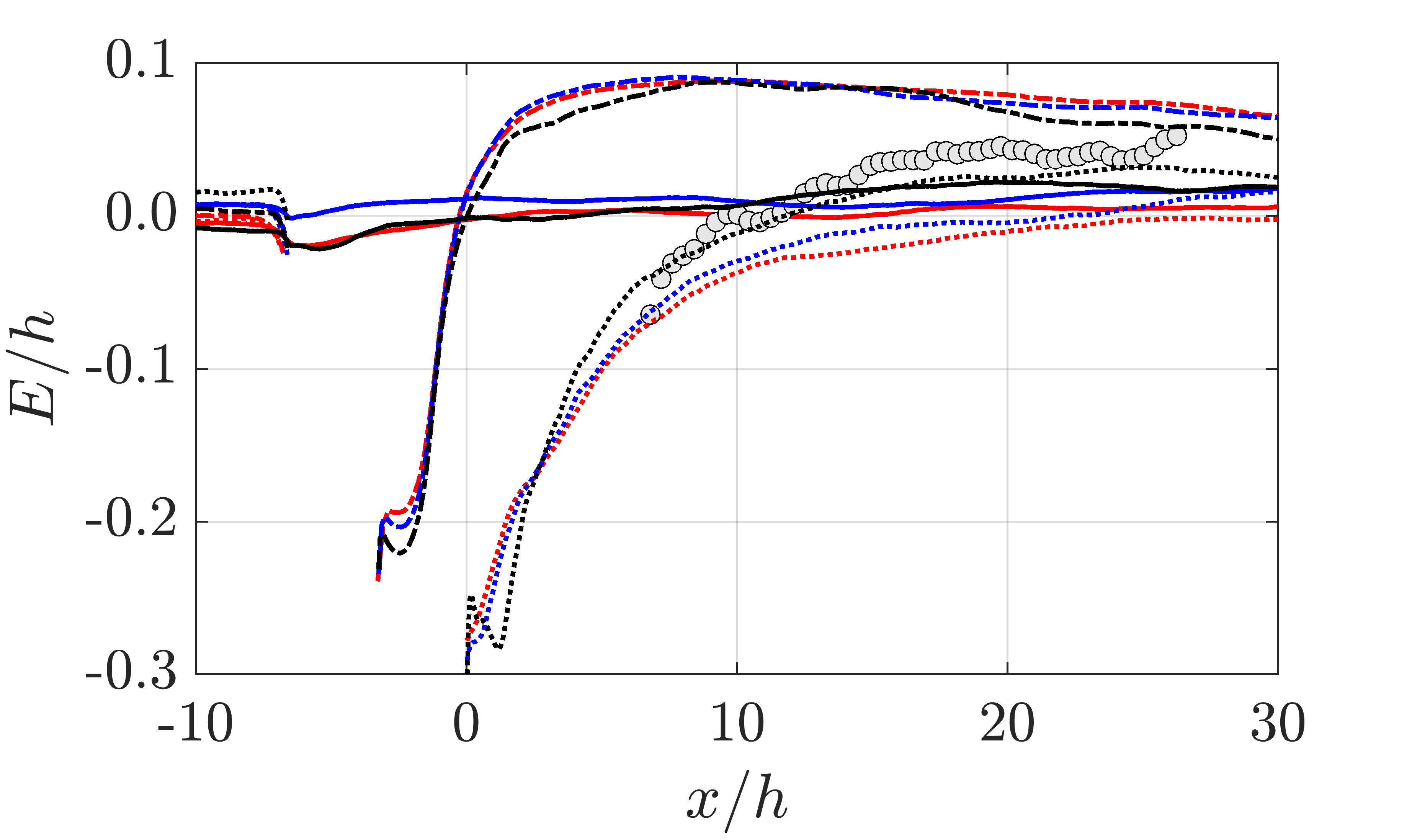} 
     }
     \caption{
     Normalised added momentum distribution (a) along the spanwise direction at $x/h = 0$ (--), $10$ (-.-), $30$ ($\cdots$); (b) along the streamwise direction at $z/h = -b/h$ (--), $-0.5 b/h$ (-.-), $0$ ($\cdots$), with $b$ being the half of the spanwise extent of the microramp. Circles denote experiments by \citet{giepman2016mach} at $M_\infty = 2.0$ and 
     $Re_\theta = 6.7 \times 10^4$.
     }
     \label{fig:E_add_lines}
\end{figure*}

To evaluate the efficiency of the momentum transfer towards the near-wall 
region induced by the presence of the microramp, 
\cite{giepman2014flow, giepman2015effects, giepman2016mach} introduced 
the so-called \emph{added momentum}. By including also the density 
in the definition, the (compressible) added momentum is defined as:
\begin{equation}
    E = \int_{0}^{0.43\,\delta_{99}} \frac{\overline{\rho u^2} - \overline{\rho u^2}_{BL}}{\rho_\infty U_\infty^2} \,\mathrm{d} y
\end{equation}
where the subscript $BL$ indicates the momentum of the uncontrolled 
boundary layer. An upper limit of integration equal to $0.43\,\delta_{99}$ 
is used, since the separation bubble of the downstream \gls{sbli} in the experiments of the cited works was found to be mostly sensitive to the momentum 
of the lower 43\% boundary layer. 
Similarly to the other wake features, $E$ is typically scaled by the ramp height. 

Contrary to the previous experimental works, 
which were limited to the symmetry plane only, our numerical database 
makes it possible to examine
the distribution of the added momentum on the entire 
surface of the bottom wall, allowing us to characterise also the spanwise 
behaviour of the momentum transfer towards the wall. 

At the sides of the ramp and close to the symmetry plane, 
the upward motion induced by the primary and secondary vortices 
removes momentum from the region close to the wall, thus generating 
areas with negative $E$. However, even at the symmetry plane, 
the added momentum ends up being larger than zero 
not far from the trailing edge of the ramp 
(only marginally for the case at low Reynolds). 
On the other hand, approximately below the primary vortices, 
a noticeable region of positive $E$ indicates that 
the helical motion related to the primary vortices 
brings fresh fluid with high momentum from the sides of the wake 
towards the plane of symmetry, from higher wall-normal locations 
to the bottom portion of the domain. 

In order to better capture the quantitative differences among the three cases, 
figure \ref{fig:E_add_lines} shows the normalised added momentum for the 
three Reynolds number along the spanwise direction at $x/h = 0$, $10$, and $30$, and 
along the streamwise direction at $z/h = -b/h$, $-0.5 b/h$, and $0$.
Although the results agree well with the experimental results of \citet{giepman2016mach}
at the symmetry plane, confirming the trend for increasing Reynolds number, 
the plots highlight again that the only added momentum at the spanwise centre is 
insufficient to portray appropriately the overall transfer to the wall 
induced by the microramp wake. In fact, we can see that in the far wake,
$E/h$ is limited at the symmetry plane but, at the same time, is  
high below the cores of the primary vortex pair, 
where the most important addition of momentum takes place. 
Concerning the differences among the cases, 
a clear trend for increasing Reynolds number 
is not easily recognisable. 
We showed in the previous sections that 
increasing the Reynolds number means dealing with stronger vortical structures, 
which lift up the wake faster and move -- reasonably -- more fluid. As a consequence, 
the region with negative added momentum at the spanwise centre shrinks progressively 
as Reynolds number is increased. However, contrary to what was expected, 
the peak added momentum at high Reynolds is smaller than the other cases.
This result could be related to the fact that if the low-momentum region rises faster
because of the increased upwash, 
the azimuthal motion of the primary vortices affects only fluid that is farther from the wall, 
and so, despite the circulation being larger, the resulting added momentum is smaller. 
Moreover, since for different Reynolds numbers, the primary vortices have different trajectories 
 (vortices close to the side walls of the ramp and more abrupt alignment to the streamwise direction 
for larger Reynolds numbers) and the wakes have different streamwise evolutions, 
comparing the added momentum for the three conditions at fixed $x/h$ and $z/h$ can be misleading. 
For this reasons, further analysis in future work is needed to better understand 
the balance and scaling of the different contrasting aspects.

\subsubsection{Relative friction coefficient}
\begin{figure*}
     \centering
     \subfloat{
     \includegraphics[width=0.33\textwidth]{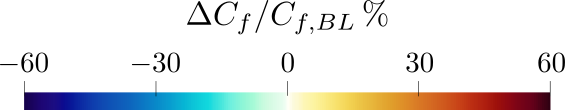} 
     }\\
     \setcounter{subfigure}{0}
     \subfloat[ \label{fig:Ldeltarel_friction}]{
     \includegraphics[width=0.9\textwidth]{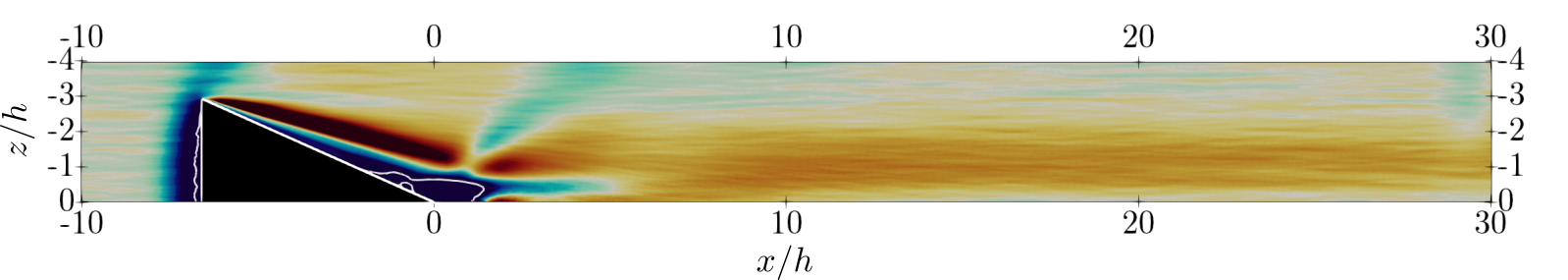} 
     }\\
     \subfloat[ \label{fig:Mdeltarel_friction}]{
     \includegraphics[width=0.9\textwidth]{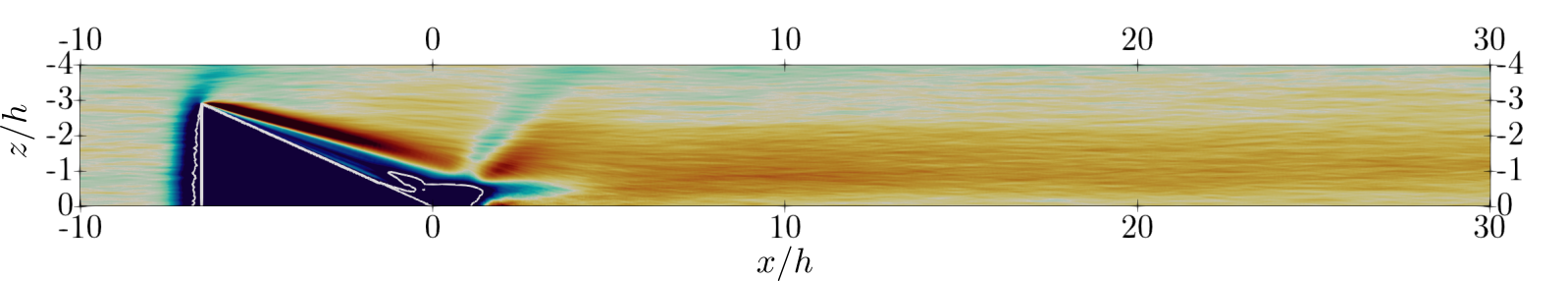} 
     }\\
     \subfloat[ \label{fig:Hdeltarel_friction}]{
     \includegraphics[width=0.9\textwidth]{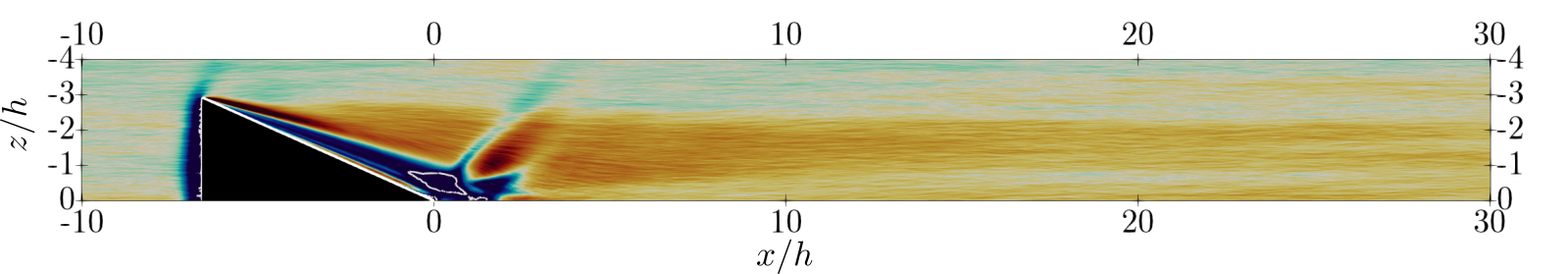} 
     }
     \caption{Percentage difference of the skin friction coefficient between the controlled and uncontrolled boundary layer. White lines indicate the points for which $(\partial \tilde{u}/\partial y)_w = 0$.}
     \label{fig:delta_cf_rel}
\end{figure*}
%

Figure~\ref{fig:delta_cf_rel} reports the percentage relative difference
of the Favre-averaged skin friction coefficient at the wall of the flat plate 
($C_f = 2 \tilde{\tau}_w/\rho_\infty U_\infty^2$) 
between the controlled and the uncontrolled cases ($\Delta C_f = C_f - C_{f,BL}$). 
In particular, since the spanwise friction component 
$\tilde{\tau}_z = (\partial \tilde{w}/\partial y)_w$ 
was far smaller in magnitude than the streamwise component 
$\tilde{\tau}_x = (\partial \tilde{u}/\partial y)_w$, 
we consider in the following the approximation for the controlled case that 
$\tilde{\tau}_w \approx \tilde{\tau}_x$.
In the $C_f$ contours, we find many flow features also observed 
in the wake analysis, confirming the insights from the added momentum. 
The primary and secondary vortices at the sides of the ramp are 
highlighted by regions of enhanced and reduced friction, whereas
the trailing edge shock is visible 
in the transversal low-friction region right after the trailing edge of the ramp. 
After the near wake, the entrainment of high-momentum fluid  
induced by the primary vortices is responsible of a
prolonged region of higher friction, which persists even 30$h$
downstream the microramp. 
The contours point out the significant streamwise and spanwise 
variation of the boundary layer properties, which plays a major role 
in the \gls{sbli} control mechanism. In this regard, 
\citet{babinsky2009microramp} showed that an important effect of microramp arrays
upstream of the interaction region in 2D \glspl{sbli} consists  
in the modulation of the separated flow area, which is broken up in 
multiple individual pockets of three-dimensional reverse flow, 
separated by regions of attached flow. 

Finally, since the wake rises faster at higher Reynolds number, 
we note that increasing $Re_\tau$ the relative skin friction
increase becomes weaker in the far wake. 
\section{Conclusions}
\label{sec:conclusions}

Microramp \glspl{mvg} are promising devices proposed to passively control flow separation and mitigate \gls{sbli} adverse effects.
%
This work presents a campaign of \glspl{dns} which aims at examining the effects of a microramp immersed in a supersonic turbulent boundary layer for different Reynolds numbers, approaching the values achieved in experiments. Three conditions are considered, with friction Reynolds number $Re_\tau$ equal to 500, 1000, and 2000, fixed free-stream Mach number $M_\infty$ equal to $2$, and fixed shape and relative height of the ramp with respect to the uncontrolled boundary layer thickness ($h/\delta_{99} = 2/3$).

The high fidelity of the \gls{dns} database and the opportunity to access completely the flow field make it possible to observe in detail the instantaneous organisation of the turbulent structures, disclosing the formation of primary, secondary, and almost-toroidal vortices in the wake, but also of the system of shock waves interacting with the wake, hardly addressed in the literature. We observe in particular two shocks nearly conical in shape, which originate at the foot and trailing edge of the microramp and leave a visible signature on the flow close to the wall, as revealed by the distributions of the skin friction coefficient and of the added momentum.

In order to describe the evolution of the boundary layer and that of the wake, and to assess the control efficiency of the device, we take into consideration the streamwise development of the streamwise and wall-normal velocity components at the symmetry plane. We find important Reynolds number effects in the range considered, although the highest Reynolds number case shows good agreement with recent experiments at $\Rey_\tau\approx 5000$, confirming the relevance of our high-Reynolds number data set.

To complete the picture of the wake development, 
we then assess the self-similarity of the velocity profiles in the far wake,
following the work of \citet{sun2020wake}. The results 
show that self-similarity is not strictly recovered close 
to the wall, where the bottom secondary vortices have a prolonged influence
on the streamwise velocity. Similarly, self-similarity breaks
down in the outer part of the wake, 
where shock waves leave their imprint.

The three-dimensional distribution of the Favre-averaged flow statistics 
highlights then the presence of a complex organisation of the vortical structures 
inside the low-momentum region, especially in the near field. 
On the basis of the mean vorticity components, we propose a model 
of the internal structure of the primary vortices close to the trailing edge,
which is fundamental to understand the following development in the far field. 
According to the model, primary vortices are initially convoluted 
at the sides of the ramp and joins naturally the \gls{kh} vortices that 
take place on top of the wake. 
The model is also able to explain the experimental observations of internal 
\gls{kh} vortices inside the wake in wall-parallel planes in \cite{sun2012three}, 
and adds further details to the understanding of the almost-toroidal 
vortices around the low-momentum area. 

We also analyse the Reynolds stress components, which indicate regions of 
intense turbulent activity and highlight the annular shear layer
around the wake. The distribution of the Reynolds stresses
also confirms that the legs of the external \gls{kh} vortices 
do not close at the bottom, but rather join the motion induced 
by the primary vortices inside the wake. 

Finally, we focus on the changes in the boundary layer 
close to the wall, which allows us to assess the spanwise flow modulation
induced by the microramp.
The added momentum, introduced by \citet{giepman2014flow}, 
and the relative difference in skin friction coefficient between the controlled 
and uncontrolled cases, confirm the control mechanism of the microramp, 
which induces two counter-rotating vortices bringing
high-momentum fluid closer to the wall thus increasing the local
skin friction coefficient.

%
Moreover, results allow us to advance a prediction about the control effectiveness of microramps. 
Even though vortical structures are those steering momentum transfer, 
shape factors show no sign of recovery in the mean boundary layer properties even 
after $x/h=30$, where the intensity of the vortices is low. 
This observation suggests that the effectiveness in controlling \gls{sbli} 
could be rather insensitive to the position of the microramp, for large relative distances. 
Hence, we expect good control performance in a fairly wide range of shock positions. 
Simulations in the presence of \gls{sbli} are currently ongoing to verify this conjecture. 


As anticipated above, almost all the quantities considered in this study 
reveal a non-negligible dependence on the Reynolds number. 
To sum up, according to the results obtained, the main effects due 
to low Reynolds number observed are:
\begin{enumerate}
    \item turbulent structures have a less coherent roll-up around the wake deficit, 
    leading to weaker almost-toroidal vortices.
    \item Wake features do not scale with the geometric size of the ramp 
    and thus depend also on the incoming flow properties.
    If one uses the height $h$ to scale typical wake features, results indicate a 
    slower and weaker development of the wake for lower Reynolds numbers. 
    \item The self-similarity of the wake, especially in its top portion, 
    is less strict. 
    \item Smaller momentum captured by the microramp at lower Reynolds number means smaller 
    circulation of the side vortices, which leads to a reduction of the induced upwash and thus a decrease in the lift-up of the wake.
    \item Side vortices become parallel more gradually. At higher Reynolds number, 
    the primary vortices remain close to the side walls of the ramp longer, 
    the separation near the trailing edge of the ramp starts more downstream and is 
    slightly more prolonged in the streamwise direction. 
    \item Weaker vortices at low Reynolds number, with a slower wake development, lead also 
    to a slower transfer of momentum towards the wall, which is visible in the smaller added momentum especially at the symmetry plane. 
    At the same time, weaker vortices means slower lift-up, 
    which in the end prolongs the region for which the wake is sufficiently close to the wall 
    to influence the increase in momentum transfer and in skin friction behind the ramp.
\end{enumerate}

%

\backsection[Acknowledgements]{We acknowledge CINECA Casalecchio di Reno (Italy) for providing us with the computational resources 
required by this work.}

\backsection[Funding]{This research received no specific grant from any funding agency, commercial or not-for-profit sectors.}

\backsection[Declaration of interests]{The authors report no conflict of interest.}

\backsection[Data availability statement]{The full data set of the DNS simulations is on the order of several thousands of gigabytes. By contacting the authors, a smaller subset can be made available.}

\backsection[Author ORCID]{
G. Della Posta, \url{https://orcid.org/0000-0001-5516-9338}; 
M. Blandino, \url{https://orcid.org/0009-0008-3478-1446}; 
D. Modesti, \url{https://orcid.org/0000-0003-2214-5799}; 
F. Salvadore, \url{https://orcid.org/0000-0002-1829-3388}; 
M. Bernardini, \url{https://orcid.org/0000-0001-5975-3734}
}

\bibliographystyle{jfm}
\bibliography{ms}


\end{document}